\documentclass[a4paper,fleqn,usenatbib]{mnras}
\pdfoutput=1
\usepackage{hyperref}
\usepackage{amsmath}
\usepackage{graphicx}
\usepackage{color}
\usepackage{epstopdf}
\usepackage{txfonts,amssymb}
\usepackage{journals}
\usepackage{txfonts}
\usepackage[T1]{fontenc}
\usepackage{ae,aecompl}

\newcommand{\be}{\begin{equation}}
\newcommand{\ee}{\end{equation}}
\newcommand{\bea}{\begin{eqnarray}}
\newcommand{\eea}{\end{eqnarray}}
\newcommand{\bfig}{\begin{figure}}
\newcommand{\efig}{\end{figure}}
\newcommand{\bfigs}{\begin{figure*}}
\newcommand{\efigs}{\end{figure*}}
\newcommand{\bt}{\begin{table}}
\newcommand{\et}{\end{table}}
\renewcommand{\vec}[1]{ {\bf #1} }

\def\mnras{MNRAS}
\def\apj{ApJ}
\def\apjl{ApJL}
\def\apjs{ApJ Sup.}
\def\aj{AJ}
\def\aap{AAP}

\def\nat{Nature}

\newcommand{\grale}{{\tt{Grale}}}

\definecolor{mylabelcolor}{rgb}{0.5,1,1}

\title[The Frontier Fields Lens Modeling Comparison Project]{The Frontier Fields Lens Modeling Comparison Project}

\author[M. Meneghetti et al.]{M. Meneghetti$^{1,2}$\thanks{E-mail: massimo.meneghetti@oabo.inaf.it}, 
P. Natarajan$^3$,
D. Coe$^4$ ,
E. Contini$^5$,
G. De Lucia$^6$, 
C. Giocoli$^7$,
 \newauthor A. Acebron$^7$,
S. Borgani$^{19,6,20}$,
M. Bradac$^8$,
J. M. Diego$^9$,
A. Hoag$^8$,
M. Ishigaki$^{10,17}$,
\newauthor  T. L. Johnson$^{11}$,
E. Jullo$^7$,
R. Kawamata$^{18}$,
D. Lam$^{12}$,
 M. Limousin$^{7}$,
 J. Liesenborgs$^{22}$,
\newauthor M. Oguri$^{10,15,16}$,
K. Sebesta$^{13}$,
 K. Sharon$^{11}$,
 L. L. R. Williams$^{13}$,
 A. Zitrin$^{14,21}$
  \\ \\ $^{1}$Osservatorio Astronomico di Bologna, INAF, via Ranzani 1,
  40127, Bologna, Italy \\
  $^{2}$INFN, Sezione di Bologna, viale Berti Pichat 6/2,
  40127, Bologna, Italy \\
  $^3$Department of Astronomy, Yale University, New Haven, CT 06511, USA \\
  $^4$Space Telescope Science Institute, Baltimore, MD, USA \\
  $^5$Purple Mountain Observatory, the Partner Group of MPI f\"ur Astronomie, 2 West Beijing Road, Nanjing 210008, China  \\
  $^6$INAF - Astronomical Observatory of Trieste, via G.B. Tiepolo 11, I-34143 Trieste, Italy \\
  $^7$Aix Marseille Universit\'e, CNRS, LAM (Laboratoire d'Astrophysique de Marseille) UMR 7326, 13388 Marseille, France \\
  $^8$Department of Physics, University of California, One Shields Ave, Davis, CA 95616, USA \\
  $^9$IFCA, Instituto de Fisica de Cantabria (UC-CSIC), Av. de Los Castros s/n, 39005 Santander, Spain \\
  $^{10}$Department of Physics, Graduate School of Science, The University of Tokyo, 7-3-1 Hongo, Bunkyo-ku, Tokyo 113-0033, Japan \\
  $^{11}$University of Michigan, Department of Astronomy, 1085 South University Avenue, Ann Arbor, MI 48109, USA \\
  $^{12}$Department of Physics, The University of Hong Kong, 0000-0002-6536-5575, Pokfulam Road, Hong Kong \\
  $^{13}$School of Physics and Astronomy, University of Min- nesota, 116 Church Street SE, Minneapolis, MN 55455, USA \\
  $^{14}$Cahill Center for Astronomy and Astrophysics, California Institute of Technology, MC 249-17, Pasadena, CA 91125, USA \\
  $^{15}$Research Center for the Early Universe, The University of Tokyo, 7-3-1 Hongo, Bunkyo-ku, Tokyo 113-0033, Japan \\
  $^{16}$Kavli Institute for the Physics and Mathematics of the Universe (Kavli IPMU, WPI), University of Tokyo, Chiba 277-8582, Japan \\
  $^{17}$Institute for Cosmic Ray Research, The University of Tokyo, 5-1-5 Kashiwanoha, Kashiwa, Chiba 277-8582, Japan \\
  $^{18}$Department of Astronomy, Graduate School of Science, The University of Tokyo, 7-3-1 Hongo, Bunkyo-ku, Tokyo 113-0033, Japan\\
  $^{19}$Astronomy Unit, Department of Physics, University of Trieste, via Tiepolo 11, I-34131 Trieste, Italy \\
  $^{20}$INFN-National Institute for Nuclear Physics, Via Valerio 2, I-34127 Trieste, Italy \\
  $^{21}$Hubble Fellow \\
  $^{22}$Expertisecentrum voor Digitale Media, Universiteit Hasselt, Wetenschapspark 2, B-3590, Diepenbeek, Belgium 
  }
  
\date{\today}
\pubyear{2016}

\begin{document}
\label{firstpage}
\pagerange{\pageref{firstpage}--\pageref{lastpage}}
\maketitle

\begin{abstract}
Gravitational lensing by clusters of galaxies offers a powerful probe of their structure and mass distribution. Deriving a lens magnification map for a galaxy cluster is a classic inversion problem and many methods have been developed over the past two decades to solve it. Several research groups have developed techniques independently to map the predominantly dark matter distribution in cluster lenses. While these methods have all provided remarkably high precision mass maps, particularly with exquisite imaging data from the Hubble Space Telescope (HST), the reconstructions themselves have never been directly compared. In this paper, we report the results of comparing various independent lens modeling techniques employed by individual research groups in the community. Here we present for the first time a detailed and robust comparison of methodologies for fidelity, accuracy and precision. For this collaborative exercise, the lens modeling community was provided simulated cluster images -- of two clusters {\em Ares} and {\em Hera} -- that mimic the depth and resolution of the ongoing HST Frontier Fields. The results of the submitted reconstructions with the un-blinded true mass profile of these two clusters are presented here. Parametric, free-form and hybrid techniques have been deployed by the participating groups and we detail the strengths and trade-offs in accuracy and systematics that arise for each methodology. We note in conclusion that lensing reconstruction methods produce  reliable mass distributions that enable the use of clusters as extremely valuable astrophysical laboratories and cosmological probes.
\end{abstract}

\begin{keywords}
clusters of galaxies --- gravitational lensing -- mass distributions -- simulations
\end{keywords}

\section{Introduction}

Gravitational lensing has become an increasingly popular method to constrain the matter distribution in clusters. Strong lensing, as it turns out, is particularly suited to probing the dense central regions of clusters. Constraining the structure of the cluster cores and their density profiles is critical to our understanding of structure formation; probing the nature of dark matter and fully comprehending the interplay between baryons and dark matter. Lensing by massive clusters has proved to be an invaluable tool to  study their properties, in particular the detailed dark matter distribution within the cluster, as well as the faint, distant background population of galaxies that they bring into view. The magnification provided by lensing therefore affords the determination of the luminosity function of these high-redshift sources down to faint luminosities, thus helping inventory and identify galaxies that might have re-ionized the universe \citep{2012MNRAS.tmpL.474V,2014ApJ...783L..12V,2014ApJ...793..115B,2015ApJ...802L..19R,2015ApJ...811..140B,2015A&A...576A.116V,2016arXiv160200688V,2016ApJ...823L..14H,2016arXiv160406799L}. 

Over the past two decades the Hubble Space Telescope (HST) has revolutionized the study of cluster lenses; and, with the deployment of ever more sensitive cameras from the Wide-Field-Planetary-Camera2 (WFPC2) to the Advanced-Camera-for-Surveys (ACS), the data have become exquisite in terms of resolution. By 2005, mass distributions derived from lensing data were available for about 30 clusters. More recently, galaxy clusters were the primary targets of two multi-cycle treasury programs of the Hubble Space Telescope (HST) aiming at finding signatures of strong gravitational lensing in their cores. These are the ``Cluster Lensing And Supernova survey with Hubble" \citep[CLASH, PI M. Postman (GO 12065); see][]{2012ApJS..199...25P}) and the ongoing Frontier Fields Initiative (FFI, PI: Lotz). 

As part of the Frontier Fields program, HST is currently collecting data of unprecedented depth on fields that harbor six massive clusters that act as powerful gravitational lenses. This program utilizes orbits under the Director's Discretionary (DD) observing time. The FFI is a revolutionary deep field observing program aimed at peering deeper into the universe than ever before to not only help understand better these dramatic lenses and their properties, but also simultaneously bring into view faint, distant background galaxies that would otherwise remain unseen without the magnification provided by the foreground lens. These high redshift sources that can be accessed due to gravitational lensing provide a first glimpse likely of the earliest galaxies to have formed in the universe, and offer a preview of coming attractions that await unveiling by the upcoming James Webb Space Telescope. These Frontier Fields uniquely combine the power of HST with that of nature's gravitational telescopes -- the high-magnifications produced by these massive clusters of galaxies. 

Utilizing both the Wide Field Camera 3 (WFC3) and ACS in parallel in this current program, HST has been producing the deepest observations of clusters and the background galaxies that they lens; as well as observations of flanking blank fields that are located near these selected clusters. These images have revealed the presence of distant galaxy populations that are $\sim 10-100$ times fainter than any previously observed \citep{2016arXiv160406799L}. The magnifying power of these clusters is proving to be invaluable in helping improve our statistical understanding of early galaxies that are likely responsible for the re-ionization of the universe, and are providing unprecedented measurements of the spatial distribution of dark matter within massive clusters. These six clusters span the redshift range $z = 0.3 - 0.55$. The program devotes 140 orbits to each cluster / blank field pair, achieving a limiting AB magnitude of $M_{\rm AB} \approx 28.7-29$ mag in the optical (ACS) and near-infrared (WFC3) bands. 

The fundamental ingredient for exploiting the science outlined above is the construction of robust and reliable lens models. The ongoing FFI is an unprecedented test-bed for lens modeling techniques. Given the depth of these HST observations, hundreds of  multiple images, covering a broad redshift range, have been newly unveiled behind each of the observed clusters \citep[][]{2014MNRAS.443.1549J,2015MNRAS.452.1437J,2015ApJ...800...38G,Diego15b,wang15,2016ApJ...819..114K,2016arXiv160300505H}. In a rather unique case, even time delay measurements from a serendipitously multiply imaged  supernova ``Refsdal" observed by the GLASS team \citep{2015ApJ...812..114T} in the FFI cluster MACSJ1149.5+2223 became available for testing and refining the lens models \citep{2015Sci...347.1123K,2016ApJ...817...60T,2016ApJ...820...50R}. Most importantly, FFI data were made publicly available immediately. Five teams were contracted by STScI to produce gravitational lensing models for all six Frontier Fields clusters to be made available to the astronomical community at large to enable wide use of this incredible data-set. All teams share the latest observational constraints, including positions and redshifts of multiple images\footnote{The redshifts are mainly obtained in the framework of the GLASS and CLASH-VLT programs \citep{2015ApJ...812..114T,2015ApJ...800...38G} and with the intergral field spectrograph MUSE on the VLT \citep[see e.g.][]{2015A&A...574A..11K}}  before working independently to produce lensing models which are also made publicly available.\footnote{https://archive.stsci.edu/prepds/frontier/lensmodels/} Several additional groups have also been working on the data and producing mass models. In short, the whole community of strong lensing modelers has been actively collaborating to maximally exploit the FFI data.  

The process of converting the observed strong lensing constraints into matter distributions is called {\em lens inversion}. Several groups have developed algorithms which perform the lens inversion employing different methodologies and using various combinations of input constraints. These include other tracers of the cluster gravitational potential, such as weak lensing, galaxy kinematics, and the X-ray emission from the Intra-Cluster-Medium \citep[see e.g.,][]{bradac04a,2011A&A...528A..73D,2013ApJ...777...43M,2013ApJ...765...24N,2013ApJ...769...13U,2014ApJ...795..163U,2015ApJ...806....4M}. Over the years, it has become clear that while all methods are equally well motivated, they do not always converge to consistent reconstructions, even when applied to the same lens system \citep[e.g.,][]{2009ApJ...703L.132Z,2009ApJ...707L.163S}. In several cases strong-lensing masses for the same cluster lens were found to be in tension (by a factor 2-3) with other independent measurements, based e.g on the modeling of the X-ray emission by the intra-cluster gas \citep{EB09.1,RI10.1,2014ApJ...794..136D}. The constraints from strong lensing need to be combined and fit simultaneously with stellar kinematic data and with weak lensing measurements \citep{NE11.1} to improve accuracy. Using constraints on the mass profile arising from probes other than lensing also helps break the mass-sheet degeneracy. Finally, in several clusters, lensing data alone seems unable to discriminate between various density profiles \citep{SH08.2}. Therefore, in some clusters the data favors steep inner density profile slopes, while, in others it favors extremely shallow density profiles. This is in contrast with the predictions from the cold-dark-matter paradigm \citep[][but see also \citealt{BA04.1,ME07.2}]{SA05.1,2013ApJ...765...24N} where a universal density profile is expected with minor modification due to the aggregation of baryons in the inner regions. 

In this paper, we challenge these lens inversion methods to reconstruct synthetic lenses with known input mass distributions. The goals of this exercise are twofold. Firstly, we aim to provide concrete feedback to the lens modelers on how they may improve the performance of their codes. And secondly, we aim to provide potential users of the FFI models and the astronomical community at large a sharper, more quantitative view of how robustly specific properties of lenses are recovered and the sources of error that plague each method. Such a comparison with numerical simulations and contrasting of lens mapping methodologies has not been undertaken before. 

The outline of the paper is as follows. In Sect.~\ref{sect:compa}, we outline the lens modeling challenge. In Sect.~\ref{sect:techniques}, we briefly introduce the various lens modeling techniques that were employed by participants in this study. In Sect.~\ref{sect:results}, we discuss the results of the reconstructions. Sect.~\ref{sect:reconmetrics} is dedicated to the detailed comparison of the independent modeling techniques through suitably defined metrics. Finally, in Sect.~\ref{sect:summary}, we summarize the main results of this study and present our conclusions.

\section{The challenge}
\label{sect:compa}

The challenge that we presented to various groups of lens modelers comprised of analyzing simulated  observations of two mock galaxy clusters and producing magnification and mass maps for them. In generating these simulated (mock) clusters, we attempted to reproduce the depth, color, and spatial resolution of HST observations of the FFI cluster images including the gravitational lensing effects. While the comparison of lensing reconstructions of real clusters using the same input observational constraints strongly indicate that currently developed lens inversion techniques are robust  \citep{2015ApJ...800...38G}, the analysis of simulated data involving a large degree of realism where the true underlying mass distribution is known can help the lens reconstruction community to greatly improve their understanding of the modeling systematics. This view of using mocks to calibrate methodologies is widely supported by a number of extensive investigations carried out in the last few years. 

There are multiple advantages to such calibration exercises. First of all, we are able to produce reasonably realistic cluster mass distributions in simulations (although up to some limit) that can be used as lensing clusters. Building on an extensive analysis of N-body/hydrodynamical simulations to improve the knowledge of strong lensing clusters, we have  identified the important properties of the lenses which need to be taken into account during the construction of a lens model: cluster galaxies \citep{ME00.1,ME03.1}, ellipticity and asymmetries \citep{ME03.2}, substructures \citep{ME07.1}, baryonic physics \citep{PU05.1,2012MNRAS.427..533K}, and the dynamical state \citep{TO04.1}. 
In fact, we can simulate the lensing effects of galaxy clusters accounting for all these important properties, using both state-of-the-art hydrodynamical simulations and semi-analytic models.   
Second, we have developed tools to produce mock observations of these simulated lenses. Our image simulator {\tt SkyLens} \citep{2008A&A...482..403M,2010A&A...514A..93M} can mimic observations taken with virtually any telescope, but here we have used it primarily to produce simulations of HST images taken with both the ACS and the WFC3. In a small scale realization of the experiment that we  present here, we applied the lens inversion techniques to a limited number of simulated observations of our mock lenses. By doing so, we highlighted some key limits of the strong lensing methods. For example, we note that strong lensing alone is powerful at constraining the cluster mass within the Einstein radius ($\sim 100$ kpc for a massive cluster) but the addition of further constraints at larger radii are required in order to appropriately measure the shape of the density profiles out to the cluster outskirts \citep{2010A&A...514A..93M,2012NJPh...14e5018R}. In what follows, we describe in detail how we generate the mock data-set for the challenge, and what kind of high-level products were distributed to the participants.

\subsection{Generation of mock cluster lenses}

For the exercise reported here, we generated mass distributions for two massive cluster lenses. These two lenses are generated following substantially different approaches, as outlined below. In order to easily distinguish them, we assigned them names -- {\em Ares} and {\em Hera}. 

\subsubsection{ {\em Ares} }

The mass distribution of the first simulated galaxy cluster, {\em Ares}, is generated using the semi-analytic code {\tt MOKA}\footnote{http://cgiocoli.wordpress.com/research-interests/moka} \citep{2012MNRAS.421.3343G}.  This software package builds up mock galaxy clusters by treating them as being comprised of three components: (i) the main dark matter halo --  assumed to be smooth, triaxial, and well fit with an NFW profile,  (ii)  cluster  members  --  subhaloes, distributed to follow  the main halo and to have a truncated Singular Isothermal Sphere profile \citep{2001ApJ...563....9M} -- and (iii) the brightest cluster galaxy (BCG) modeled with a separate \citet{2001ApJ...563....9M} profile. The axial  ratios, $a/b$  and $a/c$,  of the  main halo  ellipsoid are randomly  drawn   from  the  \citet{JI02.1}   distributions  requiring $abc=1$. 
We note that the observed FFI clusters typically consist of merging sub-clusters that cause them to be particularly efficient and spectacular lenses. 

In the attempt to generate a mass distribution that adequately replicates the complexity of the Frontier Fields clusters, {\em Ares} was produced by combining two large scale mass distributions at $z=0.5$. The two clumps have virial masses $M_1=1.32\times10^{15}h^{-1}M_\odot$ and $M_2=8.8\times 10^{14}h^{-1}M_\odot$ and their centers are separated by $\sim 400\;h^{-1}$ kpc.  In each of the two cases, we start by assigning the same projected ellipticity to the smooth component, to the stellar density and to the subhalo spatial distribution.  This is motivated by the hierarchical clustering   scenario  wherein  the  BCG   and  the substructures are related to the cluster  as a whole and retain memory of  the  directions  of  the  accretion  of  repeated  merging  events
\citep{2004ApJ...611L..73K,2008ApJ...688..254K,2009ApJ...700.1896K,2010MNRAS.404.1490F}. In order to introduce some level of asymmetry, we then added in a small twist to the surface density contours. The degree of twisting adopted reproduces variations of the orientation of iso-surface density contours measured in numerically simulated galaxy clusters \citep[see e.g.][]{ME07.1}. The two large-scale halos combined to create {\em Ares} are nearly aligned. The difference between the position angles of the two clumps is $\sim 21$ degrees. The central region of {\em Ares} contains large baryonic concentrations to mimic the presence of BCGs. We  account for the possible adiabatic contraction of the dark matter  caused  by  the presence of BCGs for Ares \citep[altough several empirical studies find no evidence of adiabatic contraction on these scales, see e.g.][]{2013ApJ...765...24N,2014MNRAS.438.3594D}.  The  adiabatic contraction   as   described   by  \citet{2001ApJ...549L..25K}   for \citet{1990ApJ...356..359H} was implemented. For further details of the {\tt MOKA}  code we  refer to  \citet{2012MNRAS.421.3343G,2012MNRAS.426.1558G}. {\tt MOKA} also takes into account the correlation between assembly history and various halo properties that are expected in CDM: (i) less massive haloes typically tend to be more concentrated than the  more massive ones, and (ii) at fixed mass,  earlier   forming  haloes  are  more  concentrated   and contain fewer substructures. These  recipes have  been implemented in consonance with 
recent results  from numerical simulations.  In  particular, we assume
the \citet{2009ApJ...707..354Z} relation to link  the concentration to mass and the
\citet{2010MNRAS.404..502G}   relation   for   the  subhalo   abundance.    When
substructures are included we define  the smooth mass as $M_{\rm smooth} =
M_{vir} - \sum_i m_{\mathrm{sub},i}$ and its concentration $c_{\rm s}$
are defined such that  the total  (smooth+clumps) mass  density
profile has a concentration $c_{vir}$, equal to that of the total virial mass of the halo. 

Throughout the  paper the quoted masses and concentrations are evaluated at the virial radius, $M_{vir}$ and $c_{vir}$. For these definitions we adopt derivations from the spherical collapse model:
\begin{equation}
M_{vir} = \dfrac{4 \pi}{3} R_{vir}^3 \dfrac{\Delta_{vir}(z)}{\Omega_m(z)}
\Omega_0 \rho_c\,,
\end{equation}
where  $\rho_c=2.77  \times  10^{11}\,  h^2\,  M_{\odot}/\mathrm{Mpc}$
represents     the    critical     density     of    the     Universe,
$\Omega_0=\Omega_m(0)$ is  the matter  density parameter  at  the present
time, $\Delta_{vir}$  is the virial  overdensity \citep{1996MNRAS.282..263E,1998ApJ...495...80B}
and $R_{vir}$ symbolizes the virial  radius of the halo, i.e.
the distance  from the halo  centre that encloses the  desired density
contrast; and:
\begin{equation}
c_{vir} \equiv \dfrac{R_{vir}}{r_s} \;,
\end{equation}
with  $r_s$  the  radius  at   which  the  NFW  profile  approaches  a
logarithmic  slope of $-2$. The concentrations assigned to the two main mass components of {\em Ares} are $c_1=5.39$ and $c_2=5.46$, respectively.

{\em Ares} is generated in a flat $\Lambda$CDM cosmological model with matter density parameter $\Omega_{m,0}=0.272$. The Hubble parameter at the present epoch is $H_0=70.4$ km/s/Mpc.

In the left panels of Fig.~\ref{fig:convmaps}, we show the convergence maps of {\em Ares}, calculated for a source redshift $z_s=9$. The cluster is elongated in the SE-NW direction and contains several massive substructures. Since {\em Ares} was generated using semi-analytical methods, the small scale substructures of its mass distribution are very well resolved, as 
shown in the bottom-left panel. The substructure mass function is shown in the right panel of Fig.~\ref{fig:convprofs}. As expected, this scales as $N(M)\propto M^{-0.8}$, consistent with results of numerical simulations \citep{2010MNRAS.404..502G} of the CDM model. The convergence profile, measured from the center of the most massive clump, is shown in the left panel of Fig.~\ref{fig:convprofs}. 

In the image simulations described later we also include the light emission from cluster members.  {\tt MOKA} populates the dark matter sub-halos with galaxies using the Halo Occupation Distribution (HOD) technique. Stellar masses and $B$-band luminosities are subsequently assigned to each galaxy accordingly to the mass of the dark matter (sub-)halo within which it formed, following \cite{2006MNRAS.371..537W}. The morphological type and the SED of each galaxy is then defined on the basis of the stellar mass so as to reproduce the observed morphology-density and morphology-radius relations in galaxy clusters \citep[e.g.][]{2008ApJ...675L..13V,2010MNRAS.406..121M}.  
  
\subsubsection{{\em Hera}}

The mass distribution of the second galaxy cluster, {\em Hera}, is instead directly derived from a high-resolution N-body simulation of a cluster-sized dark matter halo. More precisely, {\em Hera} is part of the set of simulated clusters presented in \cite{2014MNRAS.438..195P}. The cluster halo was first identified in a low--resolution simulation box with a periodic co-moving size of 1~h$^{-1}$ Gpc for a flat $\Lambda$CDM~
model with present matter density parameter $\Omega_{m,0}=0.24$ and baryon density parameter $\Omega_{b,0}=0.04$. The Hubble constant adopted was $H_0=72$km/s/Mpc and the normalisation of the matter power spectrum $\sigma_8=0.8$. The primordial power spectrum of the density fluctuations is $P(k) \propto k^{n}$ with $n=0.96$. The parent simulation followed 1024$^{3}$ collision-less particles in the box. {\em Hera} was identified at $z=0$ using a standard {\it Friends-of-Friends} (FoF) algorithm, and its Lagrangian region was re-simulated at higher resolution employing the {\it Zoomed Initial Conditions} code \citep[ZIC;][]{1997MNRAS.286..865T}. The resolution is progressively degraded outside this region to save computational time while still providing a correct description of the large--scale tidal field. The Lagrangian region was taken to be large enough to ensure that only
high-resolution particles are present within five virial-radii of the cluster.

The re-simulation was then carried out using the TreePM--SPH {\small
  GADGET--3} code, a newer version of the original {\small GADGET--2}
code by \cite{SP05.1} that adopted a more efficient domain
decomposition to improve the work-load balance. Although, the parent {\em Hera} halo 
exists in several flavors in various simulation runs (several assumptions for the nature of dark matter particles), including several baryonic processes, the simulation used in this paper uses only the version that utilized collisionless dark matter particles. This has allowed us to increase the mass resolution by about an order of magnitude compared to the hydro-dynamical versions of the simulation. The particle mass is $m_{DM}=10^{8}h^{-1}M_\odot$. Therefore,  the virial region of {\em Hera} is resolved with $\sim 10$ million particles, with a total cluster mass of $M=9.4\times10^{14}\;h^{-1}M_\odot$, comparable to that inferred for observed cluster lenses. The redshift of this halo is $z_l=0.507$. During the re-simulation, the Plummer--equivalent co-moving softening length for gravitational force in the high--resolution region is fixed to $\epsilon_{Pl}=2.3 h^{-1}$ kpc physical at $z<2$ while being fixed to $\epsilon_{Pl} = 6.9 h^{-1}$ kpc comoving at higher redshift.

The properties of cluster galaxies used for creating the simulated observations are derived from Semi-Analytic-Methods (SAM) of galaxy formation \citep{2007MNRAS.375....2D}. The process starts by using the algorithm {\small SUBFIND} \citep{SP01.2}  to decompose each FOF group previously found in the simulation into a set of disjoint substructures. These are identified as locally over-dense regions in the density field of the background halo. Only substructures that retain at least 20 bound particles after a gravitational unbinding procedure are considered to be genuine substructures. Merging histories are constructed for all self-bound structures, using the same post-processing algorithm that has been employed for the Millennium Simulation \citep{2006Natur.440.1137S}. The merger-tree is then used to construct a mock catalog of galaxies. The evolution of the galaxy population is described by a modified version of the semi-analytic model presented in \cite{2007MNRAS.375....2D}, that included the implementation of the generation of Intra-Cluster Light described in \cite{2014MNRAS.437.3787C}, given by the combination of \emph{Model Tidal Radius} and  \emph{Merger} channels presented in that paper. 

Note that, even in the case of {\em Hera}, the galaxy positions trace reasonably well the mass.  Several reconstruction methods assume that light traces the mass, a reasonable assumption which is thus satisfied both in {\em Ares} and in {\em Hera}. To increase the level of uncertainty, the galaxy shapes and orientations are chosen to be uncorrelated with the underlying mass distribution.  

\begin{figure*}
 \centering
 \includegraphics[scale=0.42]{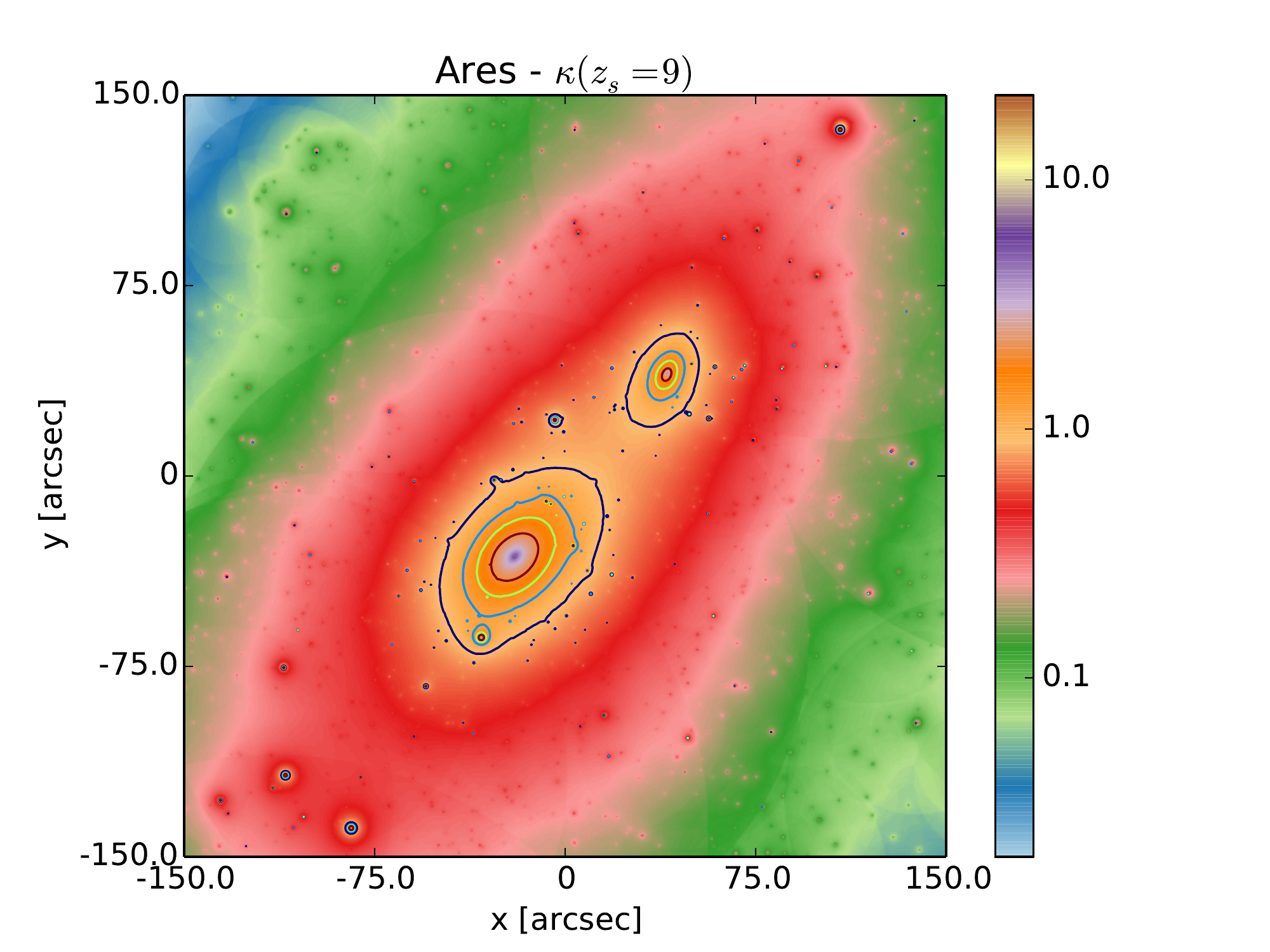}
 \includegraphics[scale=0.42]{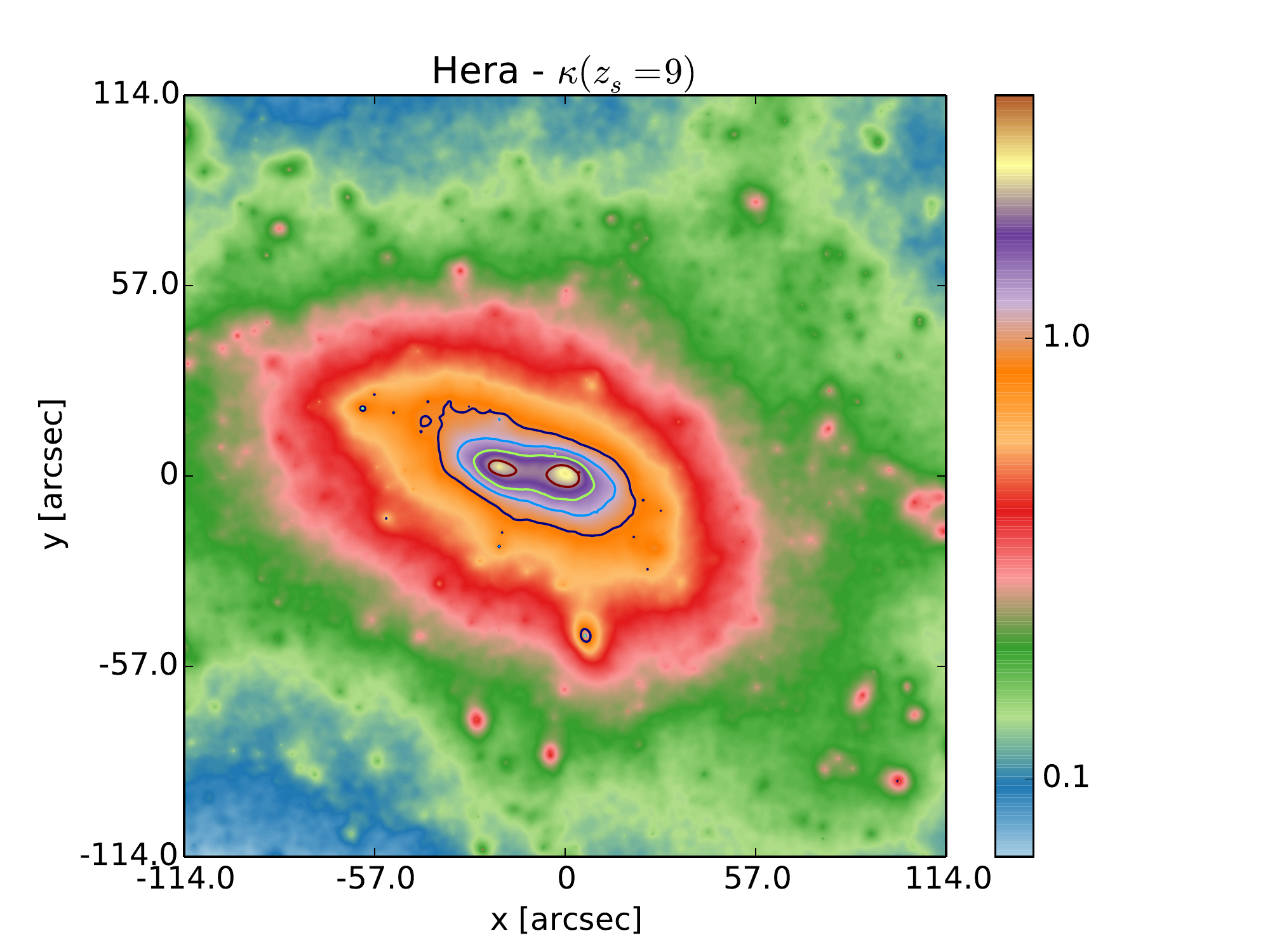}
 \includegraphics[scale=0.42]{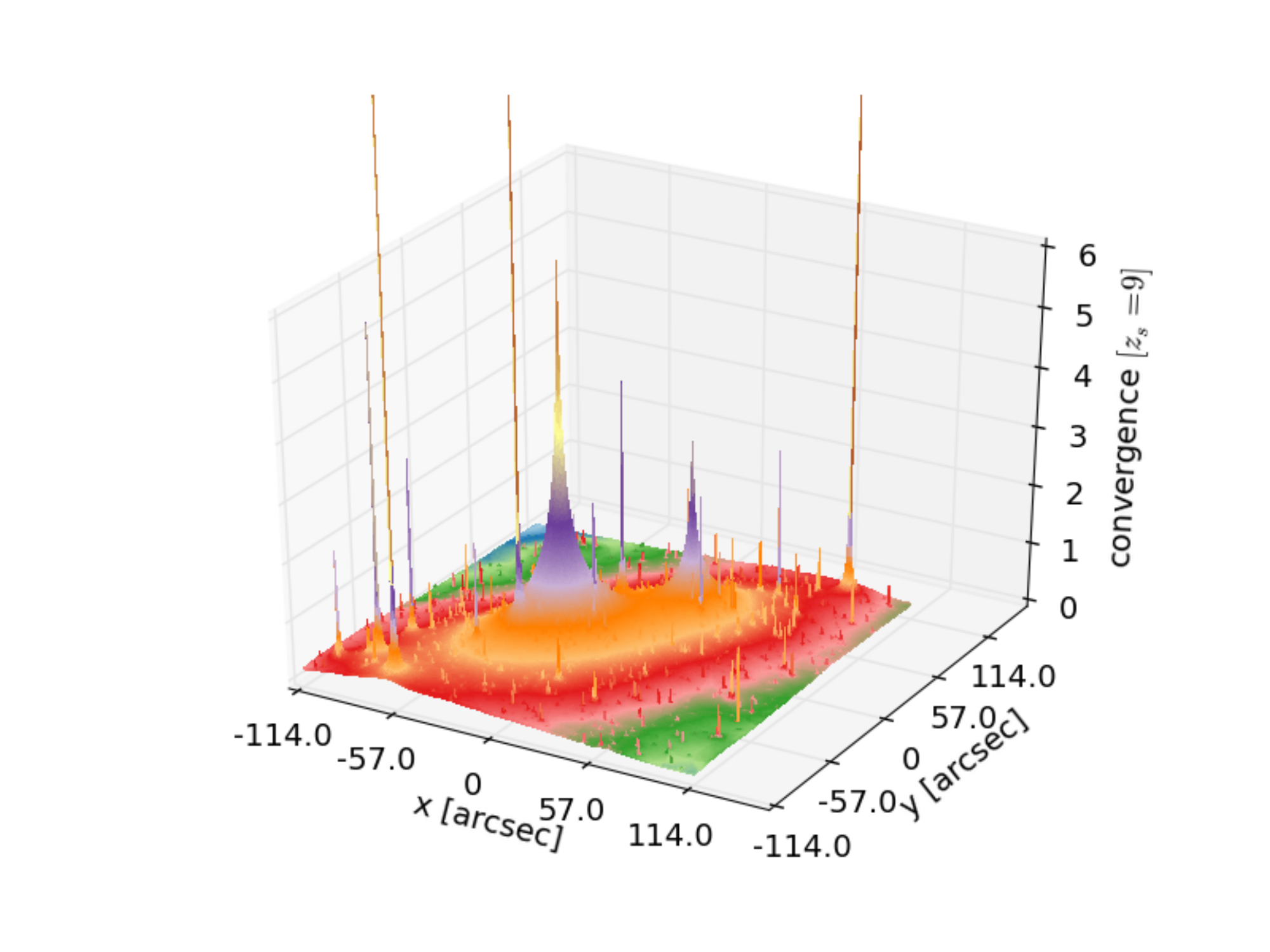}
 \includegraphics[scale=0.42]{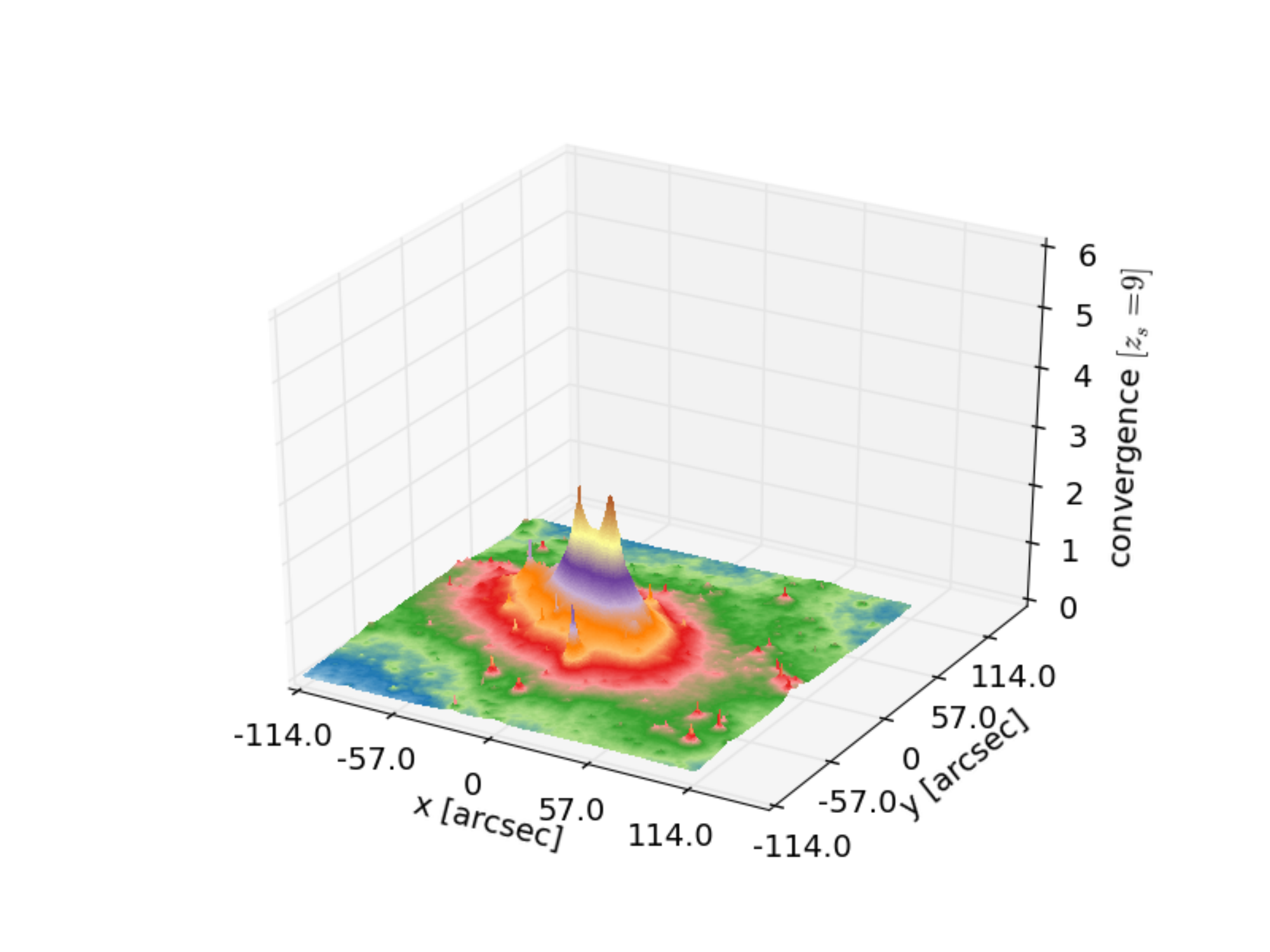}
  \caption{2D and 3D convergence maps of {\em Ares} (left panels) and {\em Hera} (right panels). The maps are normalized to the source redshift $z_s=9$}
 \label{fig:convmaps}
\end{figure*}

\begin{figure*}
 \centering
 \includegraphics[scale=0.42]{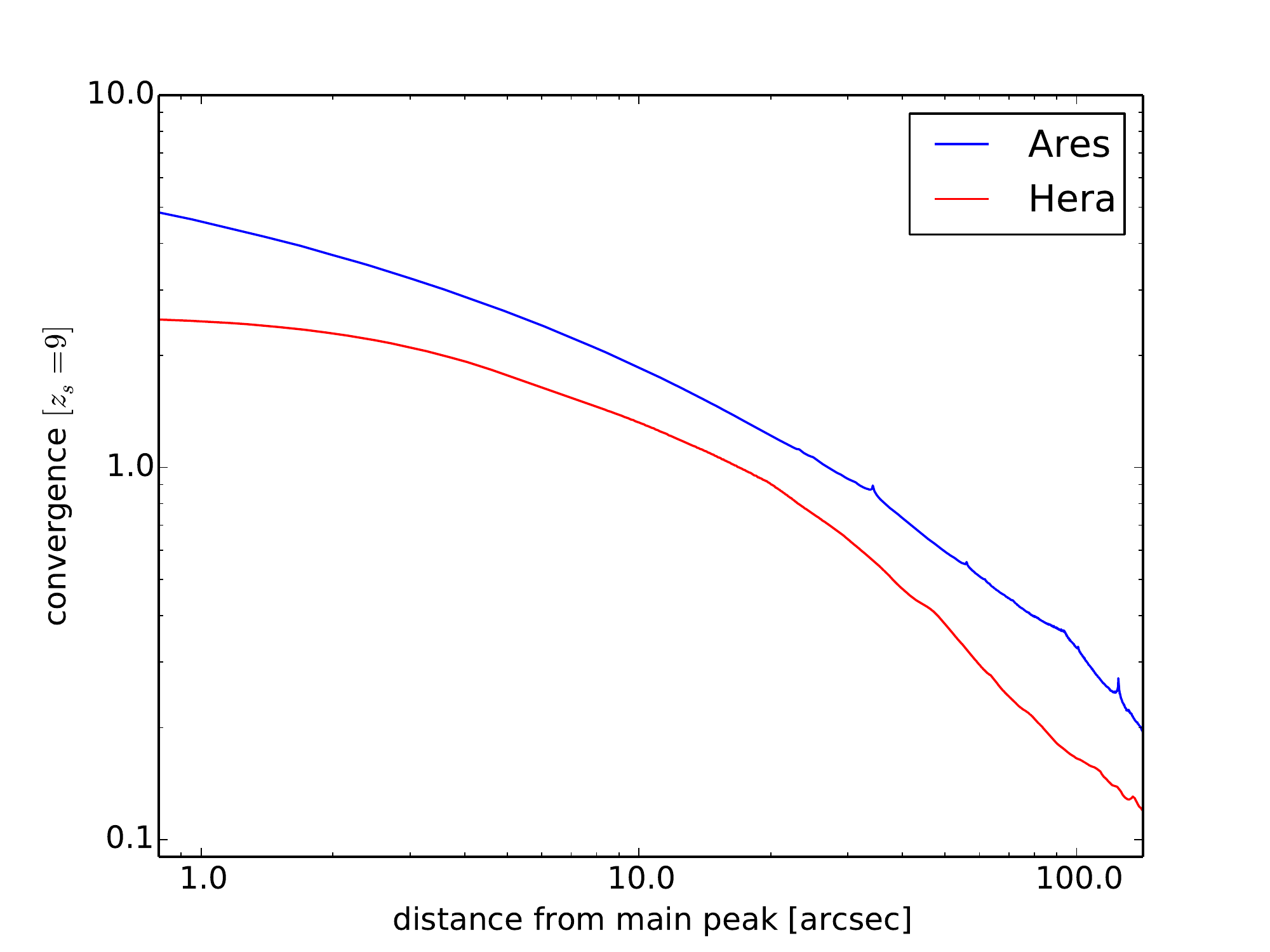}
 \includegraphics[scale=0.42]{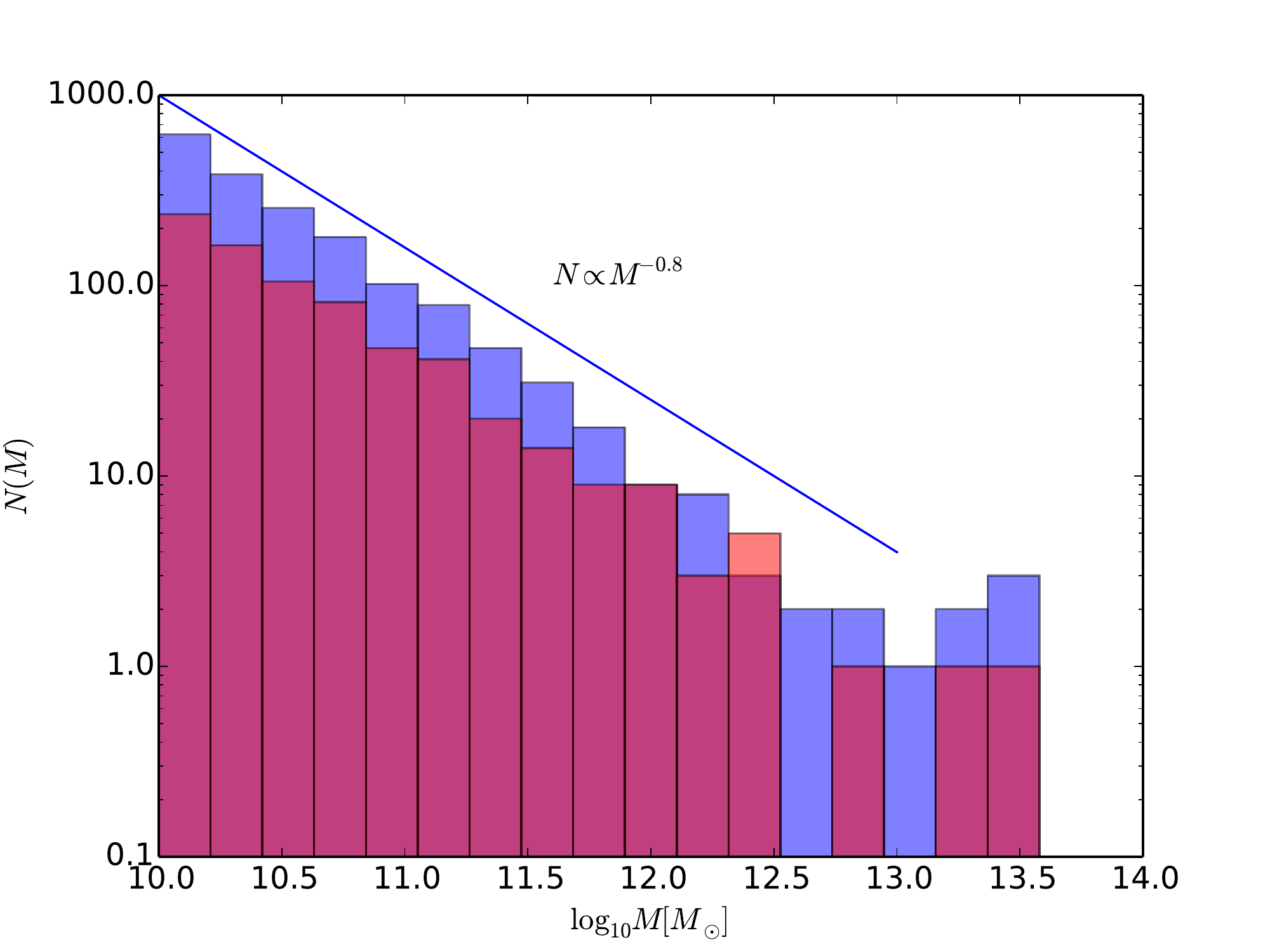}
  \caption{Key properties of {\em Ares} and {\em Hera} (blue and red colors, respectively). Left panel: Convergence profiles (for source redshift $z_s=9$). In both cases, the center has been chosen to coincide with the most massive dark matter clump in the simulation; Right panel: sub-halo mass function (built considering all sub-halos within $1 h^{-1}$Mpc from the center of the most massive clump.}
 \label{fig:convprofs}
\end{figure*}

The convergence map of {\em Hera} with its complex morphology and the abundance of substructures is shown in the right panels of  Fig.~\ref{fig:convmaps}. The convergence profile and the substructure mass function are displayed in Fig.~\ref{fig:convprofs}. Compared to {\em Ares}, the small scale structures of {\em Hera} are smoother as they are less well resolved. Nevertheless, the substructure mass function scales very similarly with halo mass. As in the case of {\em Ares}, {\em Hera} has a bi-modal mass distribution. A massive substructure ($M\sim 5 \times 10^{13}h^{-1}M_\odot$) is located $\sim 30"$ ($\sim 130 h^{-1}$kpc) from the cluster center, producing a secondary peak in the convergence map and elongating the iso-density contours in the southwest--northeasterly direction. 

\subsection{Ray-tracing}

In order to generate lensing effects in the simulated images, it is necessary to compute the deflections produced by the cluster. This allows us to then use ray-tracing methods to map the surface-brightness distribution of the sources on the camera of our virtual telescope, which is HST in this case. In practice, we shoot a bundle of light rays through a dense grid covering the field-of-view (FOV), starting from the position of the observer. Then, we use the computed deflection angles to trace the path of the light back to the sources. When simulating HST observations, we compute the deflection angles on a regular grid of $2048\times2048$ points, covering a FOV of $250"\times250"$  centered on the cluster. 

In the case of {\em Ares}, {\tt MOKA} produces a map of the convergence, $\kappa(\vec\theta)$. This can be converted into a map of the deflection angles, $\vec{\alpha}(\vec{\theta})$, by solving the Eq.
\begin{equation}
\vec{\alpha}(\vec\theta)=\frac{1}{\pi}\int d^2\theta'\frac{\vec\theta-\vec\theta'}{|\vec\theta-\vec\theta'|}\kappa(\vec\theta') \;.
\end{equation}
Since this is a convolution of the convergence, $\kappa(\vec\theta)$, with the kernel function
\begin{equation}
K(\vec\theta)=\frac{1}{\pi}\frac{\vec\theta}{|\vec\theta|} \;,
\end{equation}
this task can be achieved numerically by means of Fast-Fourier-Transform (FFT) methods. To do so, we make use of the FFT routines implemented in the {\em gsl} library.

In the case of {\em Hera}, the mass distribution of the cluster is described by a collection of dark-matter particles. Instead of mapping them on a grid to construct the convergence map,  we use our consolidated lensing simulation pipeline \citep[see e.g.][and references therein]{2010A&A...519A..90M} to compute the deflections. To briefly summarize, the procedure involves the following steps:
\begin{itemize}
\item We project the particles belonging to the halo along the desired line of sight on the {\em lens plane}. To select particles, we define a slice of the simulated volume around the cluster, corresponding to a depth of $10 h^{-1}$Mpc;
\item Starting from the position of the virtual observer, we trace a bundle of light-rays through a regular grid of $2048\times2048$ covering a region  around the halo center on the lens plane. In the case of strong lensing simulations (e.g. for HST observations) we restrict our analysis to a region of of $1\times1\;h^{-2}$Mpc$^2$. In the case of simulations extending into the weak lensing regime (e.g. for Subaru-like observations), the grid or light-rays covers a much wider area ($\sim 8\times8\;h^{-2}$Mpc$^2$);
\item Using our code {\tt GLFAST} \citep{2010A&A...514A..93M}, we compute the total deflection $\vec{\alpha}(\vec{x})$ at each light-ray position $\vec{x}$, accounting for the contributions from all particles on the lens plane. Even in the case of strong-lensing simulations, where light rays are shot through a narrower region of the lens plane, the deflections account for all particles projected out to $\sim 4 h^{-1}$Mpc from the cluster center. The code is based on a Tree-algorithm, where the contributions to the deflection angle of a light ray by the nearby particles are summed directly, while those from distant particles are calculated using higher-order Taylor expansions of the deflection potential around the light-ray positions.
\item The resulting deflection field is used to derive several relevant lensing quantities. In particular, we use the spatial derivatives of  $\vec{\alpha}(\vec{\theta})$ to construct the shear maps, $\vec{\gamma}=(\gamma_1,\gamma_2)$, defined as:
\begin{eqnarray}
\gamma_1(\vec{\theta}) & = & \frac{1}{2}\left(\frac{\partial \alpha_1}{\partial \theta_1}-\frac{\partial \alpha_2}{\partial \theta_2}\right) \;, \\
\gamma_2(\vec{\theta}) & = & \frac{\partial \alpha_1}{\partial \theta_2} = \frac{\partial \alpha_2}{\partial \theta_1} \;.
\end{eqnarray}
\end{itemize}

The convergence, $\kappa(\vec{\theta})$, may also be reconstructed as:
\begin{eqnarray}
\kappa(\vec{\theta}) & = & \frac{1}{2}\left(\frac{\partial \alpha_1}{\partial \theta_1}+\frac{\partial \alpha_2}{\partial \theta_2}\right) \;.
\end{eqnarray}

The lensing critical lines yield formally infinite magnification for a given source redshift.  They are defined as the curves along which the determinant of the lensing Jacobian is zero \citep[e.g.][]{SC92.1}:
\begin{equation}
\det A = (1-\kappa-|\gamma|)(1-\kappa+|\gamma|) = 0 \;.
\end{equation}
In particular, the {\em tangential} critical line is defined by the condition $(1-\kappa-|\gamma|)=0$, whereas the {\em radial} critical line corresponds to the line along which $(1-\kappa+|\gamma|)=0$. In the following sections, we will often use the term {\em Einstein radius} to refer to the size of the tangential critical line. As discussed in \cite{2013SSRv..177...31M}, there are several possible definitions for the Einstein radius. Here, we adopt the {\em effective} Einstein radius definition \citep[see also][]{2012A&A...547A..66R} given by, 
\begin{equation}
\theta_E \equiv \frac{1}{d_{\rm L}}\sqrt{\frac{S}{\pi}}\;,
\label{eq:einsteinr}
\end{equation}
where $S$ is the area enclosed by the tangential critical line and $d_{\rm L}$ is the angular diameter distance to the lens plane.

\subsection{SkyLens}

We simulate observations of galaxy cluster fields using the code {\tt SkyLens}, which is described in detail in \cite{2008A&A...482..403M} and in \cite{2010A&A...514A..93M}. The creation of the simulated images involves the following steps:
\begin{enumerate}
\item we generate a past light-cone populated with source galaxies resembling the luminosity and the redshift distribution of the galaxies in the Hubble Ultra-Deep-Field (HUDF; \citealt{2006AJ....132..926C});
\item we model the  morphologies of the sources using shapelet decompositions of the galaxies in the HUDF \citep{2007A&A...463.1215M}. Their spectral energy distributions were obtained as part of the photometric redshift measurements of these galaxies described in \cite{2006AJ....132..926C};
\item the deflection fields of the lensing clusters are used to trace a bundle of rays from a virtual CCD, resembling the properties of the Advanced Camera for Surveys (ACS) or of the Wide Field Camera3 (WFC3), back to the sources; 
\item  by associating each pixel of the virtual CCD to the emitting elements of the sources, we reconstruct their lensed surface brightness distributions on the CCD;
\item  we model the surface brightness distribution of the cluster galaxies using single or double Sersic models \citep{1963BAAA....6...41S}. These are obtained by fitting real cluster galaxies in a set of low to intermediate redshift clusters \citep{2005ApJ...618..195G}. The Brightest Cluster Galaxies (BCGs) all include a large scale component used to model the intra-cluster-light produced by the BCG stellar halos and by free-floating stars;
\item  the SEDs of the cluster galaxies are modeled according to prescriptions from semi-analytic models or from the Halo-Occupation-Distribution technique, as explained earlier;   
\item we convert the surface brightness distributions into counts per pixel assuming a telescope throughput curve, which accounts for the optics, the camera and the filter used in carrying out the simulated observations. In each band, we simulate the exposure times (in units of HST orbits\footnote{We assume an orbital visibility period of $2500$ sec}) used to carry out the mock Frontier Fields observations;

\item the images are then convolved with a PSF model, obtained using the Tiny Tim HST PSF modeling software \citep{2011SPIE.8127E..0JK}. Finally, realistic noise is added mimicking the appropriate sky surface brightness in the simulated bands. The noise is assumed to have a Poisson distribution, and it is calculated according to \cite{2008A&A...482..403M} Eq. 31, assuming a stack of multiple exposures, with the number varying from band to band\footnote{two exposures per orbit per filter}.      

\end{enumerate}
\begin{figure*}
 \centering
 \includegraphics[width=0.45\hsize]{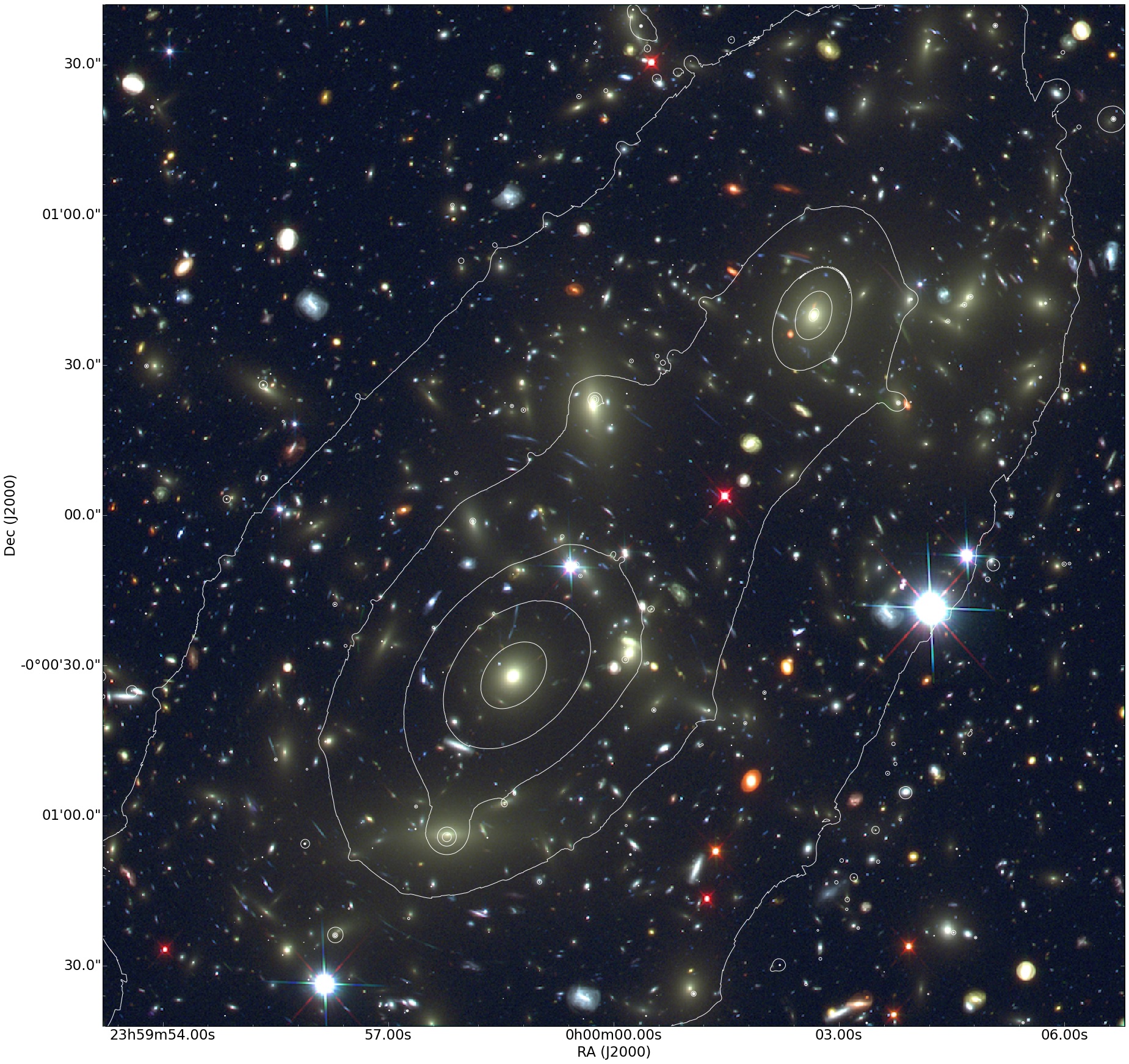}
  \includegraphics[width=0.45\hsize]{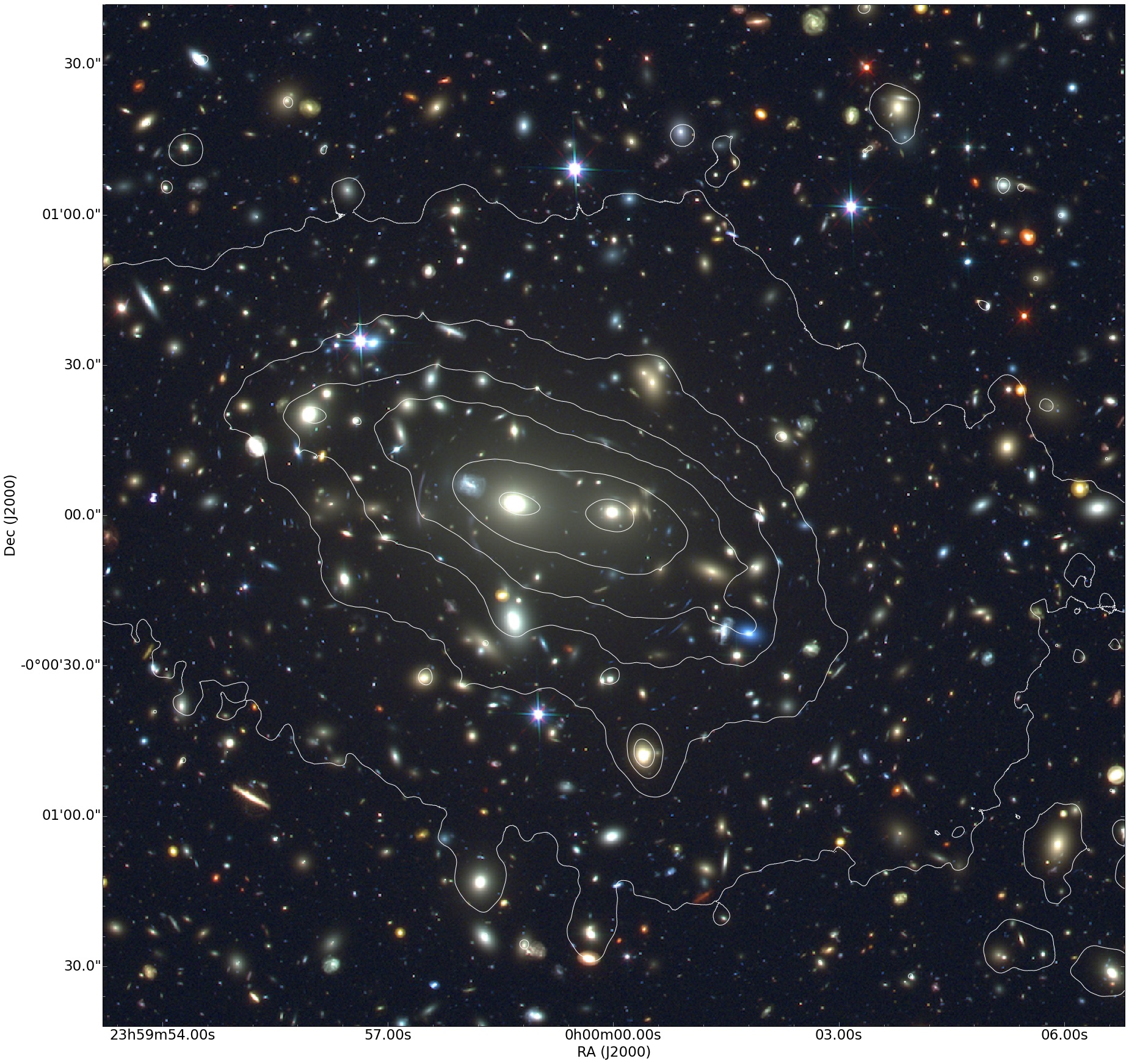} \\
 \includegraphics[width=0.45\hsize]{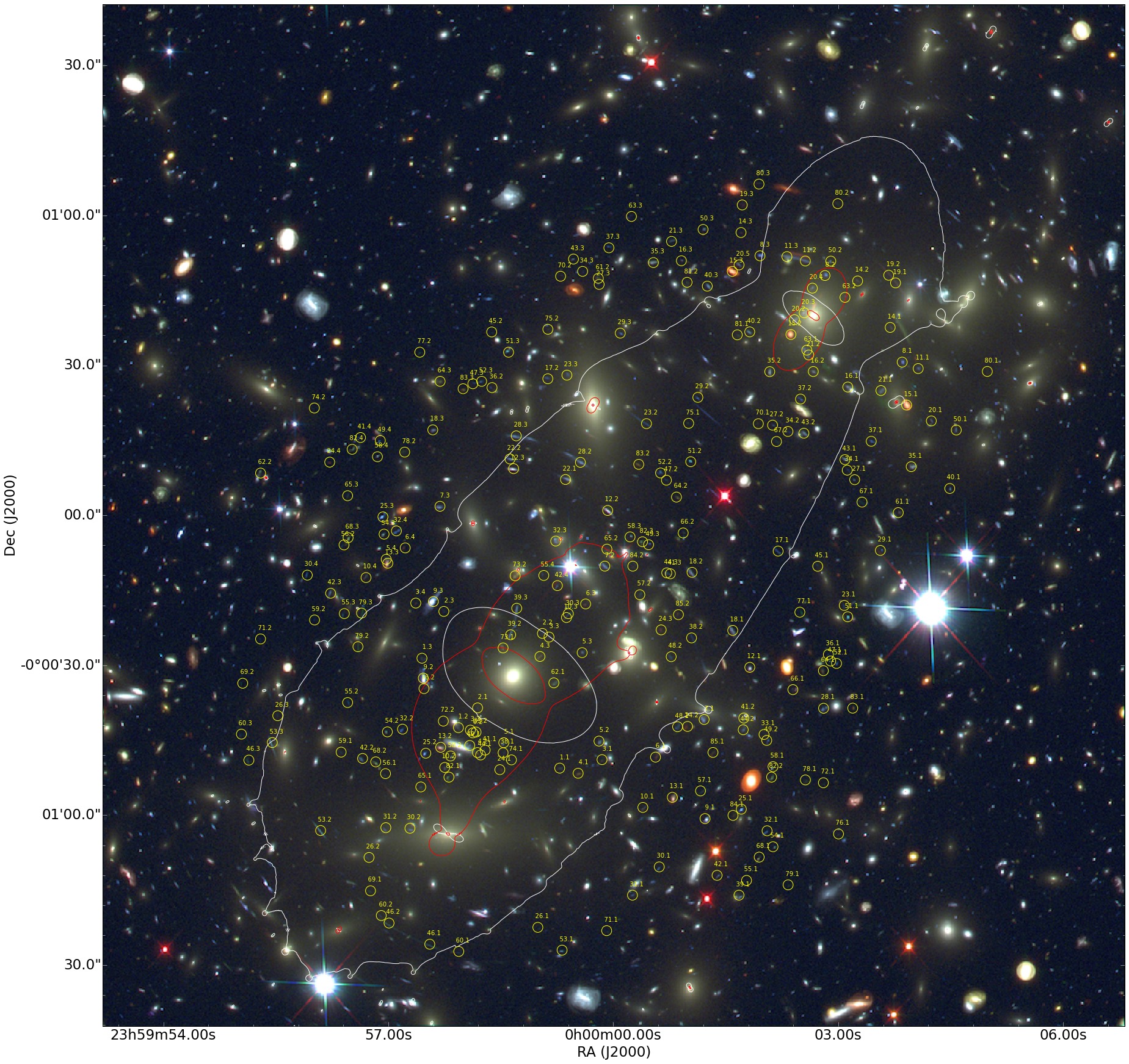} 
  \includegraphics[width=0.45\hsize]{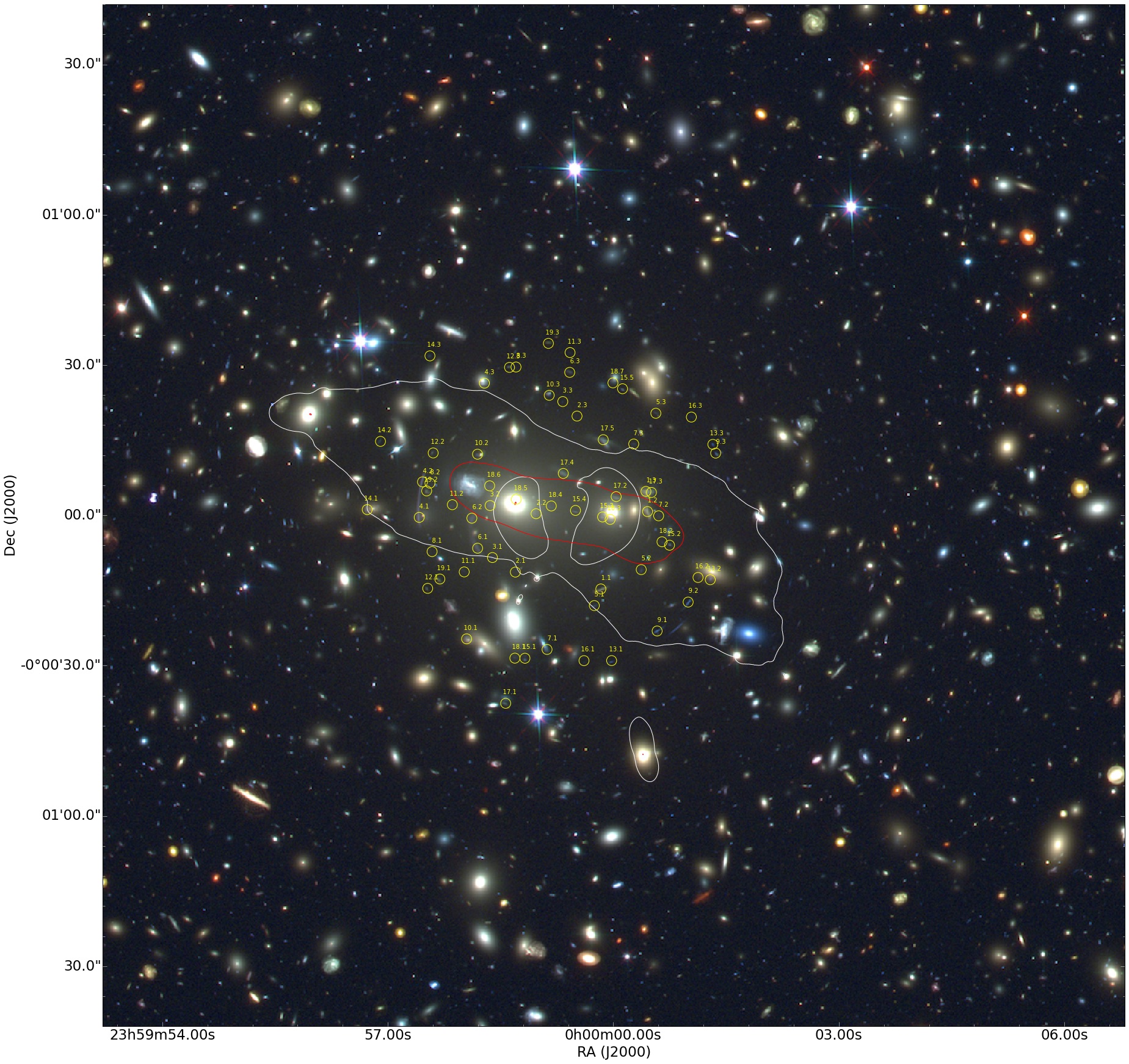} \\
 \includegraphics[width=0.45\hsize]{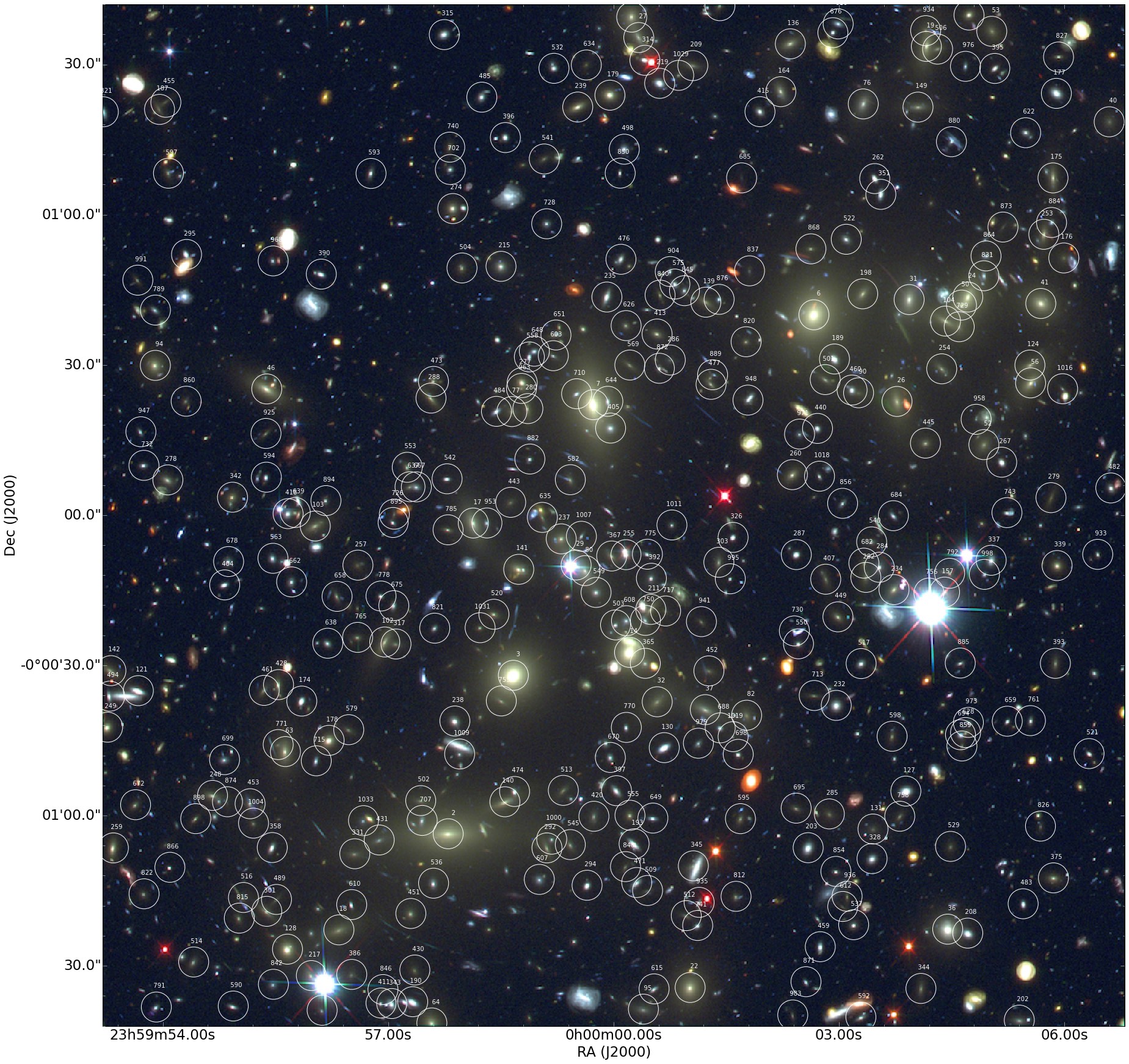}
 \includegraphics[width=0.45\hsize]{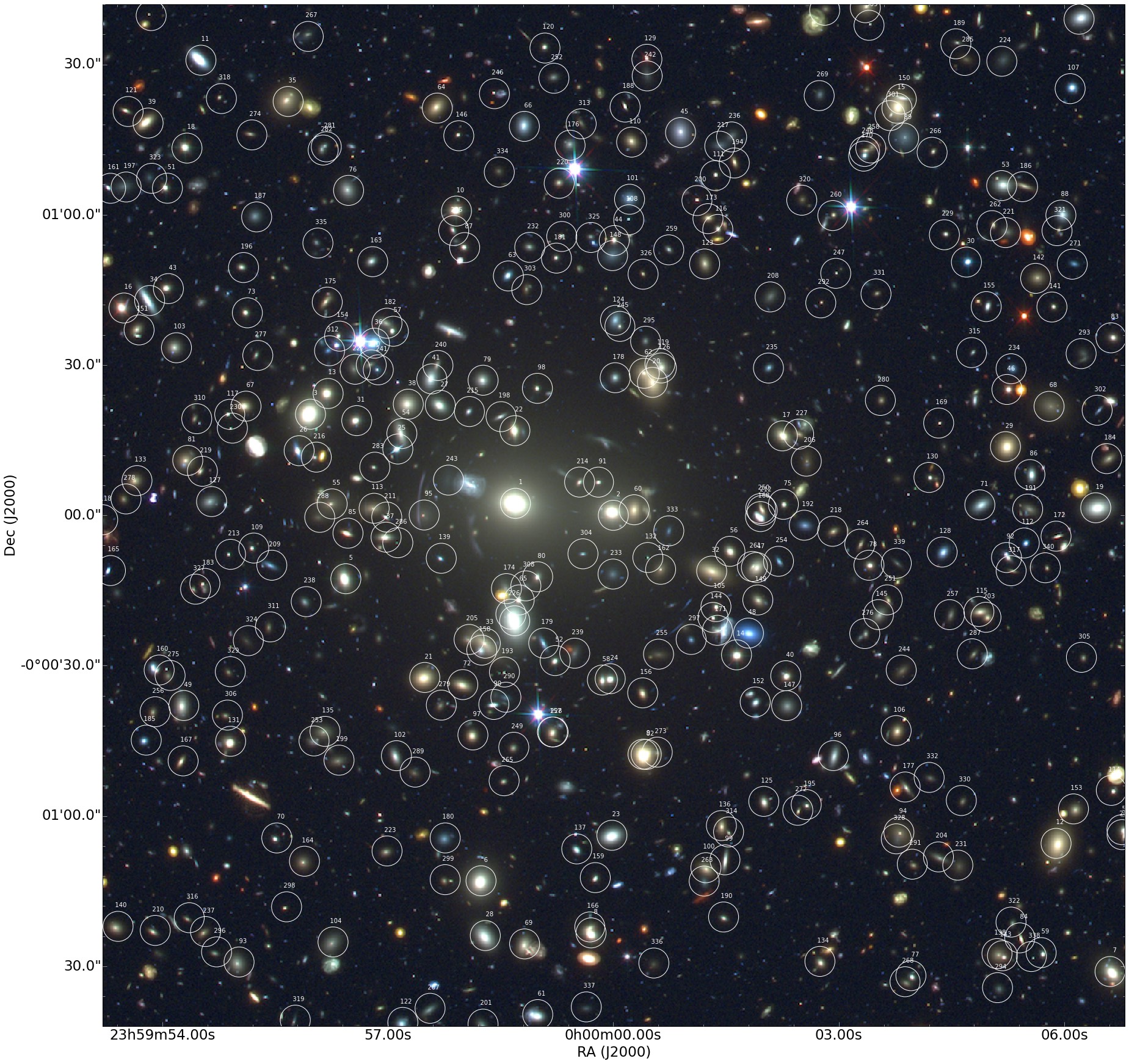} 
  \caption{Color composite images of {\em Ares} and {\em Hera} (left and right panels, respectively). In the upper panels, we overlay to the  optical images the surface density iso-contours. In the central panels, we show the critical lines for $z_s=1$ (red) and $z_s=9$ (white). In addition, we display the location of the multiple image systems (numbered yellow circles). The galaxies identified as cluster-members are indicated by white circles in the lower panels.}
 \label{fig:simobs}
\end{figure*}

\begin{table}
\begin{center}
  \begin{tabular}{ccc}
    Filter & n. of orbits & bkg level \\
    & & $[10^6~\mathrm{counts}~\mathrm{sec}^{-1}]$ \\ \hline
    F435W & 18 &  0.615 \\
    F606W & 10 & 1.821 \\
    F814W & 42 & 2.227 \\
    F105W & 24 & 1.136 \\
    F125W & 12 & 1.122 \\
    F140W & 10 & 1.371 \\
    F160W & 24 & 0.959\\
  \end{tabular}
  \caption{Details of the simulated observations. First column: filters employed; Second column: exposure time in terms of HST orbits; Third colum: level of the background (counts/sec in circles with radius 0.2 arcsec)}
  \label{tab:observations}
\end{center}
\end{table}

\subsection{Images and catalogs}

This is the first phase of a comparison project, and in the next phase we intend to include additional simulations with an even greater level of realism. For this first exercise, we proceed as follows.
\begin{itemize}
\item For both {\em Ares} and {\em Hera}, we generate simulated HST observations in all bands that are deployed for the FFI, mimicking the same exposure times (or number of orbits) as the real observations. The level of the background is set to values provided by the ACS and WFC3 exposure time calculators in each band. The details of these simulations are provided in Table~\ref{tab:observations}. Each image covers a field of view of $204\times204$ arcsec$^2$. All images are co-aligned and co-rotated. Effects like gaps between chips, pixel defects, charge transfer inefficiency, cosmic rays, etc. are not included. The resolution of the ACS and WFC3 simulations are $0.05$ arcsec/pixel and $0.13$ arcsec/pixel, respectively. These images were made available to the modelers.
\item In addition to the images, we provided the list of all multiple images obtained from the ray-tracing procedure (see Fig.~\ref{fig:simobs}, central panels). Each multiple-image system is characterized by the redshift of its source, which is also provided to the modelers. Thus, in this exercise we assume that all images can be identified without errors and that all their redshifts can be measured ``spectroscopically''. This is certainly a very optimistic assumption which will never be satisfied in the real world. In the next round of this comparison project, the assumption will be relaxed, but for the moment we decided to release this information because our objective is to determine possible systematics of the various reconstruction algorithms. Other issues related to the approaches used to search for multiple images or the impact of redshift uncertainties on the results will be studied in a future work. Some of these systematics have already been investigated for some lens modeling methods, i.e., Johnson \& Sharon (2016, submitted).
\item We also released a catalog of cluster members (circled in the right panels of Fig.~\ref{fig:simobs}), containing positions and photometry in all  bands. Several reconstruction methods (in particular those employing the parametric approach) build the lens model by combining smooth dark matter halos with substructures associated to the cluster members akin to our construction of  {\em Ares} . In this simplified test, modelers are provided with the list of all cluster members with $m_{AB,F814W}<24$. Again, this is an over-simplification which will be removed in the next round of simulations, and which implicitly favors those methods which make use of this information. In reality, such methods have to deal with the risks of misidentification of cluster members.
\item For those groups which make use of weak-lensing measurements to complement the strong-lensing constraints, we produced a single Subaru-like R-band image of both {\em Ares} and {\em Hera} covering a much larger FOV of $30\times 30$ arcmin$^2$. The provided image contained only background galaxies (i.e. lensed by the clusters) and stars, so that shape measurements could be made using any weak-lensing pipeline without worrying about the separation of background sources from the cluster members or contamination by foreground galaxies.  We also use the publicly available pipeline KSBf90\footnote{http://www.roe.ac.uk/$\sim$heymans/KSBf90/Home.html} \citep{HEY06.1} based on the Kaiser, Squires and Broadhurst method \citep{KSB95.1} to derive a catalog containing galaxy positions and ellipticities. The resulting number density of galaxies useful for the weak-lensing analysis is $\sim 14$ gal/sq. arcmin. This is significantly smaller than the number density achievable with HST.
\end{itemize}
All these data for the mock cluster lenses were shared with lens modelers participating in the project via a dedicated website\footnote{http://pico.bo.astro.it/$\sim$massimo/Public/FF}. We emphasize that the input mass distributions of the lenses and the techniques used to generate them were initially kept blinded to all groups. The strong lensing constraints amounted to 242 multiple images produced by 85 sources in the case of {\em Ares} and 65 images of 19 sources in the case of {\em Hera}.

\section{Lens modeling techniques}
\label{sect:techniques}

\subsection{Submission of the models}

A large fraction of lens modelers currently working actively on the analysis of the FFI data accepted the challenge and participated in this project. The two cluster simulations were not released simultaneously. We initially released only the data for {\em Ares}, and we received reconstructed models for this cluster from 7 groups. These groups performed a fully blind analysis of the data-set. Two additional models were submitted by A. Zitrin after the input mass distributions were already revealed, under the assurance that the reconstruction was actually performed blindly.

In a second stage of the comparison exercise, we released the simulation of {\em Hera}, and received 8 models from 6 participating groups. Also for this cluster, we received additional reconstructions after we revealed the input mass distribution of the lens. These models were submitted  by A.~Zitrin and by D.~Lam. 

There are two general classes of inversion algorithms. They comprise {\em parametric} models wherein the mass distribution is reconstructed by combining clumps of matter, often positioned where the brightest cluster galaxies are located, each of which is characterized by an ensemble of parameters including the density profile and shape. The parameter spaces of these models are explored in an effort to best reproduce the observed positions, shapes and magnitudes of the multiple images and arcs. The second approach is called {\em free-form} (a.k.a. non-parametric): wherein now the cluster is subdivided into a mesh onto which the lensing observables are mapped, and which is then transformed into a pixelised mass distribution following several methods to link the observable to the underlying lens potential or deflection field.

Both these approaches were amply represented in the challenge. A summary of all submitted models, with the indication of whether they are parametric or free-form and built before or after the input mass distribution of the lenses was revealed is given in Table~\ref{tab:models}. Each model is given a reference name used throughout the paper. Each modeling technique is briefly described below.

\begin{table*}
  \begin{tabular}{cccccc}
    Group/Author & Method & Model & Cluster & Approach & Blind \\
    \hline
    M. Bradac \& A. Hoag & SWUnited & Bradac-Hoag & {\em Ares}+{\em Hera} & free-form & yes \\
    J. Diego & WSLAP+ & Diego-multires & {\em Hera} & hybrid & yes \\
    J. Diego & WSLAP+ & Diego-overfit & {\em Hera} & hybrid & yes \\
    J. Diego & WSLAP+ & Diego-reggrid & {\em Ares}+{\em Hera}  & hybrid & yes \\
    D. Lam & WSLAP+ & Lam & {\em Hera} & hybrid & no \\
    J. Liesenborgs, K. Sebesta \& L. Williams & \grale & GRALE &  {\em Ares}+{\em Hera}  & free-form & yes \\
    D. Coe & LensPerfect & Coe & {\em Ares} & free-form & yes \\
    CATS & {\tt Lenstool} & CATS &  {\em Ares}+{\em Hera}  & parametric & yes \\
   T. Johnson \& K. Sharon & {\tt Lenstool} & Johnson-Sharon &  {\em Ares}+{\em Hera}  & parametric & yes \\  
    T. Ishigaki, R. Kawamata \& M. Oguri  & GLAFIC & GLAFIC &  {\em Ares}+{\em Hera}  & parametric & yes \\
    A. Zitrin  & LTM & Zitrin-LTM-gauss &  {\em Ares}+{\em Hera} & parametric & no \\  
    A. Zitrin  & PIEMDeNFW & Zitrin-NFW &  {\em Ares}+{\em Hera}  & parametric & no \\  
  \end{tabular}
  \caption{Models submitted by the groups participating in the project. The table lists the name of the submitting group/author of the reconstruction, the reference name of the model, the type of algorithm (free-form, parametric, or hybrid) and whether the model was submitted blind, that is before the input mass distribution of the lens was revealed.}
  \label{tab:models}
\end{table*}

\subsection{SWUnited: The Bradac-Hoag model}

The Bradac-Hoag model employs the method named {\em SWUnited: Strong and weak lensing mass reconstruction on a non-uniform adapted grid}. This combined strong and weak lensing analysis method reconstructs the gravitational potential $\psi_k = \psi(\vec\theta^k)$ on a set of
points $\vec\theta^k$, which can be randomly distributed over the
entire field-of-view. From the potential, any desired gravitational lensing
quantity (e.g. surface mass density, deflection angle
, magnification, flexion, etc.) can be readily calculated. 
Such an approach therefore does not require an assumption
of e.g. a particular model of the potential/mass distribution. The
potential is reconstructed by maximizing a log likelihood $\log{P}$
which uses image positions of multiply imaged sources and source plane
minimization (corrected by magnification); weak lensing ellipticities,
and regularization as constraints.  Current implementation also includes flexion
measurements, however the data was not used in this paper.

\subsubsection{Description of the method}

The implementation of the method follows the algorithm first proposed by
\citet{bartelmann96} and is described in detail in \citet{bradac04a}
and \citet{bradac09}.  From the set of potential values they determine
all observables using derivatives. For
example, the convergence $\kappa$ is related to $\psi$
via the Poisson equation, $2\kappa = \nabla^2\psi$ (where the physical
surface mass density is $\Sigma = \kappa \: \Sigma_{\rm crit}$ and
$\Sigma_{\rm crit}$ depends upon the angular diameter distances
between the observer, the lens, and the source). The details on how the derivatives on an non-uniform
grid are evaluated can be found in \citet{bradac09}.  By using a
reconstruction grid whose pixel scale varies across the field, the method is
able to achieve increased resolution in the cluster centre (close to
where we see strongly lensed images), and hence the magnification map
in the regions of high magnification is more accurate. The posterior peak values of the potential
$\psi_k$ are found by solving the non-linear equation $\partial
\log{P} / \partial \psi_k =0$.  This set of equations is linearized and
 a solution is reached in an iterative fashion (keeping the non-linear terms
fixed at each iteration step). This requires an initial guess for the
gravitational potential; the systematic effects arising from various
choices of this initial model were discussed in \citet{bradac06}. The choice of particular grid geometry, the regularisation parameter,
and the hyper-parameters that set the relative weighting between the
contributions to $\log{P}$ all become critical when weak lensing data
on large scales ($\gtrsim 1\mbox{Mpc}$) are included, and we need a
full-field mass reconstruction. This is not the case in this work, as
we are only interested in the magnification of the inner region.

\begin{figure}
\includegraphics[width=0.5\textwidth]{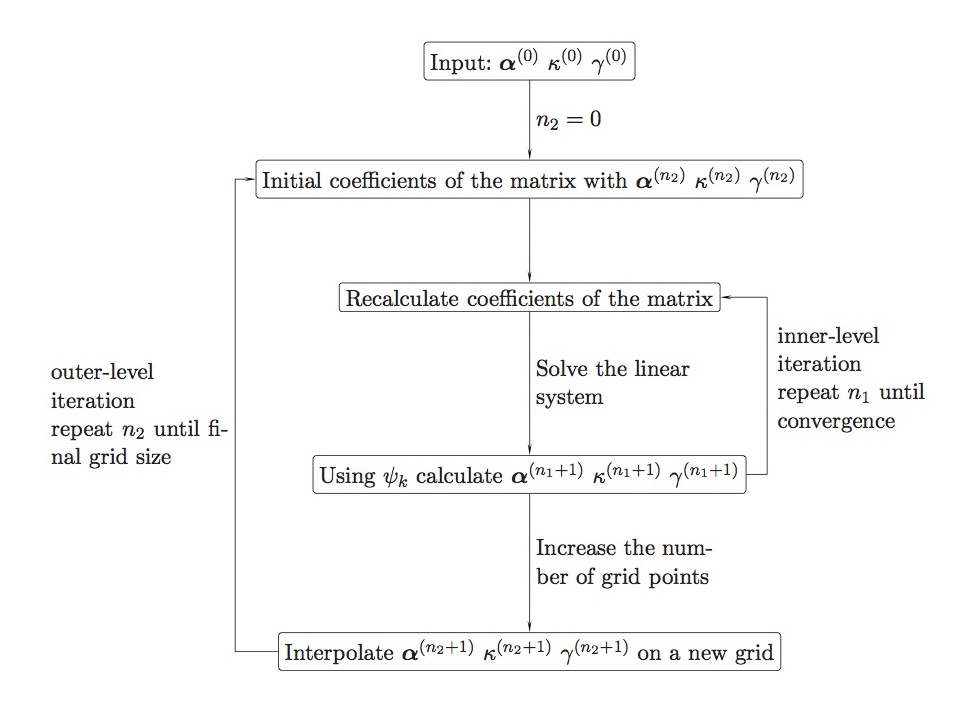}
\caption{The outline of the method used to build the Bradac-Hoag model and the two-level iteration process.}
\label{fig:method}
\end{figure}

The reconstruction is performed in a two-level iteration process,
outlined in Fig.~\ref{fig:method}.  The inner-level iteration process
described above for solving the non-linear system of equations
$\partial \log{P} / \partial \psi_k =0$ is solved in iterative fashion
and repeated until convergence of $\kappa$.  The outer-level iteration
is performed for the purpose of regularisation.  In order to penalise
small-scale fluctuations in the surface mass density,  the
reconstruction is started with a coarse grid (large cell size).  Then for each
$n_2$ step the number of grid points is increased in the field and
the new reconstructed $\kappa^{(n_2)}$ is compared with the one from the
previous iteration $\kappa^{(n_2-1)}$ (or with the initial input value
$\kappa^{(0)}$ for $n_2=0$), penalizing any large deviations. The second-level iterations are performed
until the final grid size is reached and convergence is achieved.

\subsubsection{Strengths and Weaknesses of the Method}

The main strength of the method as discussed above is that instead of
fitting a specific set of family of models to the data, the method is
free of such an assumption. Furthermore the positions of the points
where potential is reconstructed $(\vec\theta^k)$ can be chosen
arbitrarily, which allows them to use higher density of points in the
regions where signal-to-noise is the highest (i.e. where multiple
images are present), and they can employ coarser sampling in the areas
where this is not the case (e.g. at large radii from the
centre). They are also reconstructing the potential (rather than
traditionally used surface mass density), since it locally determines
both the lensing distortion (for weak lensing and flexion) as
well as the deflection (for strong lensing) and there is no  need to assume
the surface mass denisty values beyond the observed field. 

The main weakness of the method on  the other hand is the fact that they
are trying to maximize a function with a large number of parameters and the method is
inherently unstable. The inversion of the matrix satisfying the
equation  $\partial \log{P} / \partial \psi_k =0$ is also very
noisy. The method is therefore very likely to diverge or land in a
secondary minimum. Regularization needs to be employed,
which adds an additional parameters (relative weighting of regularization
term) to the rest of $\log{P}$ and a choice of regularization method
itself. The optimal choices need to be determined using simulation
data.  

\subsubsection{Improvements in progress}

A recent improvement to the method is the addition of the measurement of flexion to the input constraints. The code has been adapted (see also \citealp{cain15}) and tested on simulated data. The group is currently testing it using HST data. In the
future they plan to port the code into python to make the interface user
friendly, at which point they plan to
release it to the community.

%%%%%%%%%%%%%%%%%%%%%%%%%%%%%%%%%%%%%%%%

\subsection{WSLAP+: The Diego  and the Lam models}

All Diego models (Diego-multires, Diego-overfit and Diego-reggrid models) and the Lam model are built using WSLAP+, a free-form or non-parametric method that includes also a compact mass 
component associated to the cluster members (thus, classified as hybrid in this paper). The main part of the code is written in fortran 
and compiles with standard compilers (like gfortran) included in the most common linux 
distributions. Plotting routines written in IDL are available to display the intermediate results as the code runs. A script interface allows the user to define the input and 
output files, select the parts of the code to be run and control the plotting routines.  
A detailed description of the code and of its features can be found in the papers by \citep{Diego2005,Diego2007,Sendra2014,2016MNRAS.456..356D}. 
The code is not publicly available yet but a companion code LensExplorer is available. 
LensExplorer allows the user to easily explore the lens models derived for the Frontier Fields 
clusters, search for new counter-images, compute magnifications, or predict the location and shape 
of multiple images. Here we present a brief summary of the main code WSLAP+. Note that the code includes certain features that were not taken into account in the analysis presented in this paper but will be included in the future ``unblinded" version of this work. Among these features,  the code incorporates spatial information about knots in resolved systems greatly improving the accuracy and robustness of the results \citep[see ][ for s practical demonstration]{2016MNRAS.456..356D}. In the present work, only long elongated arcs were considered as resolved systems. 

\subsubsection{Description of the method}

The algorithm divides the mass distribution
in the lens plane into two components. The first is a compact one and
is associated with the member galaxies. The member galaxies are selected from the 
red sequence. The
second component is diffuse and is distributed as a superposition of Gaussians on a
regular (or adaptive) grid. For the compact component, the mass
associated to the galaxies is assumed to be proportional to their
luminosity. If all the galaxies are assumed to have the same
mass-to-light ($M/L$) ratio, the compact component (galaxies)
contributes with just one ($N_g=1$) extra free-parameter which
corresponds to the correction that needs to be applied to the
fiducial $M/L$ ratio. In some particular cases, some galaxies (like
the BCG or massive galaxies very close to an arclet) are allowed to
have their own $M/L$ ratio adding additional free-parameters to the lens
model but typically no more than a few ($N_g \sim$ O(1)). For this
component associated to the galaxies, the total mass is assumed to
follow either a NFW profile (with fixed concentration and scale radius
scaling with the fiducial halo mass) or be proportional to the
observed surface brightness.
The diffuse or `soft' component is described by as many free
parameters as grid (or cell) points. This number ($N_c$) varies but is
typically between a few hundred to one thousand ($N_c \sim$
O(100)-O(1000)) depending on the resolution and/or use of the adaptive
grid. In addition to the free parameters describing the lens model,
the method includes as unknowns the original positions of the lensed
galaxies in the source plane. For the clusters included in the FFI
program the number of background sources, $N_s$, is typically a few
tens ($N_s \sim$ O(10)), each contributing with two unknows ($\beta_x$
and $\beta_y$). All the unknowns are then combined into a single array
$X$ with $N_x$ elements ($N_x \sim {\rm O}(1000)$.

The observables are both strong lensing and weak lensing (shear) measurements. 
For strong lensing data, the inputs are the pixel positions of the
strongly lensed galaxies (not just the centroids). In the case of long
elongated arcs near the critical curves with no features, the entire
arc is mapped and included as a constraint. If the arclets have
individual features, these can be incorporated as semi-independent
constraints but with the added condition that they need to form the
same source in the source plane. Incorporating this information acts 
as an {\it anchor} constraining the range of possible solutions and reducing 
the risk of a bias due to the minimization being carried in the source plane. 
For the weak lensing, we use shear mesurements ($\gamma_1$ and $\gamma_2$). 
The weak lensing constraints normally complement the lack of strong lensing constraints 
beyond the central region allowing for a mass reconstruction on a wider scale. 
When weak lensing information is used, the code typically uses an adaptive grid to extend 
the range up to the larger distances covered by the weak lensing data \citep{Diego15a}

The solution, $X$, is obtained after solving the
system of linear equations
\begin{equation}
\Theta = \Gamma X
\label{eq_WSLAP}
\end{equation}
where the $N_o$ observations (strong lensing, weak lensing, time
delays) are included in the array $\Theta$ and the matrix $\Gamma$ is
known and has dimension $N_ox(N_c+N_g+2N_s)$
In practice, $X$, is obtained by solving the set of linear equations
described in Eq. \ref{eq_WSLAP} via a fast bi-conjugate algorithm, or
inverted with a singular value decomposition (after setting a threshold
for the eigenvalues) or solved with a more robust quadratic algorithm
(slower). The quadratic algorithm is the preferred method as it
imposes the physical constrain that the solution $X$ must be
positive. This eliminates un-physical solutions with negative masses
and reduces the space of possible solutions. 
Errors in the solution are derived by minimizing
the quadratic function multiple times, after varying the initial
conditions of the minimization process, and/or modifying the grid, and/or changing the fiducial 
deflection field associated to the member galaxies.

\begin{figure}  
 {\includegraphics[width=9.0cm]{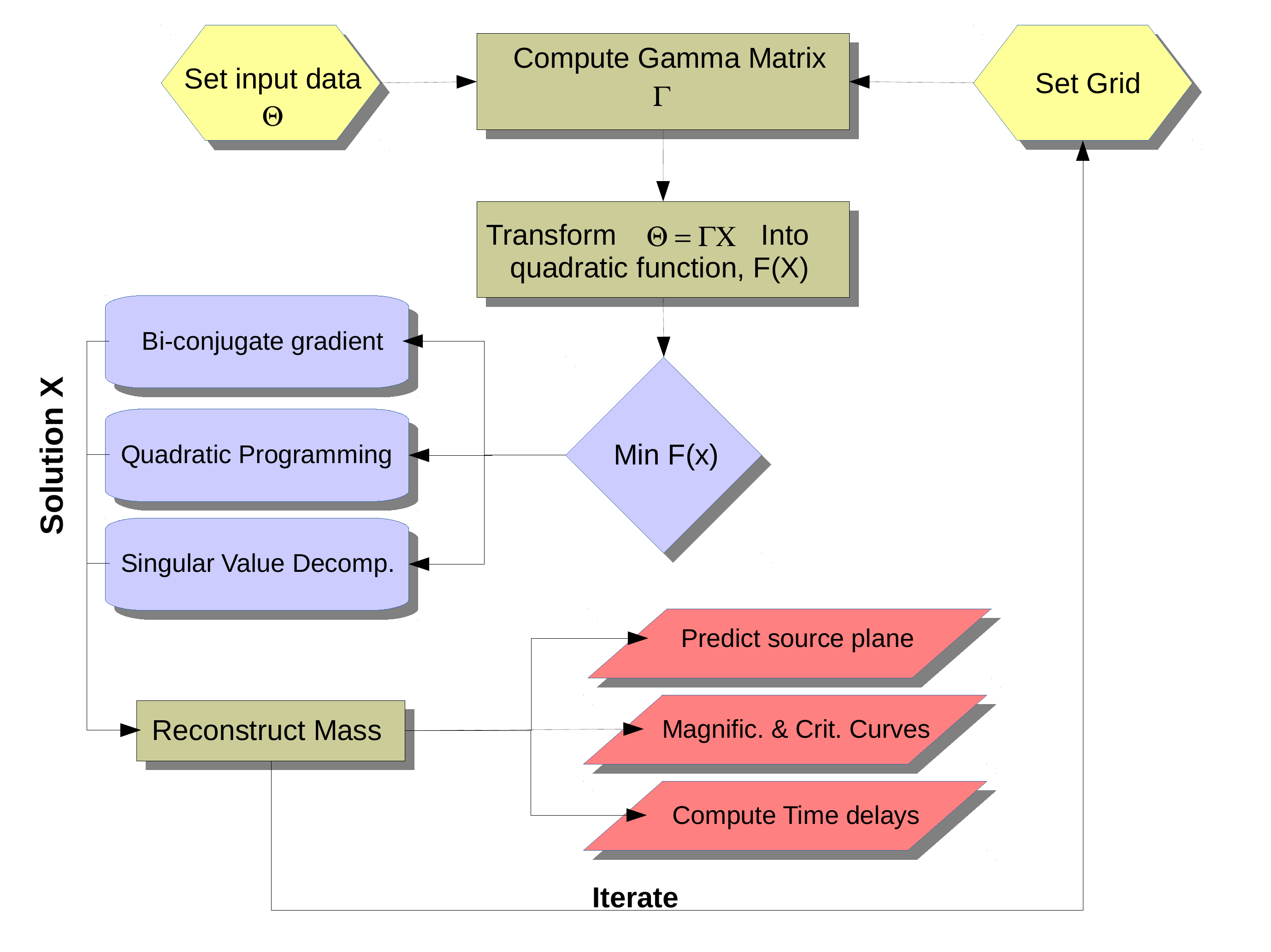}}
   \caption{Diagram showing the work-flow of WSLAP+}
   \label{fig_WSLAPp}  
\end{figure}  

\subsubsection{Strengths and weaknesses of the method}

The code implements a free-form modeling component. This implies that no strong assumptions are necessary 
about the distribution of dark matter. This is particularly useful if DM is not linked to 
the galaxies or if the baryons are also dissociated from the galaxies. The later seems to be 
the case in the FFI clusters which are in a  merging phase. Evidence that the solution 
obtained by the algorithm  may be sensitive to the mass of the X-ray emitting plasma was presented in 
\cite{2014ApJ...797...98L,Diego15b,2016MNRAS.456..356D,2015MNRAS.451.3920D}. 

The algorithm is very fast. Several methods are implemented to search for a solution. Using the 
bi-conjugate gradient algorithm a solution can be obtained in seconds. Using the slower, but more reliable, 
quadratic optimization approach a robust solution can be obained in minutes. Other fast approaches have 
been implemented as well like singular-value-decomposition. 

An adaptive grid can be used that transforms the method into a multiresolution code. Different adaptive 
grids can be implemented that introduce a small degree of freedom but also allows to explore other possible 
solutions and hence constrain better the variability of the solution. 

The code is prepared to combine weak and strong lensing. The relative weight of the two data sets is given 
by the intrinsic errors in the data sets (typically small in the strong lensing regime and larger in the 
weak lensing regime). Correlations between the lensing data can be incorporated through a covariance matrix that naturally weights the different data sets. 

The minimization is made in the source plane which may result in biases towards 
larger magnifications. To avoid this, the minimization algorithm needs to be stopped after 
a given number of iterations. Even better, including information about the size and shape of 
the sources in the source plane seems to solve this problem and the solution remains stable and 
unbiased even after a very large number of iterations. These {\it prior} information  on the size 
and shape of the source galaxies is only possible when well resolved lensed images are available 
and at least one of the multiple images is not highly magnified. 

The compact component is {\it pixelized} usually into a 512$\times$512 image that covers the 
field of view. For the small member galaxies this pixelization results in a loss of resolution that 
have a small impact on lensed images that happen to be located near this small member galaxies. A possible 
solution to alleviate this problem is to pre-compute the deflection field of these galaxies prior to the 
minimization at higher resolution and later interpolate at the position of the observed lensed galaxies. 
This approach has not been implemented yet but it is expected to eliminate this problem. 

The code can also predict more multiple images than observed. This is not being factored in at the moment 
but will be the subject of the {\it null space} implementation described in section \ref{sect_improv_Diego}. One systematic bias is know to affect the results at large distances from the centre. The reconstructed solution systematically underpredcits the mass (and magnification) in the regions where there is no lensing constraints. These regions are normally located beyond the corresponding Einstein radius for a high redshift background source. Addition of weak lensing to then constraints can reduce or eliminate this problem.

\subsubsection{Improvements in progress}\label{sect_improv_Diego}

The addition of time delays is being implemented to the reconstruction of the solution. 
Time delays will be included in a similar footing as the other observables (weak and strong lensing 
observables) with a weight that is proportional to their associated observational error. 

The addition of the {\it null space} was proven to be a useful and powerful way of improving the 
robustness of the derived solution \citep{Diego2005}. This direction has not been explored fully 
and we plan to incorporate the {\it null space} as an additional constraint. This will eliminate 
additional counterimages that are predicted by the model but not observed in the data. 

\subsubsection{Modeling of {\em Ares} and {\em Hera}}

The Diego models use both a regular grid with 32x32=1024 grid points (Diego-reggrid model) and a multi-resolution grid with approximately half the number of grid points (Diego-multires model). The compact component of the defection field is constructed based on the brightest elliptical galaxies in the cluster. We include ~ 50 such bright ellipticals for each cluster.   The mass profile for each galaxy is taken either as an NFW with scale radius (and total mass) scaling with the galaxy luminosity or directly as the observed surface brightness. This choice plays a small role in the final solution. 

Depending on the number of iterations, different solutions can be obtained. Earlier work based on simple simulations  \citep{Sendra2014}   showed how in a typical situation (similar to the one in {\em Ares} and {\em Hera}), after 10000 iterations of the code the solution converges to a stable point. The code can be left iterating longer reaching a point that we refer as "overfit" where the observed constraints are reproduced with great accuracy but sometimes at the expense of a model with fake structures.  In the case of Hera we computed the solution also in the overfit regime (90000 iterations) for comparison purposes (Diego-overfit model).

The Lam model differs from the Diego models as follows. A regular grid of Gaussian functions is used instead of a multi-resolution grid (as in diego-multires). 
In order to thoroughly explore the parameter space and to estimate the statistical uncertainty, 100 individual lens models are constructed from random initial conditions. 
Also, the submitted model is an average of these 100 individual models. 
With the exception of the 10 brightest cluster galaxies, the relative masses of all cluster galaxies are fixed, and are derived using a stellar mass-dark matter mass relation found in the EAGLE cosmological hydrodynamical simulation \citep{2015MNRAS.451.1247S}. 
The stellar masses of cluster galaxies are derived from fitting synthesized spectra to the measured photometry using FAST \citep{2009ApJ...705L..71K}. 
The contribution from cluster galaxies are parameterized by NFW halos with scale radii derived from the dark matter mass using again a relation found in the same simulation. 

%%%%%%%%%%%%%%%%%%%%%%%%%%%%%%%%%%%%%%%%

\subsection{\grale: the GRALE model}

The GRALE models are based on the reconstruction code \grale\footnote{\grale's description, software and installation instructions are available at\\ 
{\tt{http://research.edm.uhasselt.be/$\sim$jori/grale}}}.
\grale~is a flexible, free-form method, based on a genetic algorithm, that uses an adaptive grid to iteratively refine the mass model. As input it uses only the information about the lensed images, and nothing about cluster's visible mass \citep{lie06}. This last feature sets \grale~apart from many other lens mass reconstruction techniques, and gives it the ability to test how well mass follows light on small and large scales within clusters. \grale~has been used to reconstruct mass distributions in a number of clusters \citep{lie08,lie09,moh14,moh15}, quantify mass/light offsets in Abell 3827 \citep{moh14,mas15}, derive projected mass power spectra and compare to those of simulated clusters \citep{moh15b}, and to study the relation between mass and light in MACS0416 \citep{seb15}. These papers used strong lensing constraints only, and so the analysis was confined to the central regions of galaxy clusters.

\subsubsection{Description of the method}

\grale~starts out with an initial coarse uniform grid in the lens plane which is populated with a basis set, such as projected Plummer density spheres. A uniform mass sheet covering the whole modeling region is also added to supplement the basis set.  As the code runs the denser regions are resolved with a finer grid, with each cell given a Plummer with a proportionate width. The initial trial solution, as well as all later evolved solutions are evaluated for genetic fitness, and the fit ones are cloned, combined and mutated. The final map consists of a superposition of a mass sheet and many Plummers, typically several hundred to a couple of thousand, each with its own size and weight, determined by the genetic algorithm. Critical curves, caustics and magnifications for any given source redshift are automatically available.

Multiple fitness measures are used in \grale. These are: (a)~{Image positions.}  A successful mass map would lens image-plane images of the same source back to the same source location and shape. A mass map has a better fitness measure if the images have a greater fractional degree of overlap. Using fractional overlap of extended images ensures against over-focusing, or over-magnifying images. (b)~{Null space.} Regions of image plane that definitely do not contain any lensed features belong to the null space. (c)~{Critical lines.} In some cases, it is known on astrophysical grounds that a critical line cannot go through certain image regions, but must pass between them.  \grale~ can incorporate this type of constraint, but we have not used this fitness measure in the Frontier Fields work so far. (d)~{Time delay measurements.} Though not used in the present work, time delays measurements can also be incorporated into the fitness \citep{lie09,moh15}.

Each \grale~run with the same set of images will produce a somewhat different final map. The dispersion between these quantifies mass uncertainties which are due to mass degeneracies present when all image information is held fixed. The best known among these, the mass sheet degeneracy, is broken in most clusters because of the multiple redshifts of background sources. The other, more numerous and less known degeneracies---documented \citep{sah00,lie12} and not documented---are the ones that contribute to the uncertainties. 

The clusters {\em Ares} and {\em Hera} were modeled with multiple images as inputs, and using
two fitness measures: (a) image positions, and (b) null space for each
source (image set) separately. For image sets where it was not
entirely clear if or where the counterimages might be present, the
nulls were allowed to have large holes corresponding to the regions of
possible additional images. \grale~ can operate in two modes: with
lensed images represented by points, or by extended images. The
present reconstruction were done using the extended image mode.

\subsubsection{Strengths and weaknesses}

The main advantage of \grale~is its flexibility, and hence ability to explore a wide range of lensing mass degeneracies. Another important feature, which can be viewed as strength, is that \grale~does not use cluster galaxies, or any information about the distribution of luminous matter to do the mass reconstruction. This is useful if one wants to test how well mass follows light \citep{moh14,seb15}.

\grale's main weakness is that it is not an optimal tool for identifying lensed images. This is a direct consequence, or, one may say, the flip side of \grale's flexibility. A technical feature of \grale~worth mentioning is that it requires significant computational resources: \grale~runs on a supercomputer.

\subsubsection{Improvements in progress} 

The \grale~team has carried out numerous tests of the code, to optimize the set of genetic algorithm and other code parameters. In the near future \grale~will be extend to include fitness measure constraints from weak shear and flexion.

%%%%%%%%%%%%%%%%%%%%%%%%%%%%%%%%%%%%%%%%

\subsection{LensPerfect: the Coe model}

The Coe model for {\em Ares} uses 
LensPerfect\footnote{http://www.stsci.edu/$\sim$dcoe/LensPerfect/} 
\citep{Coe08, Coe10}.
LensPerfect makes no assumptions about light tracing mass.  The lens models perfectly reproduce the input observed positions of all strongly lensed multiple images.  Redshifts may be either fixed to input spectroscopic redshifts or included in the model optimization based on input photometric redshifts and uncertainties.

The image positions, redshifts, and estimated source positions define the lensing deflection field sparsely at the multiple image positions.  LensPerfect interpolates this vector field, obtaining a smooth model which exactly reproduces the image deflections at the input image positions.  Based on this 2D deflection map, the mass distribution, magnification, and all other model quantities may be derived.

\subsubsection{Description of the Method}

The curl-free vector interpolation scheme \citep{Fuselier06, Fuselier07} uses direct matrix inversion to obtain a model composed of radial basis functions (RBFs) at the positions of the input vectors (our multiple image locations).  Each 2D RBF has two free parameters -- amplitude and rotation angle -- which are determined uniquely by the matrix inversion.

After setting the width of the RBF, the free parameters are the source positions and any uncertain redshifts.  LensPerfect performs an optimization routine searching for those parameters which yield the most “physical” mass model according to a set of criteria including positive mass smoothly decreasing outward from the center on average with rough azimuthal symmetry.  See \cite{Coe08, Coe10} for more details.

\subsubsection{Strengths and Weaknesses}

In high-resolution HST ACS images, strongly lensed multiple image locations are observed and measured with accuracies of $\sim$1 pixel, or $\sim$0.05''.  By fully utilizing this information, LensPerfect is able to obtain relatively high resolution maps of galaxy cluster substructure without relying on any assumptions about light tracing mass.  Large numbers of multiple images may be input, and the number of free parameters is always roughly equal to the number of constraints.  The mass model spatial resolution increases with the density of multiple images on the sky.

Given current numbers of multiple images (up to $\sim$100 or so) for a single cluster \citep[e.g.,][]{Coe10}, LensPerfect can accurately recover cluster mass profiles along with some larger subhalos.  Magnifications, however, are influenced by local mass density gradients, which are not accurately reproduced by LensPerfect given current constraints.  Furthermore, LensPerfect mass models are only well constrained within the area enclosed by the multiple images and should generally be disregarded outside this region.

\subsubsection{Future improvements}

LensPerfect is well suited to future datasets such as JWST imaging revealing still greater numbers of multiple images.  Initial tests with hundreds to a thousand multiple images show great potential for resolving many individual cluster galaxy halos without assuming light traces mass.  The biggest hurdle (seen in tests with up to 10,000 multiple images) may be accounting for multiple lens plane deflections due to mass along the line of sight.

One potential improvement would be to develop a hybrid method combining light traces mass assumptions with LensPerfect adding deviations to the mass distribution.

%%%%%%%%%%%%%%%%%%%%%%%%%%%%%%%%%%%%%%%%

\subsection{{\tt Lenstool}: the CATS and Johnson-Sharon models}

{\tt Lenstool} as an inversion algorithm deploys both strong and weak lensing 
data as input constraints. Below, we first briefly outline the available capabilities of the {tt Lenstool} software package and then describe the specific versions and assumptions that were used to reconstruct {\em Ares} and {\em Hera} by two groups: CATS and Johnson-Sharon. The CATS collaboration developed the {\tt Lenstool} algorithm collectively over a decade. 
The code utilizes the positions, magnitudes, shapes, multiplicity and spectroscopic redshifts for the multiply imaged background galaxies to derive the detailed mass distribution of the cluster. The overall mass distribution in cluster lenses is modeled in {\tt Lenstool} as a super-position of smoother large-scale potentials and small scale substructure that is associated with the locations of bright, cluster member galaxies. Individual cluster galaxies are always described by parametric mass models, whereas the smoother, large-scale mass distribution can be flexibly modeled non-parametrically or with specific profiles. This available multi-scale approach is optimal, in as much as the input constraints required for this inversion exercise are derived from a range of scales. Further details of the methodology are outlined in Jullo \& Kneib (2009). In its current implementation in {\tt Lenstool}, the optimization of the combined parametric and non-parametric model is computationally time intensive. And some degeneracies persist, despite the large number of stringent input constraints from the positions, shapes, brightnesses, and measured spectroscopic redshifts of several families of multiple images. However, these degeneracies are well understood, in particular for specific parameters of models used to characterize the mass distribution. In order to tackle this challenge an iterative strategy has been developed wherein initial models are derived with the best-fit values solely from the parametric model, which are then optimized using the underlying
multi-scale grid. Both the multi-scale and the parametric models are adjusted in a Bayesian way, i.e.,their posterior probability density is probed with a MCMC sampler. This process allows an easy and reliable estimate of the errors on derived quantities such
as the amplification maps and the mass maps.

The CATS and the Sharon Johnson models are built using the {\tt Lenstool} public modeling software \citep[see e.g.][]{2007NJPh....9..447J}. The public version of {\tt Lenstool} deployed by Johnson-Sharon adopts the original modeling approach developed by \cite{1997MNRAS.287..833N} wherein a small-scale dark-matter clump is associated with each bright cluster galaxy and a large-scale dark-matter clump with prominent concentrations of cluster galaxies. This technique of associating mass and light has proven to be very reliable and results in mass distributions that are in very good agreement with theoretical predictions from high-resolution cosmological N-body simulations.  The Johnson-Sharon models follow the methods described in \cite{2012ApJ...746..161S,2014ApJ...797...48J}.

\subsubsection{Description of the method}

Typically, cluster lenses are represented by a few cluster-scale or group-scale halos (representing the smooth
component, with $\sigma$ in the range of hundreds to $\sim 1500$ km s$^{-1}$), with contribution from galaxy-scale halos (see below).
Large scale dark matter halos are parametrized as Pseudo-Isothermal Elliptical Mass Distribution (PIEMD), 
\begin{equation}
\rho(r) = \frac{\rho_0}{(1+r^2/r^2_{core})(1+r^2/r^2_{cut})},
\end{equation}
where $\rho_0$ is a normalization, and $r_{core}$ and $r_{cut}$ define a region $r_{core}\lesssim r \lesssim r_{cut}$ in which the mass distribution is isothermal, i.e., $\rho \propto r^{-2}$. 
In Lenstool, PIEMD has seven free parameters: $x$, $y$ are the coordinates on which the halo is centered, $e$ and $\theta$ are the ellipticity and the position angle, respectively; $r_{core}$; $r_{cut}$; and effective velocity dispersion $\sigma_0$ which determines the normalization \citep[we note that the $\sigma_0$ is not exactly the observed velocity dispersion, see][]{2007arXiv0710.5636E}.
The parameters of cluster-scale halos are kept free, with the exception of $r_{cut}$ which is usually unconstrained 
by the strong lensing data, and is thus fixed at an arbitrary value (typically 1500 kpc). 

CATS also model galaxies as PIEMD, whereas Johnson-Sharon model galaxies as circular isothermal distributions (see 3.6.5). 
To keep the number of free parameters reasonably small, the parameters of galaxy-scale halos are determined from their photometric properties through scaling relations assuming a constant mass-to-light ratio for all galaxies,
\begin{equation}
\sigma _0 = \sigma_0^*\Big(\frac{L}{L^*}\Big)^{1/4} {\rm  ~~~~ and ~~~~}
r_{cut} = r_{cut}^*\Big(\frac{L}{L^*}\Big)^{1/2}.
\end{equation}
The positional parameters, $x$, $y$, $e$, and $\theta$, are fixed to their observed values as measured from the light distribution in the imaging data. 

CATS used the simulated strong lensing catalogs and the Lenstool software to perform a mass reconstruction of both simulated clusters, assuming a parametric model for the distribution of dark matter. The model is optimized with the Bayesian Markov chain Monte-Carlo sampler, described in detail in Jullo et al. (2007). The mass distribution is
optimized in the image plane by minimizing the distance between the observed and predicted multiple image positions. Weak lensing information is not taken into account. The image-plane root mean squared (RMSi) distance of the images predicted by the model were used to compare with the observed positions as an accuracy estimator 
of the model (Limousin et al. 2007).

The CATS collaboration has modeled both clusters, {\em Ares} and{\em Hera} have been modeled as bi-modal clusters with two smooth dark matter clumps and two BCGs lying in the centre of those main clumps.
Each smooth component is modeled using a PIEMD profile.
Cluster member galaxies are taken from the given simulated catalogs up to a magnitude of $m_{f160w}$ < 22.0 for {\em Ares} and
$m_{f814w}$ < 24.0 for {\em Hera}. These are modeled with PIEMD profiles under the assumption that (i) their position, ellipticity and orientation corresponds to the brightness profile of their associated galaxy, (ii) their mass is proportional to the galaxy magnitudes in F160W band. 
In the provided models, it is assumed that they all have the same M/L ratio.  All multiple images provided were used in this model. 
In addition, a few (massive) cluster galaxies in both clusters were
more carefully modelled in order to improve the RMSi of nearby
multiple images. Four central cluster galaxies were modeled in this
way in {\em Hera} (of which one is considered to be a foreground) and three,
also central, galaxies in {\em Ares}.\\
These reconstructions have a resulting RMS in the image plane of 0.87'' for {\em Ares} and 0.95'' for {\em Hera}.\\

The Johnson-Sharon models for {\em Ares} and{\em Hera} were constructed using techniques similar to those in Johnson et al. (2014), using the catalogs of multiple images that were provided to the lens modelers as positonal constraints. The redshifts of the background sources were assumed to be known spectroscopically with no uncertainty or outliers. 
 Both clusters were modeled with two PIEMD halos, to represent the smooth dominant dark matter components, each centered close to the two peaks in the light distribution in the mock HST images with their exact positions set by the MCMC minimization process. 

Individual PIEMD halos were assigned to each galaxy in the provided catalog, with positional parameters, $x$, $y$, $e$, and $\theta$, fixed to their observed values as measured from the light distribution in the mock imaging data.
The parameters that describe the slope of the projected mass density were scaled with the light in the F125W band 
assuming a constant M/L ratio for all the galaxies, following the scaling relations in Equation (12). As both clusters are at z~0.5,  the same scaling relations were used for the cluster member galaxies: $\sigma_0^\star=120\ \mathrm{km\ s^{-1}}$, $r_\mathrm{core}=0.15$ kpc, and $r_\mathrm{cut}=30$ kpc, and $m_\star=20.00,19.87$ for {\em Ares} and Hera, respectively. 

A few galaxies located near constraints were modeled independent of the scaling relations and their core radius and velocity dispersion were left as free parameters in the lens models. This includes the two bright cluster galaxies lying at the centers of the gravitational well of both clumps in the dark matter distribution in both clusters.

We note that the PIEMD functional form of the cluster galaxies used in {\tt Lenstool} differs from the function that was used in the simulation of {\em Ares} in the treatment of the truncation radius. While the PIEMD profile transitions smoothly from isothermal and asymptotes to zero at large radii, the {\it simulated} mass distribution truncates the mass function sharply to zero at $r=r{cut}$. This discrepancy is what causes the sharp circular residuals seen in Figure~8. We thus do not expect the model to accurately reconstruct the mass distribution at radii larger than the truncation radius. 

In addition, the Johnson-Sharon model assumes that the ellipticity and position angle of the light of each mock galaxy follows the underlying mass distribution. In practice, all the galaxies in the underlying simulated mass distribution had circular geometry (i.e., no ellipticity) and the galaxies were painted on with arbitrary ellipticities and position angles. This feature of the blinded analysis contributes to residuals on small scales in the mass reconstruction. Finally, the Johnson-Sharon model does not use weak lensing information and does not include cluster-scale halos outside of the main field of view if such halos are not required by the strong lensing constraints alone. 

\subsubsection{Strength and weaknesses}

{\tt Lenstool} strengths and weaknesses are typical of parametric models. This approach is useful in the sense that it directly compares physically motivated models to data, propagating errors in a fully consistent an Bayesian manner. It allows direct comparison with simulation outputs and the assessment of possible discrepancies. 
On the other hand, parametric models can significantly differ from reality and their lack of freedom introduces biases in the estimated masses, matter densities or errors. Regarding practical aspects, errors estimation implies running MCMC sampling, which can only be performed on supercomputer. {\tt Lenstool} calculations can last for a couple of weeks on shared memory machines depending on the model complexity and the amount of multiple images. In the case of{\em Hera} and Ares, optimization lasts for about 10 hours.

\subsubsection{Improvements in progress}

CATS is currently working actively on two improvements that should significantly improve the accuracy of their mass reconstructions. First, {\tt Lenstool} in its current revision does not permit radial variation of the ellipticity for the mass distribution, and this restricts the flexibility of models that can be generated. Current code development aims to include this additional degree of freedom in the modeling. Secondly, in order to maximally extract information from the exquisite image resolution afforded by the HST FFI, flexion measurements will be included as input constraints in the modeling.  
Finally, a new MCMC engine with MPI support and a GPU-based {\tt Lenstool} are under development to decrease the computing time.

%%%%%%%%%%%%%%%%%%%%%%%%%%%%%%%%%%%%%%%%%%%%%%%%%%%%%%%%%%%%%%%%%%%%%%%%%%%%%%%%%%%

\subsection{GLAFIC: the GLAFIC models}
The publicly available GLAFIC code
\citep{2010PASJ...62.1017O}\footnote{http://www.slac.stanford.edu/\~{}oguri/GLAFIC/}
is used  
for mass modeling in the GLAFIC models.

\subsubsection{Description of the method}
GLAFIC adopts the so-called parametric lens
modeling 
in which the lens mass distribution is assumed to consist of
multiple components, each of which is characterized by a small number of
parameters such as the centroid position, mass, ellipticity, and
position angle. Mass distributions of cluster member galaxies are
modeled by the pseudo-Jaffe model. In order to reduce the number of
parameters, the velocity dispersion $\sigma$ and truncation radius
$r_{\rm trunc}$ of each member galaxy are assumed to scale 
with the galaxy luminosity L as $\sigma \propto L^{1/4}$ and $r_{\rm
  trunc}\propto L^\eta$, and the normalizations of the scaling
relations are treated as free parameters
\citep[see e.g.,][]{2010PASJ...62.1017O}. Ellipticities and position
angles of individual member galaxies are fixed to values measured in the
image. These parameters are optimized to reproduce positions of
observed multiple images, either using the downhill simplex method or
Markov-Chain Monte Carlo. Examples of detailed cluster mass modeling
with GLAFIC are found in \citet{2010PASJ...62.1017O},
\citet{2012MNRAS.420.3213O},  \citet{2013MNRAS.429..482O}, and
\citet{2015ApJ...799...12I,2016ApJ...819..114K}. GLAFIC can also simulate and fit lensed
extended sources. This functionality has been used to e.g, fit a lensed
quasar host galaxy \citep{2013MNRAS.429..482O}, estimate a selection
function of lensed high-redshift galaxies \citep{2015ApJ...799...12I},
and derive sizes of lensed high-redshift galaxies
\citep{2015ApJ...804..103K}. 

%%%%%%%%%%%%%%%%%%%%%%%%%%%%%%%%%%%%%%%%%%%%%%%%%%%%%%%%%%%%%%%%%%%%%%
\subsubsection{Strengths and weakness}
%%%%%%%%%%%%%%%%%%%%%%%%%%%%%%%%%%%%%%%%%%%%%%%%%%%%%%%%%%%%%%%%%%%%%%
An advantage of GLAFIC is a wide range of lens potential implemented
in the code, which enables flexible modeling of cluster mass
distributions. For example, in addition to the standard external
shear, one can add higher-order perturbations with arbitrary
multipole orders \citep[see][]{2010PASJ...62.1017O}. When necessary, 
in addition to observed multiple image positions, GLAFIC can also
include flexible observational constraints such as time delays and
flux ratios between multiple images, and (reduced) shear and
magnification values at several sky positions measured by weak lensing
and Type Ia supernovae, respectively. 

The source plane $\chi^2$ minimization is often adopted for efficient
model optimizations. In doing so, GLAFIC converts the distance between
observed and model positions in the source plane to the corresponding
distance in the image plane using the full magnification tensor. In
Appendix 2 of \citet{2010PASJ...62.1017O} it has been shown that this
source-plane $\chi^2$ is accurate in the sense that it is very close
to the image-plane $\chi^2$ and therefore is sufficient for reliable
mass modeling. 

Of course GLAFIC supports the image plane $\chi^2$ minimization as
well. Adaptive-meshing with increased resolution near critical curves
is used for efficient computations of multiple images for a given
source position. Multiple images and critical curves are computed for
the best-fit model from the source plane $\chi^2$ minimization to
check the robustness of the result. 

A known limitation of GLAFIC is that it can only handle single lens
planes. Lens systems for which multiple deflections at different
redshifts play a crucial role are difficult to be modeled by GLAFIC.

%%%%%%%%%%%%%%%%%%%%%%%%%%%%%%%%%%%%%%%%%%%%%%%%%%%%%%%%%%%%%%%%%%%%%%
\subsubsection{Modeling {\em Ares} and {\em Hera}}
%%%%%%%%%%%%%%%%%%%%%%%%%%%%%%%%%%%%%%%%%%%%%%%%%%%%%%%%%%%%%%%%%%%%%%
Each halo component is modeled by the elliptical NFW profile. For Ares,
five halo components are included, in addition to the member galaxies
modeled by the pseudo-Jaffe model (see above). For {\em Hera}, two NFW halo
components are placed around two brightest galaxies. These two
brightest galaxies are modeled by the Hernquist profile, separately
from the other member galaxies. Ellipticities and position angles of
the brightest galaxies are fixed to observed values. To achieve better
fit, external shear and third-order multipole perturbation are added
for {\em Hera}. In modeling member galaxies, $\eta$ in the scaling relation
of truncation radii is fixed to 0.5 for {\em Ares} and is treated as a free
parameter for {\em Hera}. Simulated F814W band images are used to measure
luminosities, ellipticities, and position angles of member galaxies
with SExtractor for both {\em Ares} and {\em Hera}. Overall, more elaborated lens
model is adopted for{\em Hera} compared with Ares, because  in the initial
exploration period of mass modeling it was found that the lens
potential of{\em Hera} appears to be much more complex. The resulting
best-fit models reproduce image positions very well, with rms of $\sim
0\farcs27$ for {\em Ares} and $\sim 0\farcs43$ for {\em Hera} .

\subsection{LTM: the Zitrin-LTM models}

The Zitrin light-traces-mass (LTM) method \citep[][]{2009MNRAS.396.1985Z,BR05.1}, was designed primarily to be a very simple, straightforward modeling method with a minimal number of free parameters, relying only on the observable light distribution of cluster members (namely their positions and relative fluxes) to supply a well-guessed and highly predictive solution to the mass distribution of the lens and the location of multiple image systems \citep[e.g.][]{2012MNRAS.423.2308Z,2013ApJ...762L..30Z}.  Previous to the design of this method, it had been shown that (a) cluster galaxies must be included, typically with a mass in proportion to their luminosity, in order the solution to have predictive power to find multiple images and that (b) a dark-matter component should be added \citep[see][]{KN04.1,BR05.1}. This simple parametrization, as we detailed further below, has allowed to identify systems of multiple-images in an unprecedented number of clusters, where the images are physically matched also by the initially-guessed model (which is then refined), and are not only matched by eye based on their color information as is often accustomed. 

\subsubsection{Description of the method}

As mentioned above, this method was designed to include both a galaxy component and a dark-matter component, yet to successfully do so with a minimal number of free parameters. To form the galaxy component, cluster galaxies (found following the red sequence in a color-magnitude diagram) are assigned each with a power-law mass density distribution, where the normalization of each galaxy's weight is proportional to its (relative) flux, and the exponent is the same for all galaxies and is the first free parameter of this method. The superposition of all galaxy power-law mass distribution then constitutes the lumpy, galaxy component of the model. To describe the dark-matter distribution, the galaxy component is smoothed with either a Spline interpolation or usually a Gaussian kernel whose degree or width, is the second free parameter of this method. The smoothing yields a diffuse, smooth dark-matter component that depends on the initial light distribution; therefore the method is dubbed Light-Traces-Mass as both the galaxy \emph{and} dark-matter components roughly follow simply the light distribution. Next, the two components are  added with a relative weight (typically around few to a couple dozen percents for the galaxies), which is another free parameter in the modeling. The fourth parameter is  an overall normalization of the lens model to a certain redshift or multiple-image system. In addition to the four parameters, we often introduce several other parameters that add some flexibility and help in refining the final solution given the set of input multiple images. These include a core and two-parameter ellipticity for the BCG(s), two parameter external shear (which mimics ellipticity for the critical curves), and chosen galaxies whose weights (or fluxes) are left free to be optimized in the modeling, meaning that they are allowed to deviate from the adopted mass-to-light relation. The minimization for the best-fit solution and related errors, given a set of multiple-images (often found with the aid of the initially guessed map from this method), is performed with a $\chi^{2}$ criterion comparing the positions of multiple-images with the predicted ones, in the image plane, via a few-dozen thousand MCMC steps with Metropolis-Hastings algorithm. 

\subsubsection{Strengths and weaknesses}

The resulting lens model from this procedure, as its name suggests, is strongly coupled to the input light distribution of the lens (cluster members positions and luminosities). This entails various strengths and weaknesses. The fact that the solution is coupled to the light distribution is what grants this method with the unprecedented prediction power to delineate the critical curves and locate multiple images in advance, even if no multiple-image system is used as constraint \citep[see ][]{2012MNRAS.423.2308Z}. In fact, most of the free parameters in the initial solution are relatively well known, so that as a first step (i.e. to find multiple images) we can reduced these to one free parameter - namely the normalization of the lens, and obtain a well-guessed solution, that we have shown is not much different that the resulting solution for the same clusters when using many multiple images as constraints \citep{2012MNRAS.423.2308Z}. This means that the method is capable to supply a well guessed solution also in cases where HST high quality data is lacking. 

The simplistic nature of this method also means that the solution is often faster to converge and compute than other grid-based methods or parametrizations, allowing the analysis of many dozens of cluster lenses in a relatively short time.  

Another  advantage that this method encompasses is that the same very simple procedure applies to all clusters - from relaxed, small clusters and groups \citep[such as the relatively smaller cluster lenses A383, MS2137 or A611, see ][for recent modeling]{2015ApJ...801...44Z}, to the most complex merging clusters such as M0416, M1149 or M0717 \citep{2015ApJ...801...44Z}, that often require multi-halo fits in other parametric methods..  

But the coupling to the light distribution also means that the spatial flexibility of the model is small. While our parameterization does allow for a flexible mass profile in the sense that it is not limited to a certain analytic form, the solution is limited spatially by the light distribution. This means that the multiple-image reproduction accuracy is often smaller than in other more flexible parametric methods that model the dark-matter independently of the light (such as other well-known methods listed in this work including our own second method listed below). This is manifested usually in clusters that have a large number of multiple-images spread across the field; for these the LTM method often reaches a finite \emph{rms} value of $\sim 1-2"$ on which it cannot improve further. 

A second disadvantage stemming from our parameterization -- since we do not model the dark matter independently of the light and since the critical curve's ellipticity in our modeling 
is for the most part generated by the external shearshear, is that there is  no ellipticity assigned directly to the mass distribution. This creates some discrepancy between the \emph{lens} and \emph{mass} models: the mass distribution can be often significantly rounder than implied by the critical curves, whose ellipticity comes from the external shear that does not contribute ellipticity to the convergence map. In simple words, this reveals a degeneracy regarding the true ellipticity of the mass distribution - as the ellipticity of the lens can be attributed to intrinsic ellipticity or to external shear. 

To summarize, we thus consider this method very \emph{reliable and robust}, supplying especially well-guessed initial maps for any given cluster regardless of its complexity, and with unprecedented prediction power to find multiple-images, but it can also be in some cases \emph{less accurate and spatially flexible}. Also, given this is a light-tracing method,  we do not expect this method to describe well numerical simulations whose mass to light relations are not completely representative.

\subsubsection{Improvements in progress}

We are always looking for ways to speed up the minimization procedure so that a larger parameter space could be for a refined final solution. We are also testing whether replacing the galaxy component with the more well-behaved PIEMD (see below), despite having a fixed isothermal slope, would be sufficient for our purposes. We have also implemented an option of smoothing with an elliptical gaussian which then introduces ellipticity into the matter distribution itself. 

Note that our calculations are performed on an input grid matching an actual image of the field, with its native pixel-scale. To speed the minimization procedure, we often reduce the resolution (especially in the case of HST which has high spatial resolution) by factors 4 to 10 on each axis. This contributes to the finite, non-negligible rms obtained often in this method (e.g. due to pixel coordinates round ups etc.). We intend to investigate this further and try to improve the resolution in the crucial places, such as near the critical curves and when delensing to the source plane, where this lower resolution might prevent a further improved solution.

\subsubsection{Modeling of {\em Ares} and {\em Hera}}
To model {\em Ares}  and {\em Hera} we use the following setting in our LTM pipeline. We  create a grid of 4080$\times$4080  pixels covering the field-of-view, with an angular resolution of $0.5"$/pixel. The calculation in practice is performed in two stages - first we run many individual random MC chains with a grid resolution lower by factor 10 on each axis. From this we find the global minimum area and extract the covariance matrix. A proper, long MCMC is then run with a grid of 4 times lower resolution than the original input image. The final solution is then interpolated to match the original pixel scale map. Errors were derived using 50 random models form the MC chain, with a positional uncertainty of 1.4" for the $\chi^{2}$ term.
We use the input list of galaxies supplied by the simulators scaled by their light. In {\em Ares}, we allow five galaxies to deviate from the nominal mass-to-light ratio and be freely weighted by the MC chain, and for two of them -- especially important where radial images are seen in the data -- we allow for a free core radius as well. In the case of {\em Hera}, only three galaxies were modeled in this way. The ellipticity (and direction) of these  bright galaxies are also left as free parameters.
As constraints, we use the full list of multiple images. No weak lensing constraints were used. The final rms of the model is 1.8" and 1.2" for {\em Ares} and {\em Hera}, respectively, which, as we mention above, is in part limited by the finite lower resolution of the grid we work on.

\subsection{PIEMDeNFW: the Zitrin-NFW models} 
\cite{2013ApJ...762L..30Z}  expanded their pipeline to also allow for a fully parametric solution. This method in essence is similar to the other parametric techniques mentioned here such as {\tt Lenstool} and GLAFIC. The main motivation for adding this parametric pipeline was to (a) allow for further flexibility and improved fits by having a semi independent solution in which the dark matter is modeled independently of the light, and (b) test for the magnitude of systematic differences between these methods \citep{2015ApJ...801...44Z}. 
  
\subsubsection{Description of the method}
As is usually accustomed in parametric modeling, in order to describe well the multiple-image positions with enough prediction power,  this method also relies on a combination of galaxy and dark matter. The red sequence cluster galaxies are modeled each as PIEMDs based on the prescription and scaling relations used in {\tt Lenstool}, and typically with a fixed mass-to-light ratio. Usually two or three parameters are left free to describe the galaxy component: the velocity dispersion, core radius and truncation radius, of an M* galaxy, which is used as reference for the scaling relations.  
The dark matter component is modeled also with an analytic, fully parametric recipe. We can choose either an elliptical \cite{NA96.1} profile (eNFW), or, also PIEMD for the cluster's dark matter halo. Therefore, in this method, the dark matter is modeled with a symmetric analytic form, independent from the light distribution. Similar to our LTM method the same minimization engine is used here: a long MCMC with a $\chi^{2}$ image-plane criteria. Also here we can add other parameters to be optimized in the minimization, such as the ellipticities of the BCGs, their mass can be allowed to deviate from the adopted scaling relation, and so forth.
  
\subsubsection{Strengths and weaknesses}

Compared to our LTM technique, for example, we have found that the fully parametric technique is more spatially flexible and can thus often supply a more accurate solution with a (somewhat) smaller image-plane rms.  On the other hand, the higher flexibility reduces the prediction power for finding multiple images (especially before the model is initially constrained), and the reliability of the results, since they can for a wider range of (not necessarily physically viable) configurations. 

In a similar sense, another main disadvantage of such parametrizations is the need to add dark matter halos to model sub halos for complex structures (such as merging clusters), without knowing if these are fundamental parameters, e.g. accounting truly for additional dark matter halos,  or just nuisance parameters that help add flexibility and refine the fit. Additionally, each such added halo adds several (usually 4-6) free parameters to the minimization procedure rendering it significantly more cumbersome.

Note that since we developed this method with the same infrastructure used for our LTM method, and in part, for comparison with it, the solutions given by the PIEMDeNFW method, despite being analytic in nature, are also calculated on a grid the size of the input *.fits image, similar to our LTM procedure. This results in a somewhat slower procedure compared, for example, to our LTM technique, and also here to achieve higher converging speed we lower the grid resolution by factors of a few on each axis. Again this leads to a finite rms due to e.g. numerical round-ups in high-magnification regions. 

\subsubsection{Improvements in progress}
The main improvement we wish to implement is to speed up the procedure. This for start can be achieved if part of the calculation is done completely analytically/numerically (say, only around the positions of multiple images) rather than on a full-frame grid. We intend to explore such possibilities.  
Also, recently we added the possibility for an external shear to allow for further flexibility.

\subsubsection{Modeling of {\em Ares} and {\em Hera}}

To model {\em Ares} and {\em Hera}, we use the following setting in our PIEMDeNFW pipeline. As done withe the LTM-gauss method,  we create a grid of 4080$\times$4080  pixels covering the cluster.  We start by running many individual random MC chains with a grid resolution lower by factor 20 on each axis. From this we find the global minimum area and extract the covariance matrix. A proper, long MCMC is then run with a grid of 4 times lower resolution than the original input image. The final solution is then interpolated to match the original pixel scale map. Errors were derived using 50 random models form the MC chain, with a positional uncertainty of 1.4" for the $\chi^{2}$ term.
We use the input list of galaxies, scaled by their light. The brightest galaxies are optimized individually as done with the LTM-gauss pipeline. In this case, however, the ellipticity (and direction) of the four brightest galaxies in both clusters are also left as free parameters.
In {\em Ares}, two cluster-scale DM halos in the form of elliptical NFW mass densities are introduced, with fixed centering on the respective BCGs. In {\em Hera}, we used three such large halos.
As constraints we use the full list of multiple images. No weak lensing constraints were used. The final rms of the model is 1.8", which as we mention above is in part limited by the finite lower resolution of the grid we work on.

\section{Results}
\label{sect:results}
In this section we describe how the different methods  perform at recovering several properties of the lenses. 

\subsection{Convergence maps}

\begin{figure*}
 \centering
 \includegraphics[width=1.0\hsize]{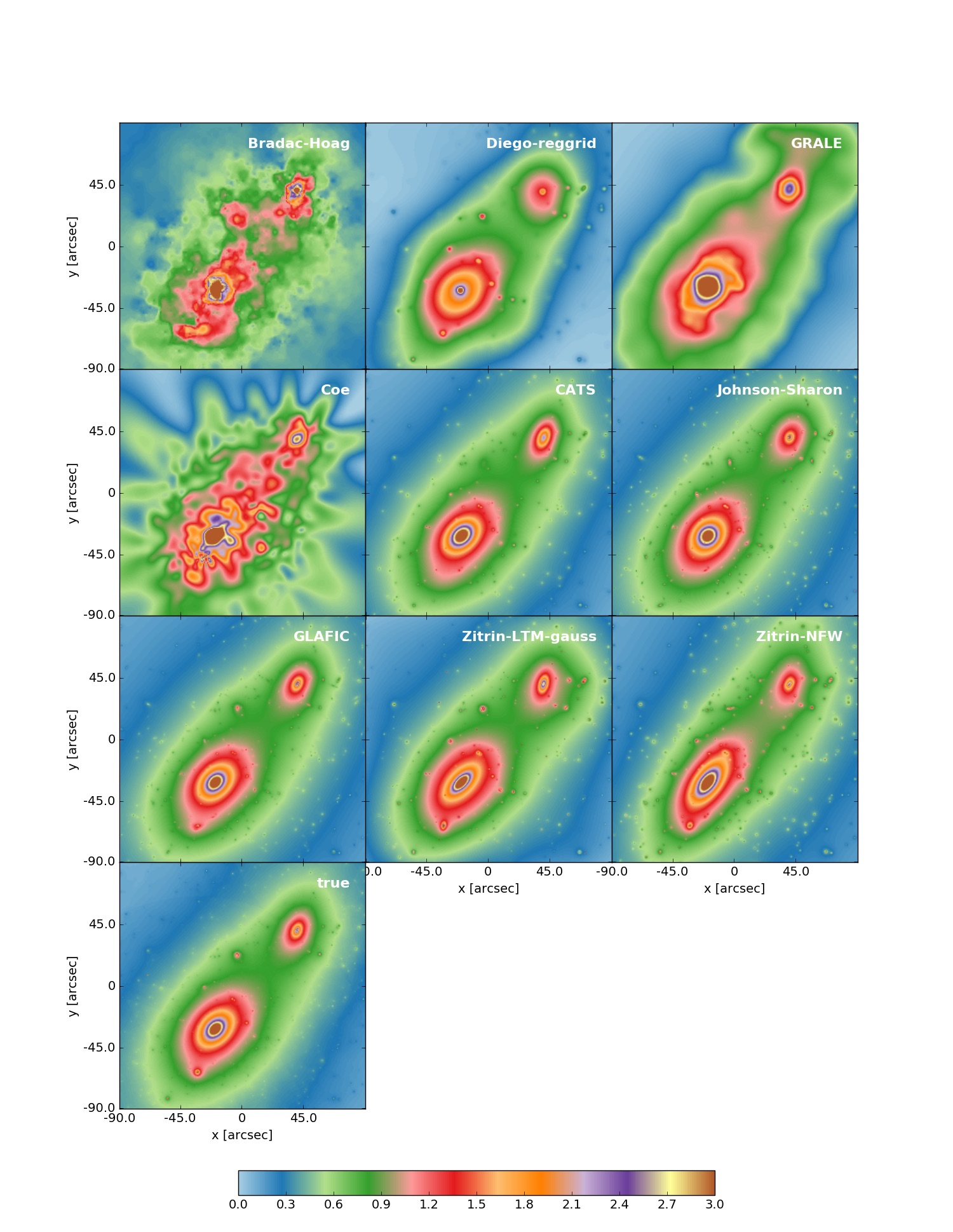}
  \caption{Convergence maps ($z_s=9$) of {\em Ares}. The first nine panels show the results of the reconstructions, beginning with the free-form methods (panels 1-4) and concluding with the parametric models (panels 5-9). The lower left panel shows the true convergence map, for comparison.}
 \label{fig:ares_kappamaps}
\end{figure*}

\begin{figure*}
 \centering
 \includegraphics[width=1.0\hsize]{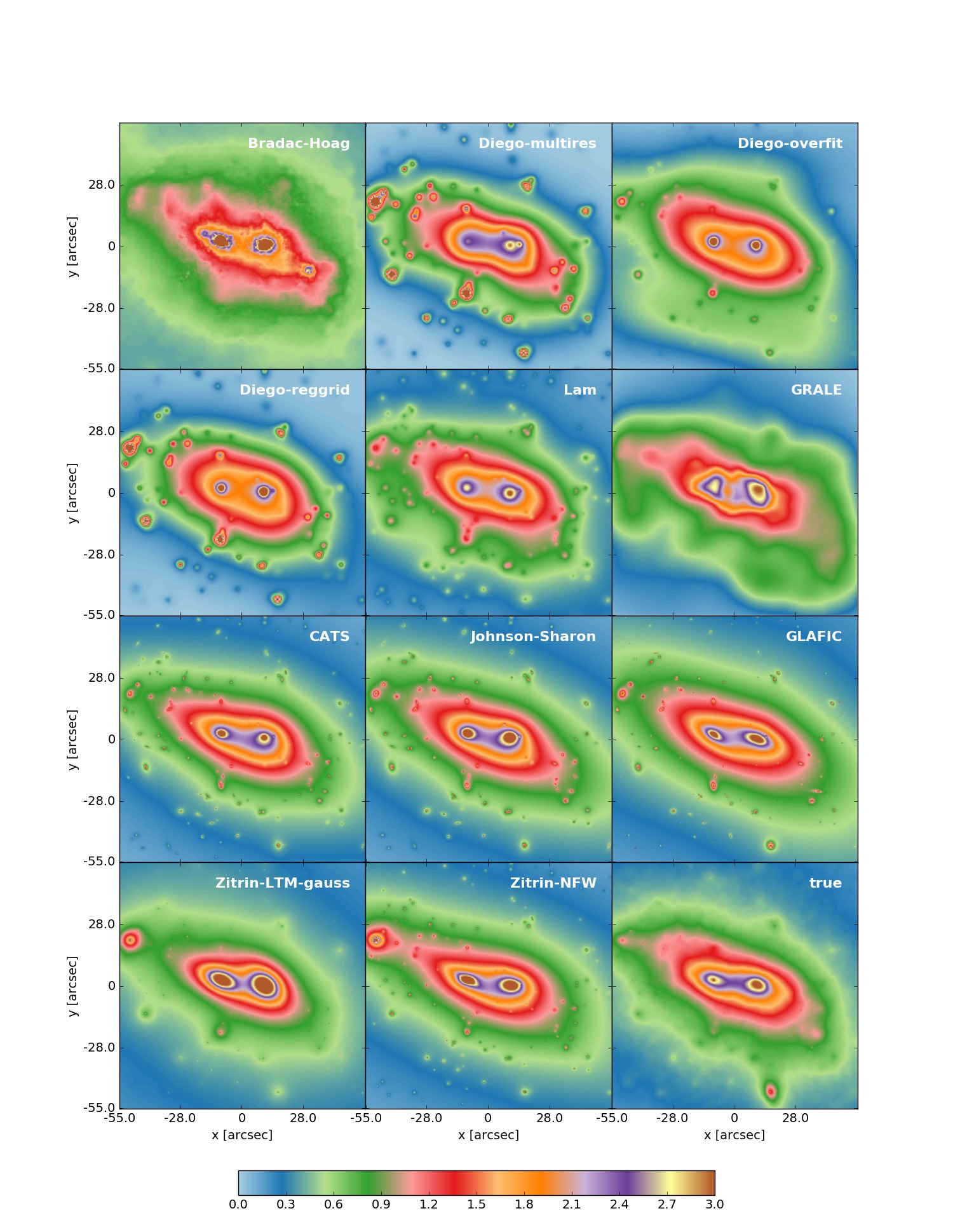}
  \caption{Convergence maps ($z_s=9$) of {\em Hera}. The first eleven panels show the results of the reconstructions, beginning with the free-form methods (panels 1-6) and concluding with the parametric models (panels 7-11). The lower right panel shows the true convergence map, for comparison.}
 \label{fig:hera_kappamaps}
\end{figure*}

The reconstructed convergence maps of {\em Ares} and {\em Hera} are shown for all models in Figs.~\ref{fig:ares_kappamaps} and \ref{fig:hera_kappamaps}, respectively. The maps are all normalized to $z_S=9$. In both figures, the maps derived from the free-form algorithms are shown first (beginning from the upper-left panel). The last panel in each figure shows the true convergence map, for easy comparison. All maps cover the same field of view. This does not correspond to the size of the simulated images that were made available to the modelers. Indeed, for several technical reasons inherent to each methodology employed, the submissions by the different groups were different in size. To carry out a proper comparison between the models, we restrict our analysis to the area around each of the two lenses, which is covered by all the reconstructions. More precisely, we used as footprints for identifying the area of analysis the submissions by the GLAFIC and by the GRALE teams for {\em Ares} and {\em Hera}, respectively. In the first case, the FOV is $\sim 180"\times 180"$. In the second, the reconstructed area is $\sim 110"\times 110"$ wide.   

Since {\em Ares} was constructed parameterically with light tracing mass, it is particularly well suited for reconstruction by parametric techniques. The parametric CATS, GLAFIC, Johnson-Sharon, and Zitrin models and the hybrid Diego model all include mass substructure at the observed positions of cluster galaxies, recovering the {\em Ares} mass distribution with high fidelity.  The free-form GRALE, Bradac-Hoag, and Coe models do not assume light traces mass, reconstructing the mass distribution solely based on the observed lensing.  They recover the main mass peaks, but smaller substructures are not constrained by the lensing data.  The GRALE model accurately reproduces the cluster bimodailty.  The Bradac-Hoag and Coe models are less smooth, including noisy smaller substructure, especially outside the region constrained by strongly lensed multiple images.

The {\em Hera} cluster, obtained from an N-body simulation, is less ideal for being reconstructed using parametric methods. Indeed, the performance of the parametric algorithms appears more consistent with that of the free-form ones. {\em Hera} is constructed assuming light traces its massive substructure, as assumed by the parametric and hybrid methods.  The Bradac-Hoag and GRALE models do not make that assumption and thus recover fewer small subhalos.

The major differences between the models and between the models and the true convergence maps are found near substructures, but also the shape of the mass distributions, especially at large distances from the center, show inconsistencies. We will discuss them in more details in the next sections.

\begin{figure*}
 \centering
 \includegraphics[width=1.0\hsize]{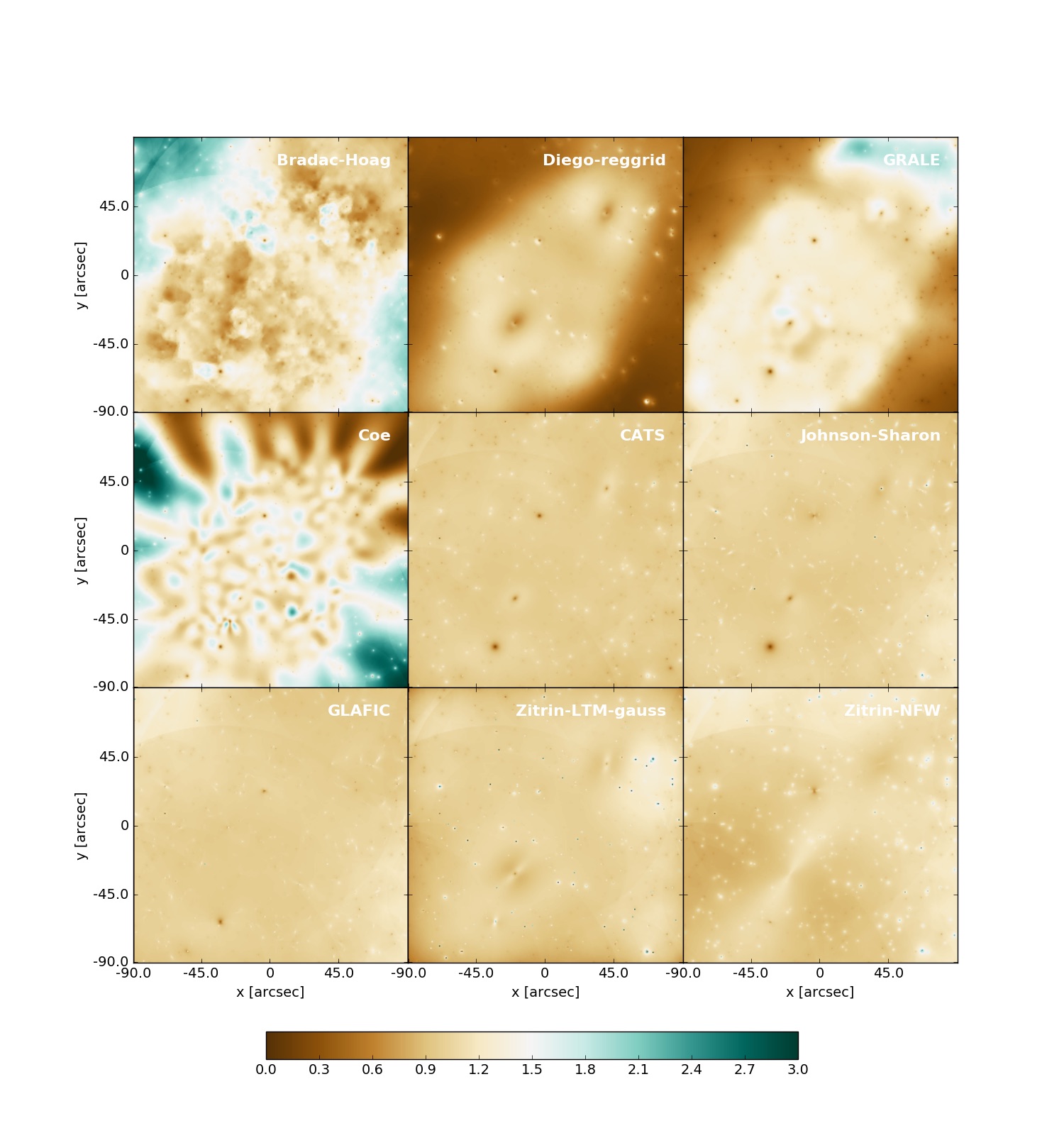}
  \caption{Ratios between the mass reconstructions and true {\em Ares} mass distribution.}
 \label{fig:ares_kapparatios}
\end{figure*}

\begin{figure*}
 \centering
 \includegraphics[width=1.0\hsize]{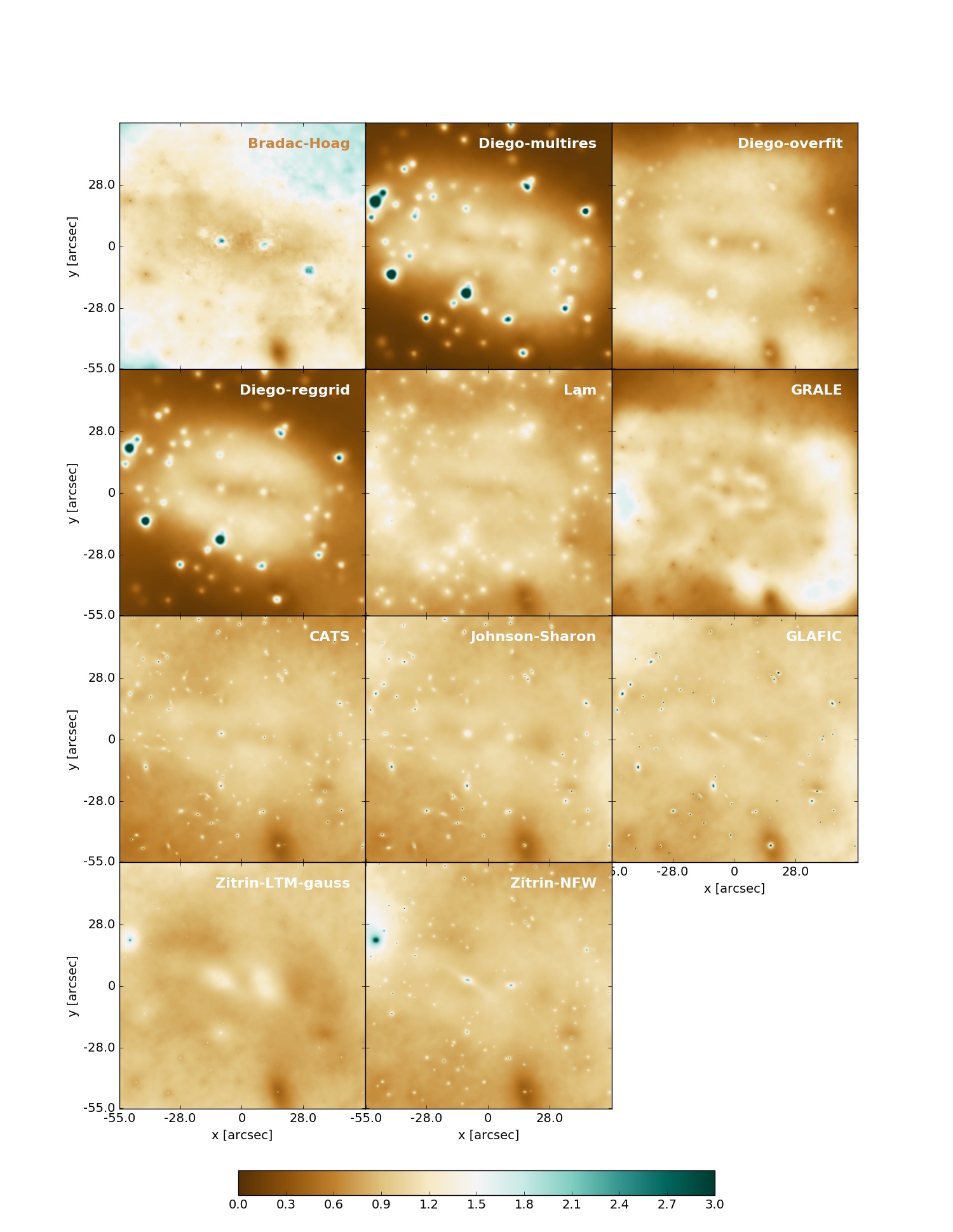}
  \caption{Ratios between the mass reconstructions and true {\em Hera} mass distribution.}
 \label{fig:hera_kapparatios}
\end{figure*}

To better highlight the differences between the maps, we show the ratios between the reconstructed and the true convergence maps for {\em Ares} and {\em Hera} in Figs.~\ref{fig:ares_kapparatios} and \ref{fig:hera_kapparatios}, respectively.

\subsection{One dimensional mass and convergence profiles}
\begin{figure}
 \centering
 \includegraphics[width=1.0\hsize]{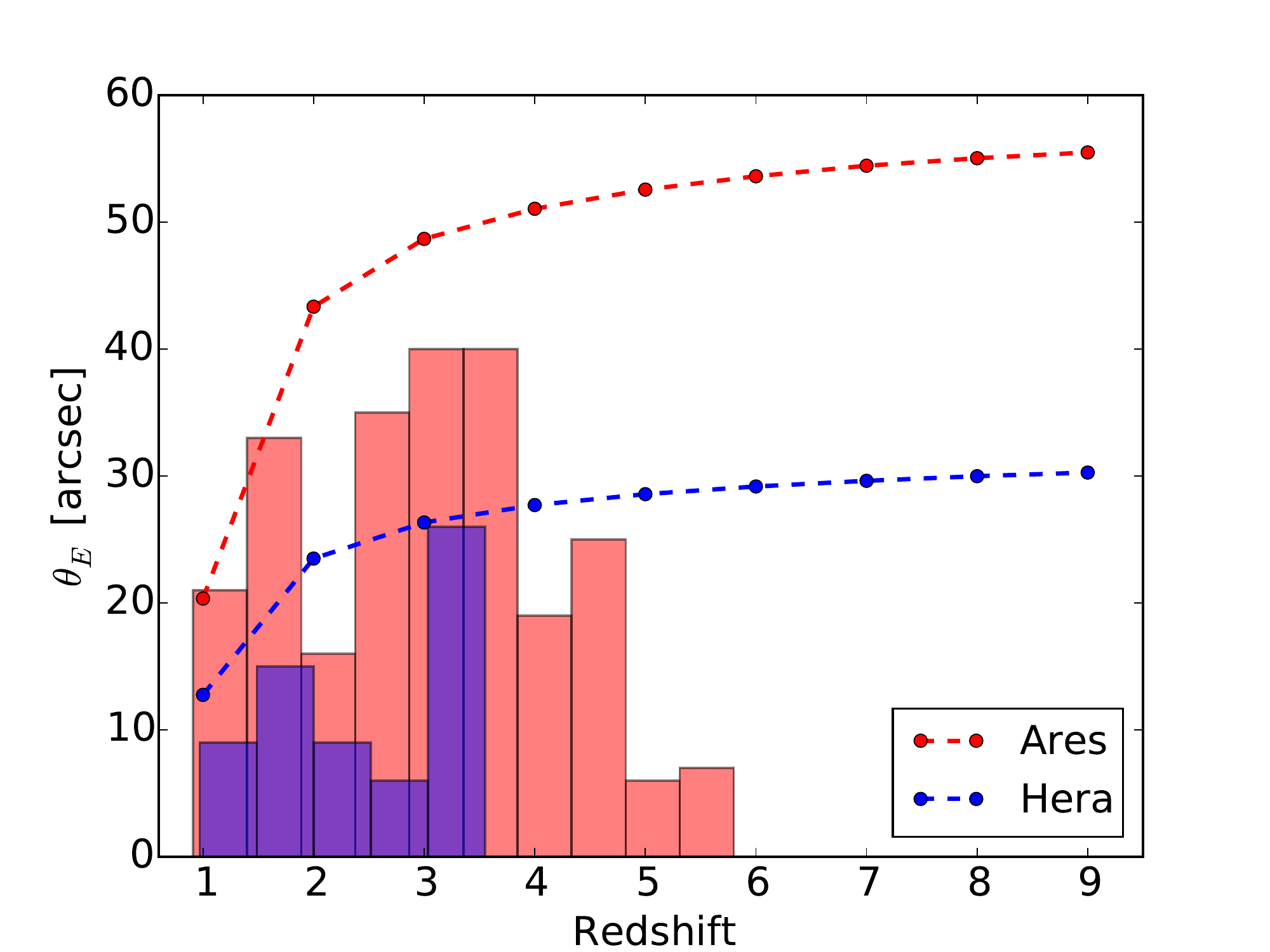}
  \caption{Size of the Einstein radii of {\em Ares} and {\em Hera} as a function of the source redshift (red and blue dashed lines, respectively). The histograms show the redshift distribution of the multiple images for the two clusters. }
 \label{fig:einst_radii}
\end{figure}

We begin discussing the results on the mass and convergence (or surface density) profiles. \cite{2010A&A...514A..93M} already showed using only one of the methods employed in this paper ({\tt Lenstool}, employed by both the CATS and the Johnson-Sharon teams) that strong lensing can potentially measure the mass inside the Einstein radius with an accuracy of the order of a few percent. In the cases of {\em Ares} and {\em Hera} , the sizes of the Einstein radii are significantly different. In Fig.~\ref{fig:einst_radii}, we show how $\theta_E$ grows as a function of the source redshift $z_s$. The Einstein radius of {\em Ares} is $\sim 20$ arcsec at $z_s=1$. Its size at $z_s\sim 2$ is more than doubled and  it grows asymptotically to $\sim 55$ arcsec at higher redshift. The reason of the steep rise between $z_s=1$ and $z_s=2$ is that {\em Ares} has a bi-modal mass distribution. For sources at low redshift ($z_s\sim 1$), each of the two mass clumps have their own critical lines. These are shown by the red curves in the upper middle panel of Fig.~\ref{fig:simobs}. To draw the plot in Fig.~\ref{fig:einst_radii}, we use the center of the most massive mass clump as reference, and only the critical line enclosing this point is used to measure $\theta_E$. By increasing the source redshift, the critical lines around the two mass clumps merge into a single, very extended critical line  (see the white line in the upper middle panel of Fig.~\ref{fig:simobs}, which shows the critical line for sources at $z_s=9$). 

In the case of {\em Hera}, the Einstein radius grows from $\sim 12$ arcsec at $z_s=1$ to $\sim 30$ arcsec at $z_s=9$. The critical lines for these two source redshifts are shown in the lower central panel of Fig.~\ref{fig:simobs}. 

In Fig.~\ref{fig:einst_radii} we also show the redshift distributions of the multiple images identified in the background of the two clusters (red and blue histograms). These multiple images are marked with numbered circles in the central panels of Fig.~\ref{fig:simobs}. The labels of each image are constructed as X.Y, where X is the ID of the source and Y is the ID of of the multiple images belonging to the same system. Being such a powerful lens, {\em Ares} produces many more multiple images than {\em Hera}, some of which originate from galaxies at redshift $z_s\sim 6$. The most distant multiple image system in the field of {\em Hera} is only at $z_s\sim 3.5$. In both cases, however, the redshift distribution of the multiple images overlaps with the redshift range where the size of the Einstein radii have the strongest growth. Indeed, the relative variation of $\theta_E$ between $z_s=3$ and $z_s=9$ is only $\lesssim 10\%$. Thus, we expect that the models constructed using these constraints can be safely used to trace the growth of the cluster strong lensing region up to very high redshifts. Anologously, we expect that the mass profiles are recovered with higher precision in the radial ranges $20 \lesssim \theta \lesssim 60$ arcsec and $10 \lesssim \theta \lesssim 30$ arcsec for {\em Ares} and {\em Hera}, respectively.

\begin{figure*}
 \centering
 \includegraphics[width=1.0\hsize]{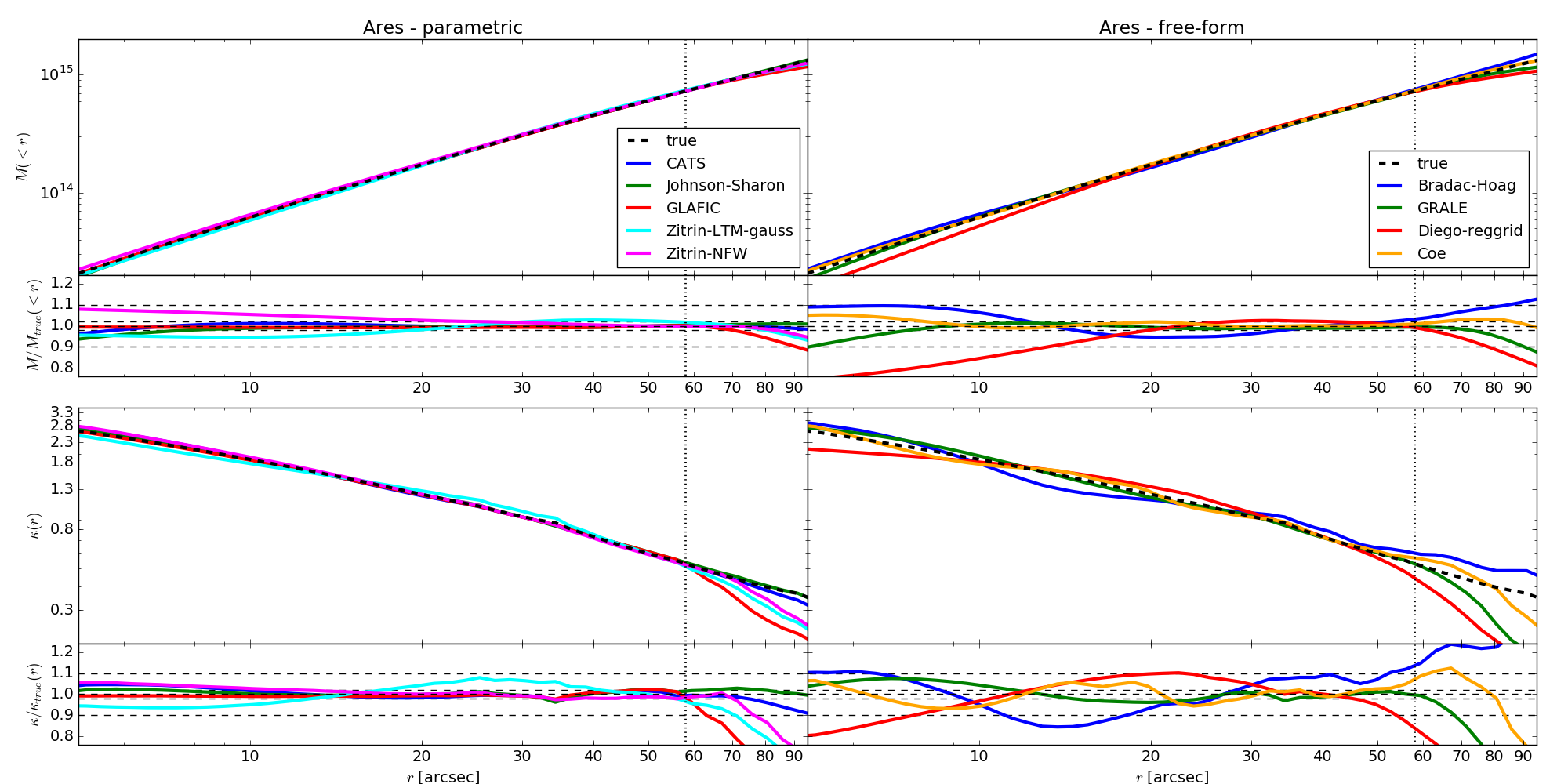}
  \caption{Mass profiles in the inner $100$ arcsec of Ares: enclosed mass (upper panels) and mass surface density (lower panels). Results for parametric and free-form methods are shown in the left and in the right panels, respectively. The insets on the bottom of each panel show the ratio between the reconstructed and the true mass profiles. The horizontal dashed lines correspond to $\pm2\%$ and $\pm 10\%$ differences between lens models and input mass distribution.}
 \label{fig:mass_profs_ares}
\end{figure*}

\begin{figure*}
 \centering
 \includegraphics[width=1.0\hsize]{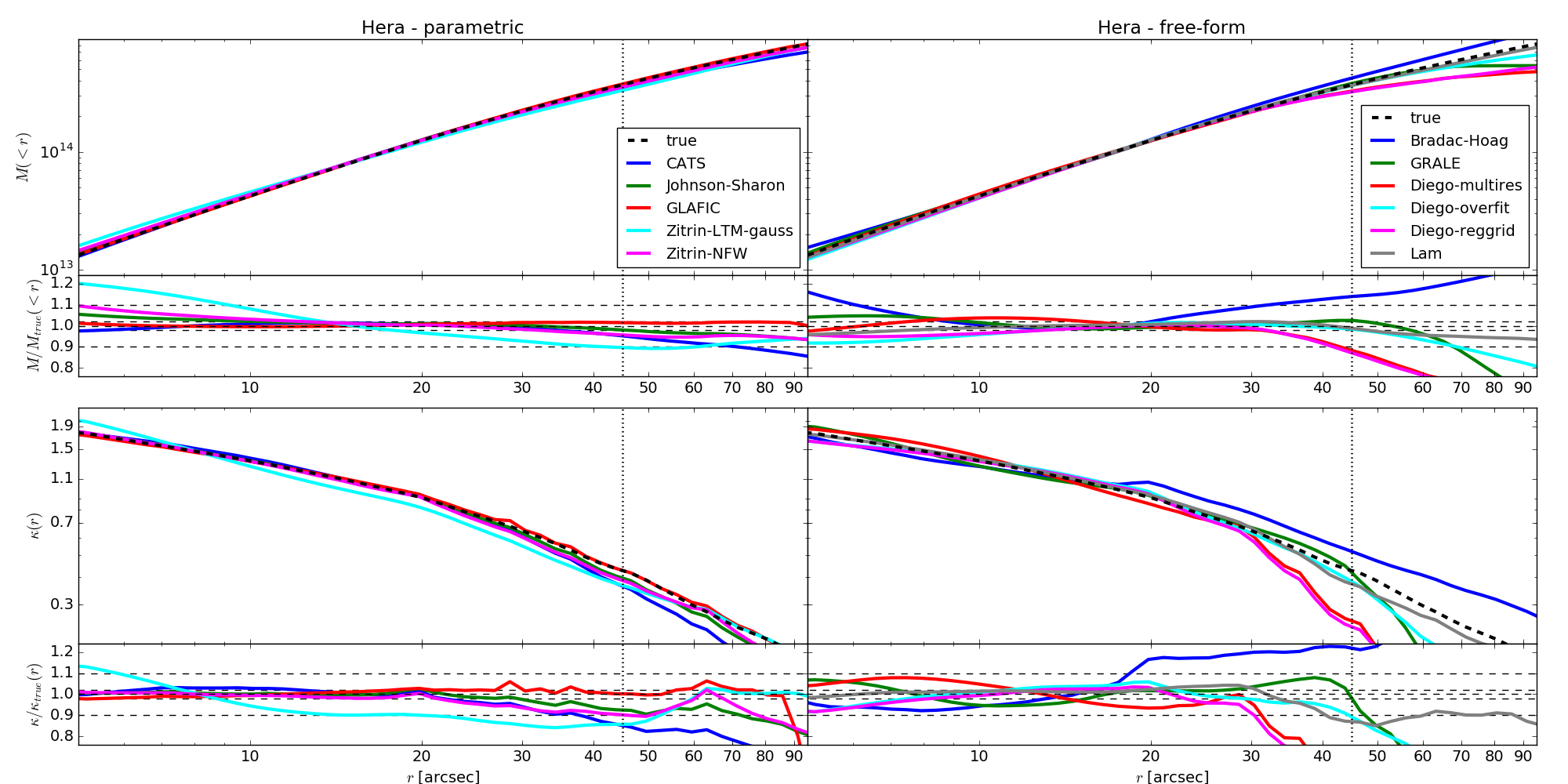}
  \caption{Mass profiles as in Figure \ref{fig:mass_profs_ares} but for {\em Hera}.}
 \label{fig:mass_profs_hera}
\end{figure*}

This is consistent with our findings. The upper panels of Figs.~\ref{fig:mass_profs_ares} and \ref{fig:mass_profs_hera} show the projected enclosed mass profiles of {\em Ares} and {\em Hera}, respectively. The bottom panels show the projected mass density profiles in units of convergence $\kappa$ for $z_S=9$. The profiles are computed with respect to the center of the most massive sub-clump in each cluster field. To facilitate the comparison between parametric and free-form methods, we show the results for these two classes of models separately (left and right panels). 

The mass distribution of {\em Ares} is generated in a very similar manner as used by the parametric techniques (except Zitrin-LTM) to model the lenses -- as a combination of parametrized mass components, including subhalos at the positions of cluster galaxies. Therefore, it is not surprising that these methods recover the true mass profile of {\em Ares} with very good accuracy. For example, the CATS, Johnson-Sharon, and GLAFIC profiles differ from the true mass profile by  $\lesssim \pm 2\%$. Larger differences are found for the Zitrin-LTM-gauss and the Zitrin-NFW approaches (perhaps because these are calculated on a lower resolution grid), see discussion in Section 3.8 and 3.9), but even for these models, in the region probed by strong lensing, the deviations from the true mass profiles are within $\sim \pm 10\%$.

It is noteworthy that neither the CATS nor the Johnson-Sharon teams used the NFW density profile to model the smooth DM halos of the two main mass components of Ares. On the contrary, they used cored isothermal profiles, which can of course be tweaked to match the lensing properties of NFW halos. This is consistent with the findings of \cite{SH08.2}, who showed that, in several cases, strong lensing clusters are equally well modeled with cored-isothermal and NFW density profiles. The additional constraints provided by complementary analysis, such as stellar-kinematics in the BCG could helping to break this degeneracy \citep{2013ApJ...765...24N}. Moreover, the adoption of an isothermal profile with core instead of the NFW profile does not prevent several models from recovering the correct slope of the surface mass density (i.e. convergence) profile over a relatively broad range of distances from the cluster center. The constraints available to carry out the reconstructions include both radial and tangential features, with the former particularly sensitive to the slope of the projected density profile.

Among the free-form methods, the reconstructed profiles generally deviate by $\lesssim 5-15\%$ from the true mass and convergence profiles. Some models (e.g. GRALE) have a very similar performance to parametric methods. The best agreement between the true profile and the models is found between $20"$ and $60"$ from the lens center, which nicely corresponds to the size of the Einstein radius, as shown in Fig.\ref{fig:einst_radii}. 

{\em Hera} is a less idealized test case for most of the parametric methods, but it still assumes light traces the mass substructure. So also for this lens, the parametric models reproduce the input mass profiles  more closely than the free-form methods, though the differences between the two approaches are now reduced. We find that the mass profiles obtained with the parametric methods  differ from the input mass profile by less than $10\%$ within $\sim 80$ arcsec from the assumed center. The same level of accuracy is reached by the free-form methods within $10\lesssim r \lesssim 30$ arcsec. This radial range corresponds to the size of the region probed by strong lensing. Both parametric and free form methods clearly converge to the true mass profiles within this range of distances, where the relative differences are of the order of few percent.

\subsection{Shape and orientation}

Having quantified the performance of the methods to reconstruct one-dimensional mass profiles, we discuss now their ability to recover the two-dimensional mass distributions of the lenses. 

\begin{figure}
 \centering
 \includegraphics[width=1.0\hsize]{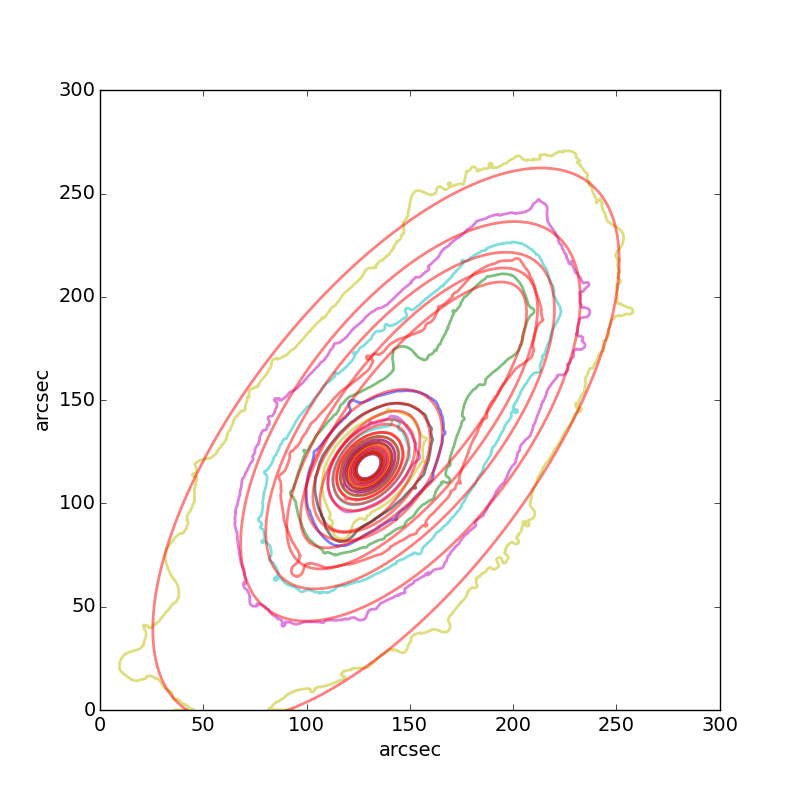}
  \caption{{\em Ares} mass iso-surface-density contours $\kappa =$ 0.5, 1.0, 1.5, 2.0, 2.5, 3.0 for $z_s = 9$ (jagged lines) and elliptical fits in red.}
 \label{fig:kappaell}
\end{figure}

To be more quantitative about how well the methods employed  recover the true shape and orientation of the two clusters, we consider the projected mass distributions of the lenses in terms of their iso-surface-density (or convergence, $\kappa$) contours. We use the following procedure:
\begin{enumerate}
\item From the convergence maps, we extract the contours corresponding to $\kappa$-levels in the range 0.5-3.0. Since both {\em Ares} and {\em Hera} have bi-modal mass distributions, we use the center of the largest mass clump as the reference center for this analysis and we consider only the contours enclosing it.
\item We fit an ellipse to each contour and measure its ellipticity and position angle. We also measure the size of each contour by means of an {\em equivalent radius} $r_\kappa$, defined as
\begin{equation}
r_\kappa=\sqrt{ab} \;,
\end{equation}
where $a$ and $b$ are the semi-axes of the best fitting ellipse. 
\item Finally, we draw the radial profiles of both the ellipticity and the position angle. The radius used to produce the profiles is the equivalent radius of the iso-density contours.
\end{enumerate}
The procedure outlined above is shown in Fig.~\ref{fig:kappaell} for the cluster {\em Ares}.  

\begin{figure*}
 \centering
 \includegraphics[width=1.0\hsize]{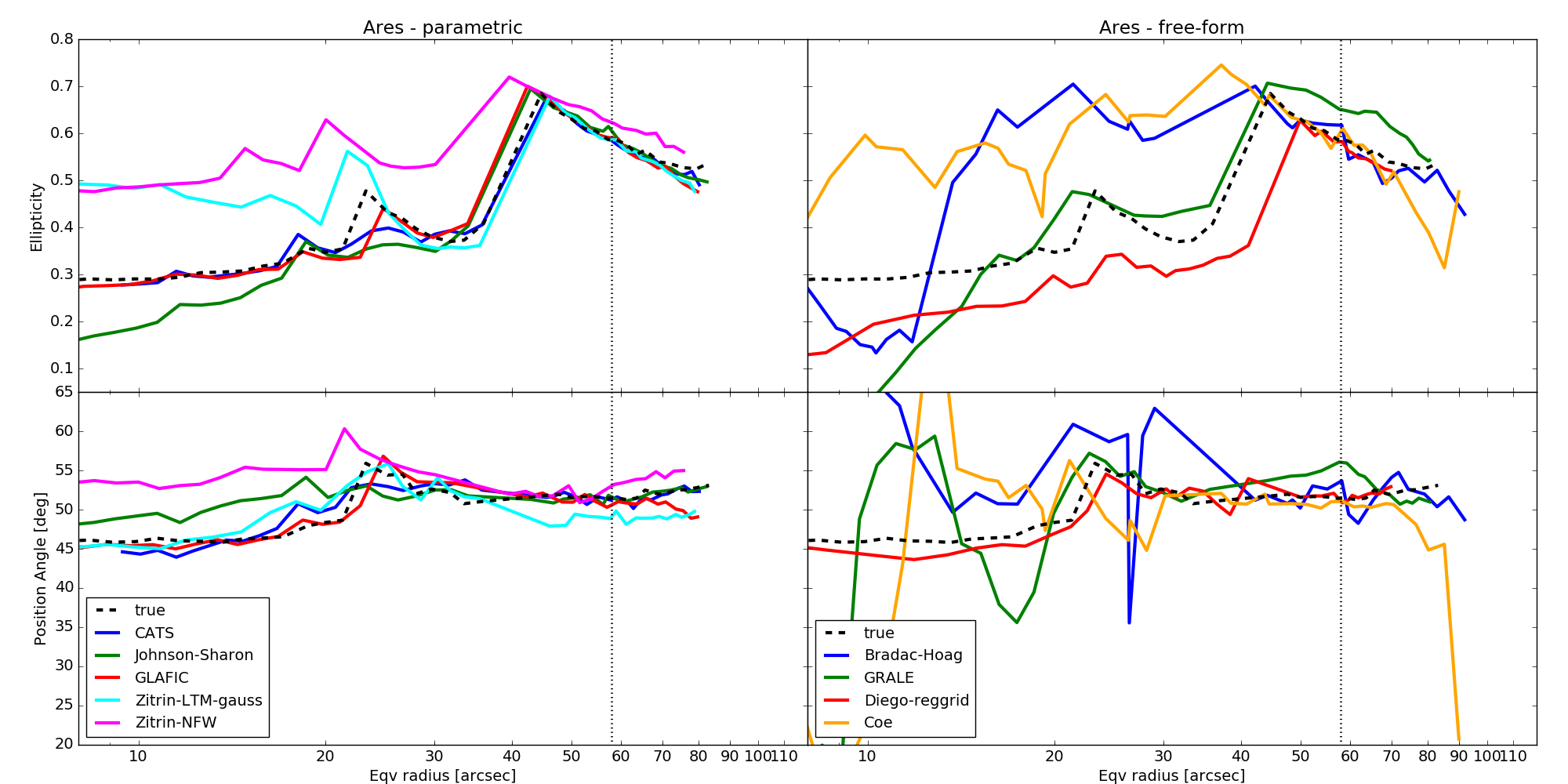}
  \caption{Ellipticity (upper panels) and position angle (lower panels) as a function of the equivalent radius of the convergence contours of {\em Ares}. In each plot we show the true profile as a dashed black line. The profiles obtained from the reconstructions are given by the solid colored lines. The left and the right panels refer to the parametric and to the free-form methods, respectively. The vertical dotted lines indicate the maximum radius covered by all reconstructions.}
 \label{fig:ares_shape_profs}
\end{figure*}

\begin{figure*}
 \centering
 \includegraphics[width=1.0\hsize]{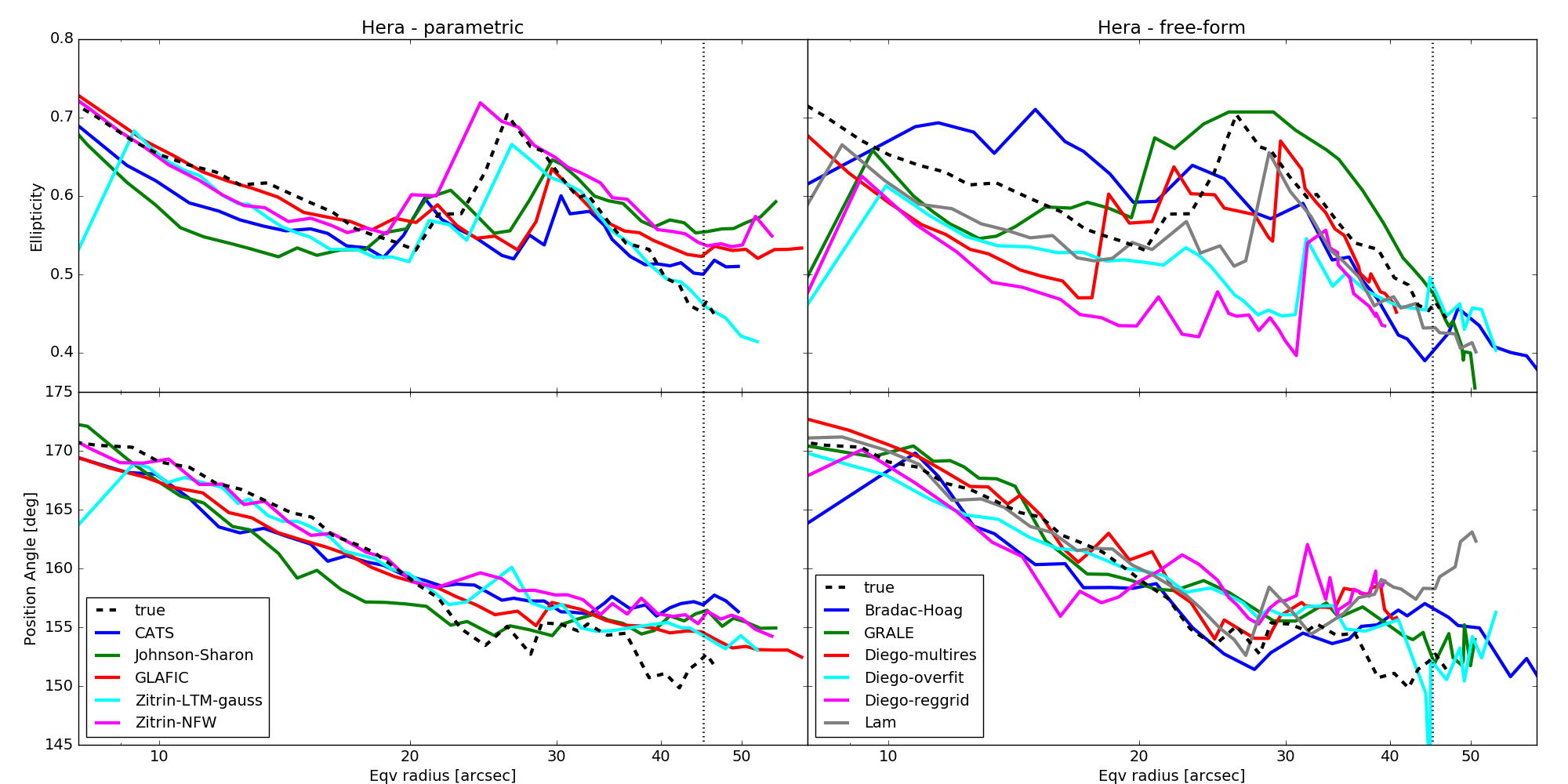} 
  \caption{Ellipticity (upper panels) and position angle (lower panels) profiles as in Figure \ref{fig:ares_shape_profs} but for {\em Hera}.}
 \label{fig:hera_shape_profs}
\end{figure*}

The radial profiles of the ellipticity and of the position angle for the two clusters are shown in Figs.~\ref{fig:ares_shape_profs} and \ref{fig:hera_shape_profs}. As done in Figs.~\ref{fig:mass_profs_ares} and  \ref{fig:mass_profs_hera}, the results for parametric and free-form methods are displayed separately (left and right panels, respectively). 

In each panel, the true profile is given by the black dashed line. 
The two clusters investigated in this work exhibit quite different ellipticity profiles. Indeed, due to the larger spatial separation between the two mass clumps, {\em Ares} has a less elongated inner core ($e=1-b/a \sim0.3$) compared to {\em Hera} ($e\sim0.7$).  {\em Ares}'s ellipticity increases with radius, while {\em Hera} shows the opposite trend. 

Despite the fact that we have introduced some modest radial variation of the ellipticity of two main mass clumps in {\em Ares}, the largest jumps in the ellipticity profile of this cluster are produced by massive substructures. These variations of ellipticity are generally well reproduced in the parametric reconstructions, and, to some extent, also in the free-form model of GRALE.  
Clearly, the parametric techniques produce better measurements of the core shapes, both in the cases of {\em Ares} and {\em Hera}. Indeed, due to resolution limits, the convergence maps produced by the free-form methods are noisier, resulting in more irregular iso-density contours. Under these circumstances, the ellipticity measurements are more uncertain. 

Among the parametric reconstructions of {\em Ares}, the largest deviations from the true ellipticity profile are found for the Zitrin-NFW and for the Zitrin-LTM-gauss models within $\sim 40"$ and $\sim 20"$, respectively. 
Interestingly, these same algorithms provide some of the most accurate measurements of the core shape in the case of {\em Hera}. These algorithms generally find higher halo ellipticities compared to the other parametric methods. Such behavior is consistent with the results of \cite{2015ApJ...801...44Z}, where the Zitrin-NFW and Zitrin-LTM-gauss methods are both employed in the  reconstruction of the galaxy clusters in the CLASH sample. As shown in their Fig.~3, the first of these two methods leads to more elliptical mass distributions. The most likely interpretation of this behavior is that external shear compensates the smaller ellipticity of the LTM models.

All parametric methods except the Zitrin-LTM-gauss tend to over-estimate the ellipticity of the mass distribution at large radii in the case of {\em Hera}. We shall recall that all these algorithms fit the data by combining multiple mass components, each of which has a fixed  ellipticity. The results show that, within the region probed by strong lensing ($\lesssim 40"$ for {\em Hera}), the combination of multiple mass clumps is effective in reproducing the overall ellipticity of the cluster. At larger radii, though, the models are unconstrained and the ellipticity is extrapolated from the inner region. Free-form methods do not show the same trend; their ellipticity profiles are more noisy. 

Also the orientation angles of the iso-density contours in the parametric reconstructions deviate from {\em Hera}'s true orientations at large radii. Being a numerically simulated cluster, {\em Hera} is characterized by asymmetries and twists of the iso-density contours that result to be much stronger than in {\em Ares}. For example, the position angle of the iso-density contours changes by $\sim 20$ degrees between the very inner region of the cluster and a distance of $\sim 50"$. 

As a result of the not perfectly reproduced shape and orientation of the cluster at large radii, the CATS, Johnson-Sharon, GLAFIC and Zitrin-NFW models have an excess of mass  along the major axis of the cluster with respect to the true mass distribution of {\em Hera} (and consequently they lack mass in the perpendicular direction). Such peculiarities can be seen in Fig.~\ref{fig:hera_kapparatios}, where the ratios between reconstructed and true convergence maps of {\em Hera} are shown.

\subsection{Substructure}

Figs.~\ref{fig:ares_kapparatios} and \ref{fig:hera_kapparatios} show that significant differences exist between the models near sub-structures. Measuring the mass of substructures is an important task that several authors have performed via strong lensing \citep[see e.g.][ and references therein]{2007MNRAS.376..180N, 2009ApJ...693..970N, 2015ApJ...800...38G}. Therefore it is interesting to quantify the lens model precision near these secondary mass clumps.

From the perspective of strong lensing, substructures are often identified as massive halos around cluster galaxies. This is particularly true for parametric methods: they use the luminous galaxies as tracers of the underlying mass distribution. Instead, free-form methods can in principle detect any kind of mass substructure, even if not traced by light. However, they cannot distinguish between the projected mass belonging to the cluster halo and bound to the substructures.

Indeed, as part of their submissions, the groups did not provide estimates of the masses in substructures, nor substructure catalogs. Here, we perform the following analysis:
\begin{itemize}
\item We start from the assumption that galaxies trace the substructures. This is not a strong assumption given the method employed to generate the galaxy populations of  {\em Ares} and {\em Hera}. In both cases, galaxies tend indeed to coincide with dark matter substructures. In the case of {\em Ares} there is a one-to-one correspondence between luminous galaxies and  dark matter sub-halos. In the case of {\em Hera}, we have excluded from the image simulations those galaxies which had their dark matter halos stripped off in the course of the cluster evolution.
\item We create apertures centered on the cluster galaxies with $m_{AB,F814W}<24$, with radii equal to twice the effective radius of the  galaxies, and we measure the projected mass within each aperture from both the reconstructed and  the true convergence maps.
\item In the following, we will refer to these masses as substructure masses, keeping in mind that these are however the sum of the substructure mass and of the projected mass of the underlying cluster dark-matter halo.
\end{itemize}

\begin{figure*}
 \centering
 \includegraphics[width=1.0\hsize]{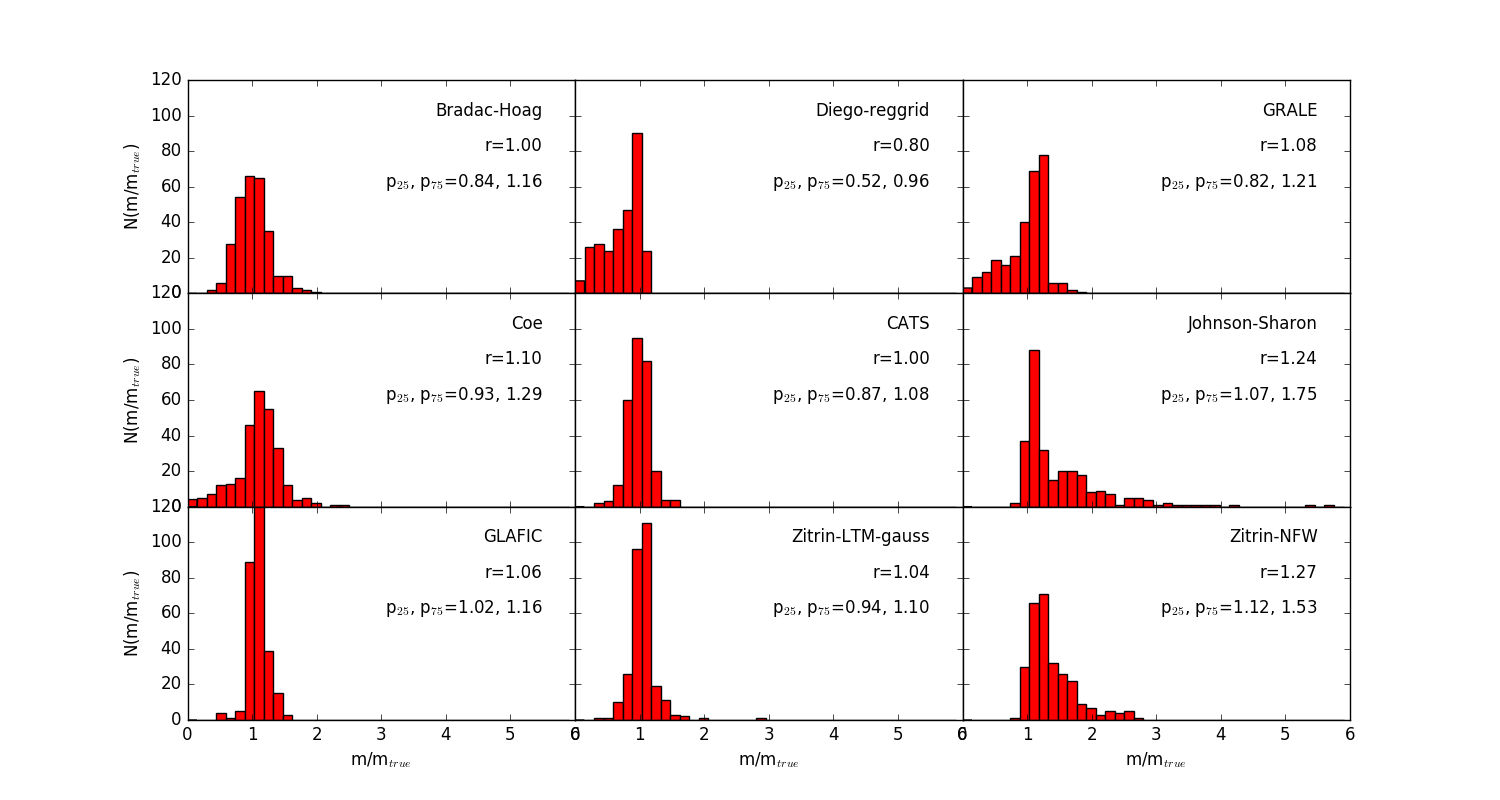}
  \caption{Distributions of the ratios between measured and true substructures masses for {\em Ares}. The total number of substructures is 282. Each panel shows the results for a mass model. In each panel, we indicate the median $r$ and the $25-th$ and  $75-th$ percentiles of the distribution.}
 \label{fig:ares_subs}
\end{figure*}

\begin{figure*}
 \centering
 \includegraphics[width=1.0\hsize]{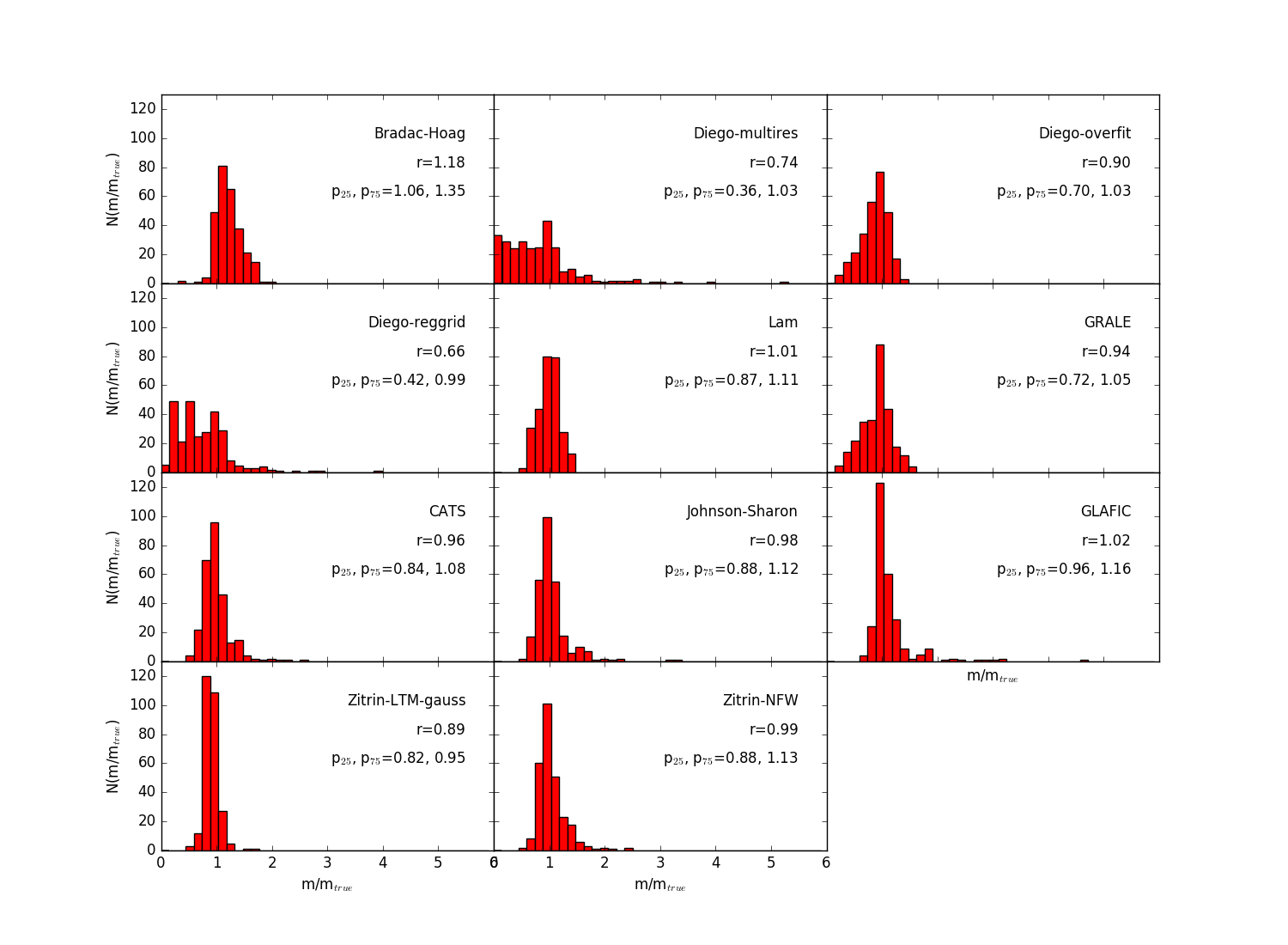}
  \caption{Distributions of the ratios between measured and true substructures masses for {\em Hera}. The total number of substructures is 282. Each panel shows the results for a mass model. In each panel, we indicate the median $r$ and the $25-th$ and  $75-th$ percentiles of the distribution.}
 \label{fig:hera_subs}
\end{figure*}

In Figs.~\ref{fig:ares_subs} and \ref{fig:hera_subs}, we show the distributions of the ratios between measured and true substructure masses. The two figures refer to {\em Ares} and {\em Hera}, respectively and show the results for all the models. We characterize the distributions of the ratios $r$ by means of their median $r$ and of their $25-th$ and  $75-th$ percentiles, $p_{25}$ and $p_{75}$. The analysis is carried out on the same areas covered by the maps in Figs.~\ref{fig:ares_kappamaps} and \ref{fig:hera_kappamaps}. Therefore, the same number of substructures have been used to build the histograms ($282$ and $278$ for {\em Ares} and {\em Hera}, respectively).

The results found for {\em Ares} show that several methods recover nearly unbiased  substructure masses with  good accuracy. For example, the inter-percentile range found for the CATS model is only 0.21 and the median is $r=1$. Similar results are found for the Zitrin-LTM-gauss model, although with a median slightly larger than unity. Some parametric models, such as those of Johnson-Sharon and Zitrin-NFW, and marginally GLAFIC, have skewed distributions with tails extending towards ratios larger than unity. Interestingly, Johnson-Sharon's model is based on the same modeling software employed by the CATS group.

Among the free-form models, the distributions are generally broader than for the parametric methods. The distribution for the Bradac-Hoag model  has median $r=1$ and inter-percentile range $0.32$. Similar or slightly larger scatter is found for the GRALE and Coe models. The ratio distribution obtained for the Diego-reggrid model has a tail extending towards small values and its median is $r=0.8$.

The results found for {\em Hera} are quite in agreement with those found for {\em Ares}. Parametric methods perform very similarly among each other, providing  mass measurements accurate at the level of few percent. The Zitrin-LTM-gauss model has a median $r=0.89$. The dispersions of the ratio distributions, as quantified by the inter-percentile ranges, are $\sim 0.2-0.25$. This is quite remarkable given the very different methods used to populate {\em Ares} and {\em Hera} with substructures and the significant differences between the density profiles of the substructures themselves in the two simulations, as shown in Fig.~\ref{fig:convmaps}. This seems to indicate that the methods are flexible enough to account for even large variations in the substructure properties, provided they are traced by light. It is less surprising that the flexible free-form methods also behave so similarly in {\em Hera} and  {\em Ares}.

\subsection{Magnification maps}

\begin{figure*}
 \centering
 \includegraphics[width=1.0\hsize]{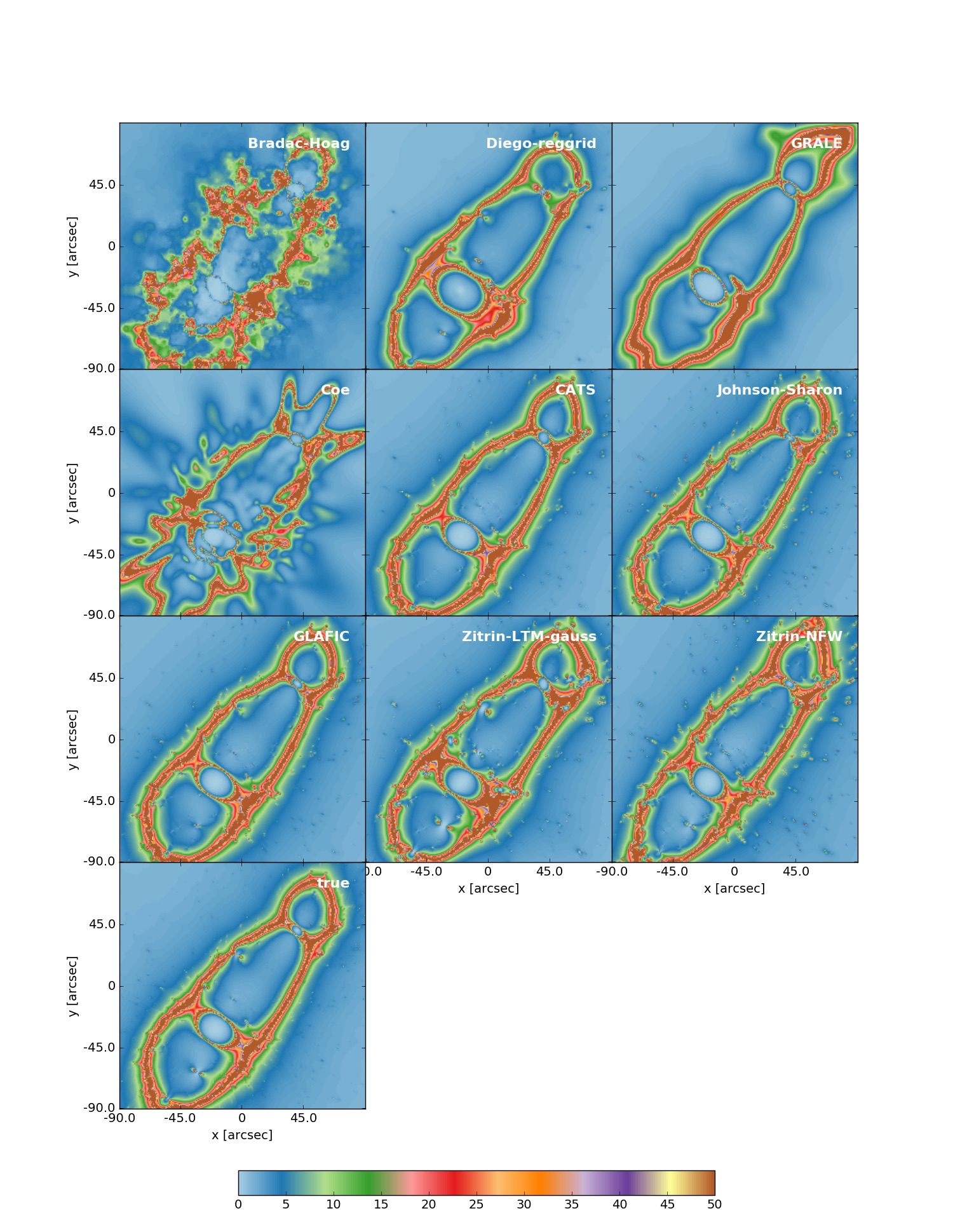}
  \caption{Magnification maps for sources at $z_s = 9$ lensed by each {\em Ares} reconstruction and, at bottom, the true simulated cluster.}
 \label{fig:ares_mumaps}
\end{figure*}

\begin{figure*}
 \centering
 \includegraphics[width=1.0\hsize]{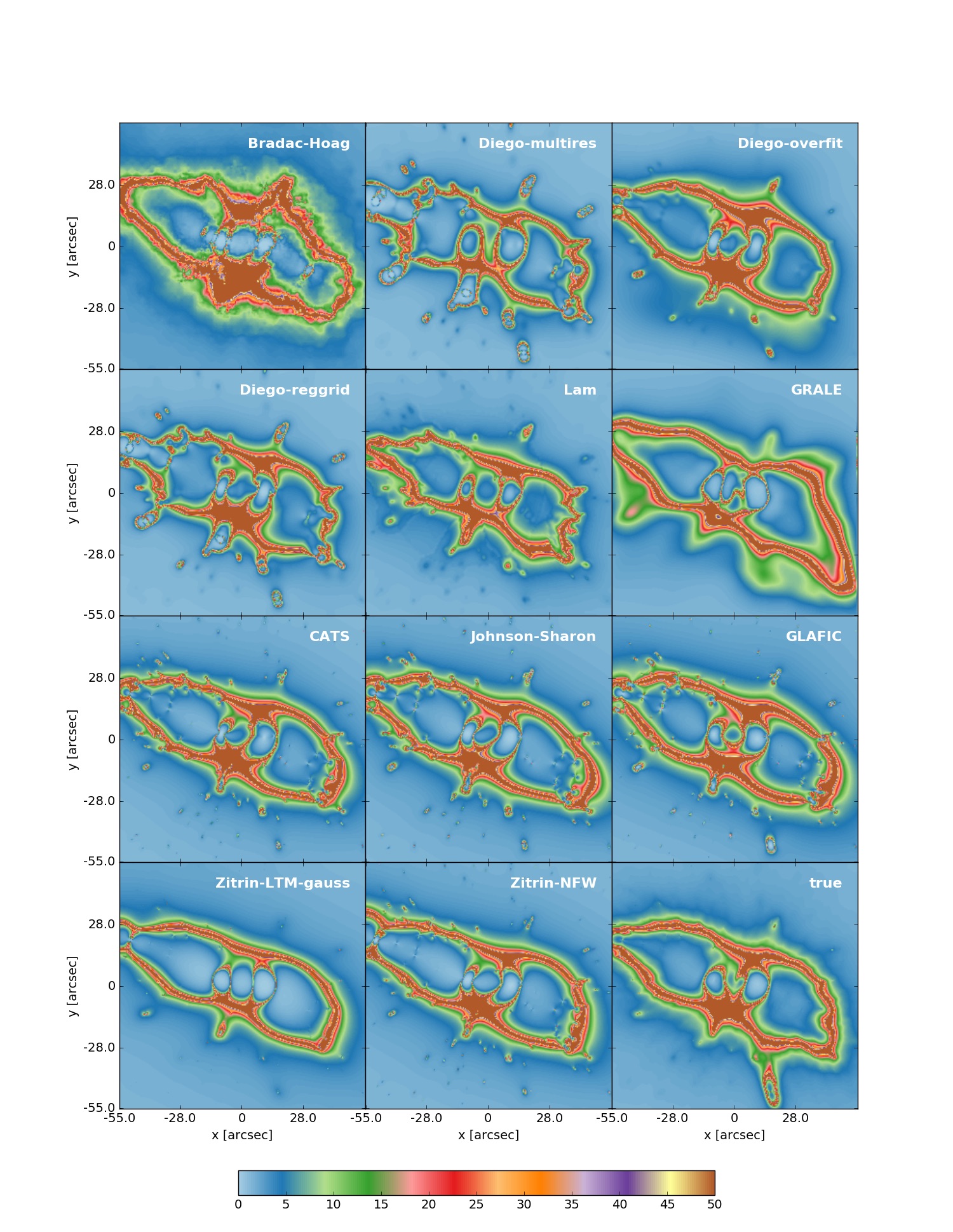}
  \caption{Magnification maps for sources at $z_s = 9$ lensed by each {\em Hera} reconstruction and the true simulated cluster (bottom right).}
 \label{fig:hera_mumaps}
\end{figure*}

As one of the major goals of the Hubble Frontier Fields is to use the lensing power of galaxy clusters to detect and characterize very high redshift galaxies, we focus now on the magnification.  Of course, the results shown in this section are not independent of those discussed earlier, since the convergence is one of the two quantities entering the definition of magnification. The other quantity is the shear, which was not discussed so far. 

\begin{figure*}
 \centering
 \includegraphics[width=1.0\hsize]{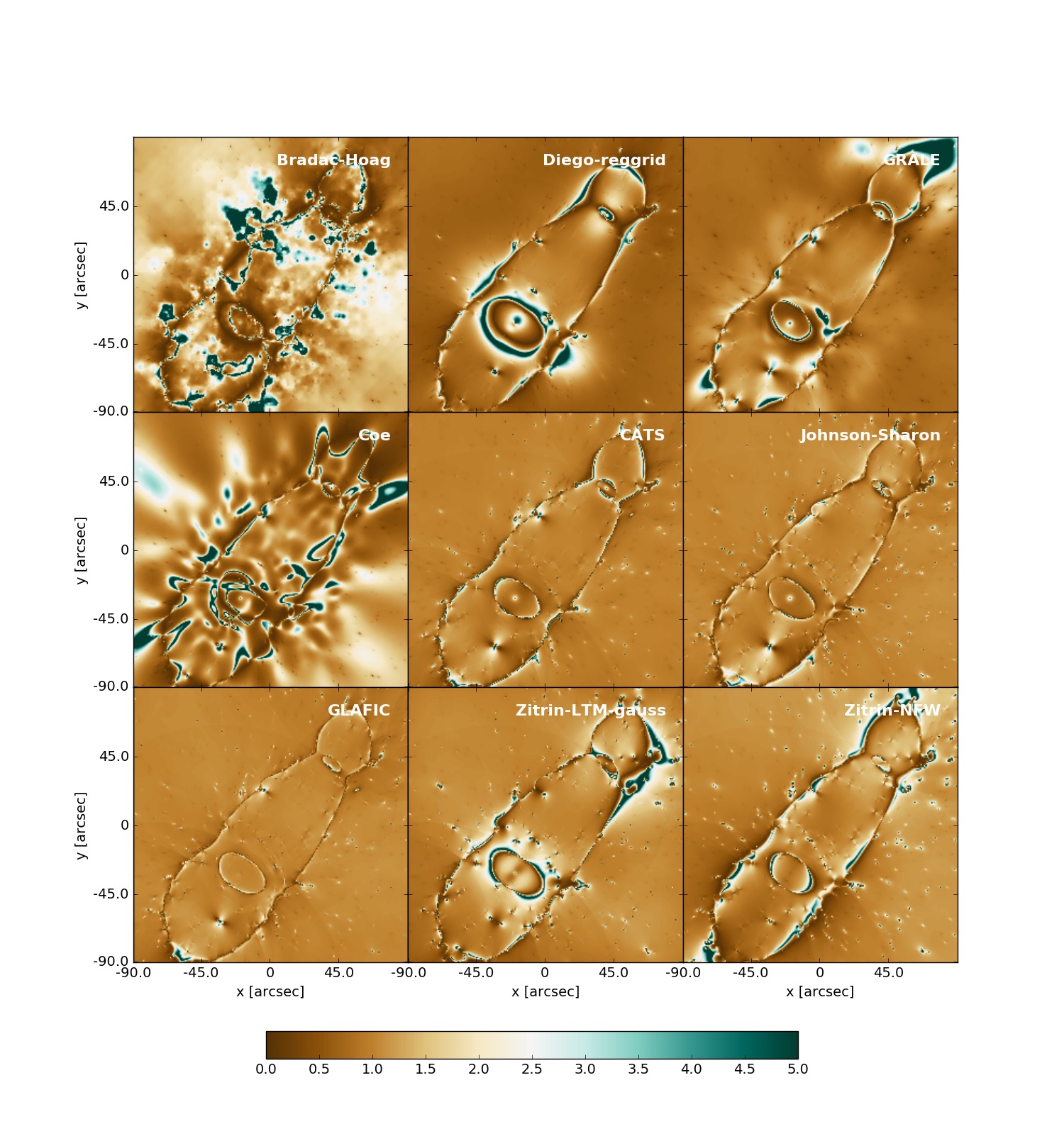}
  \caption{Ratios of model divided by true magnifications for  {\em Ares} .}
 \label{fig:ares_muratios}
\end{figure*}

\begin{figure*}
 \centering
 \includegraphics[width=1.0\hsize]{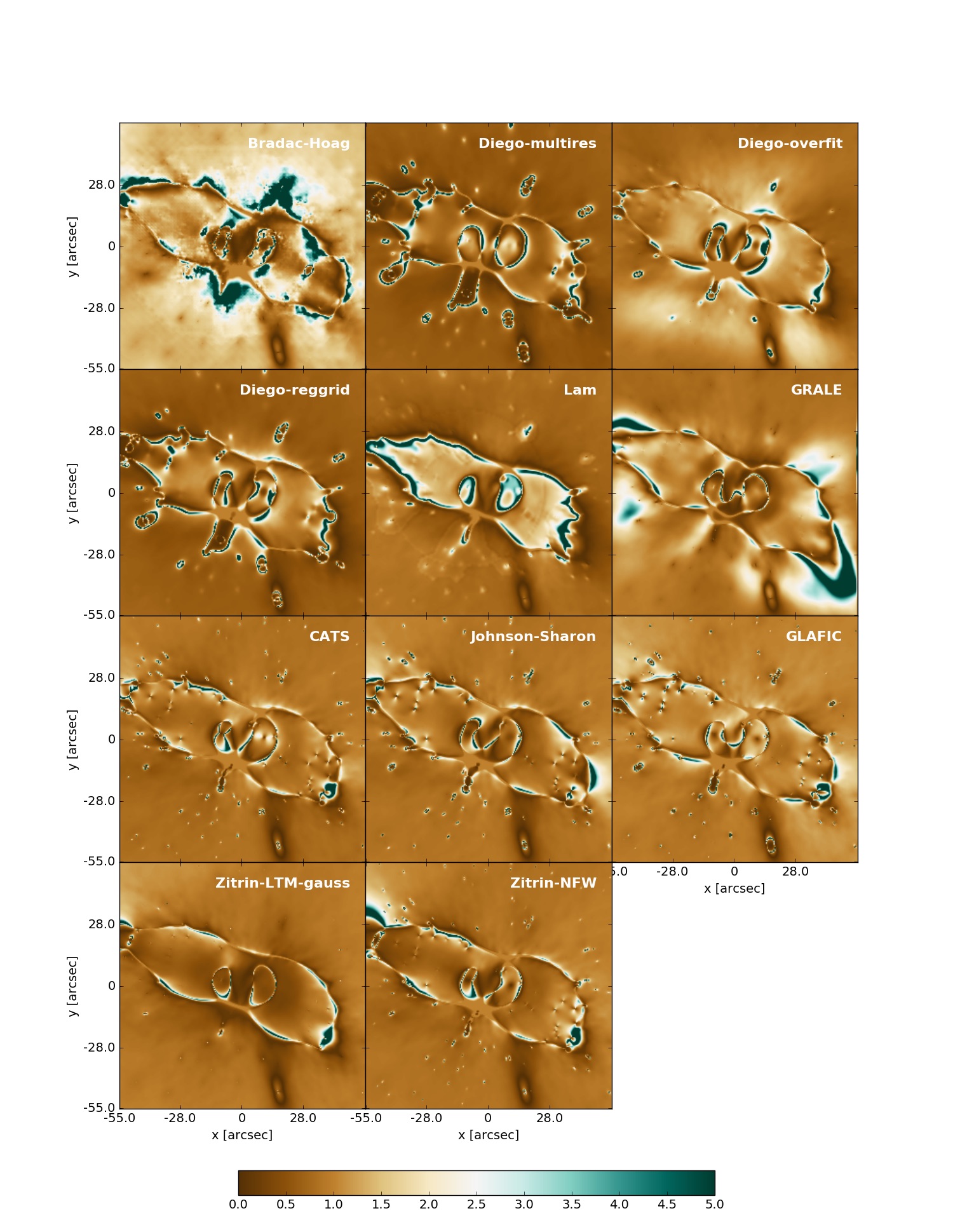}
  \caption{Ratios of model divided by true magnifications for {\em Hera} .}
 \label{fig:hera_muratios}
\end{figure*}

In Figs.~\ref{fig:ares_mumaps} and \ref{fig:hera_mumaps}, we show the magnification maps for $z_S=9$ obtained for {\em Ares} and {\em Hera}. As done previously, the results for each model are displayed in different panels. The last panel on the bottom shows the true magnification.  The ratios between each reconstructed magnification map and the true magnification maps are shown in  Figs.~\ref{fig:ares_muratios} and \ref{fig:hera_muratios}. 

The largest discrepancies between reconstructed and true magnifications appear around the lens critical lines. These are the loci where the magnification formally diverges. Therefore,  even a small misalignment of the true and reconstructed critical lines will result in  potentially large magnification differences. Most of the models recover the shape and the size of the critical lines well. Others, as the Bradac-Hoag, the Coe, and the Diego-multires models are characterized by critical lines with very irregular shapes. 

\begin{figure*}
 \centering
 \includegraphics[width=1.0\hsize]{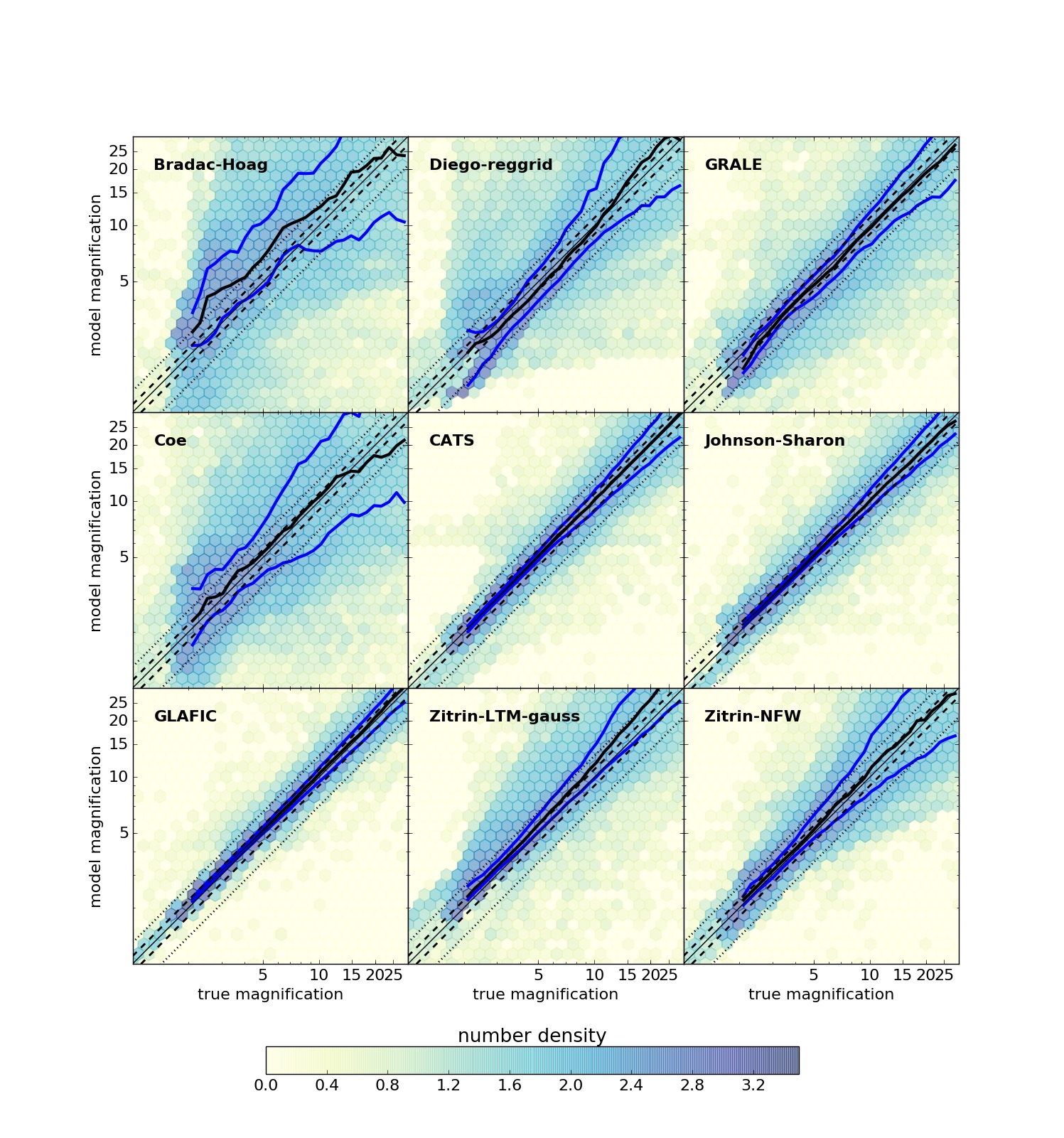}
  \caption{Model vs.~true magnifications ($z_s=9$) for  {\em Ares} .  The underlying 2D histograms show the distributions of the pixel values on the $\mu-\mu_{true}$ plane after sampling the magnification maps on a grid of $256\times256$ pixels. The black and the blue solid lines show the median and the $25-th$ and $75-th$ percentiles of the measured magnifications in bins of $\mu_{true}$. The dashed and the dotted lines parallel to the diagonal in each panel denote the limits of $\pm 10\%$ and $\pm 30\%$ deviations from the relation $\mu=\mu_{true}$.}
 \label{fig:ares_muvsmu}
\end{figure*}

\begin{figure*}
 \centering
 \includegraphics[width=1.0\hsize]{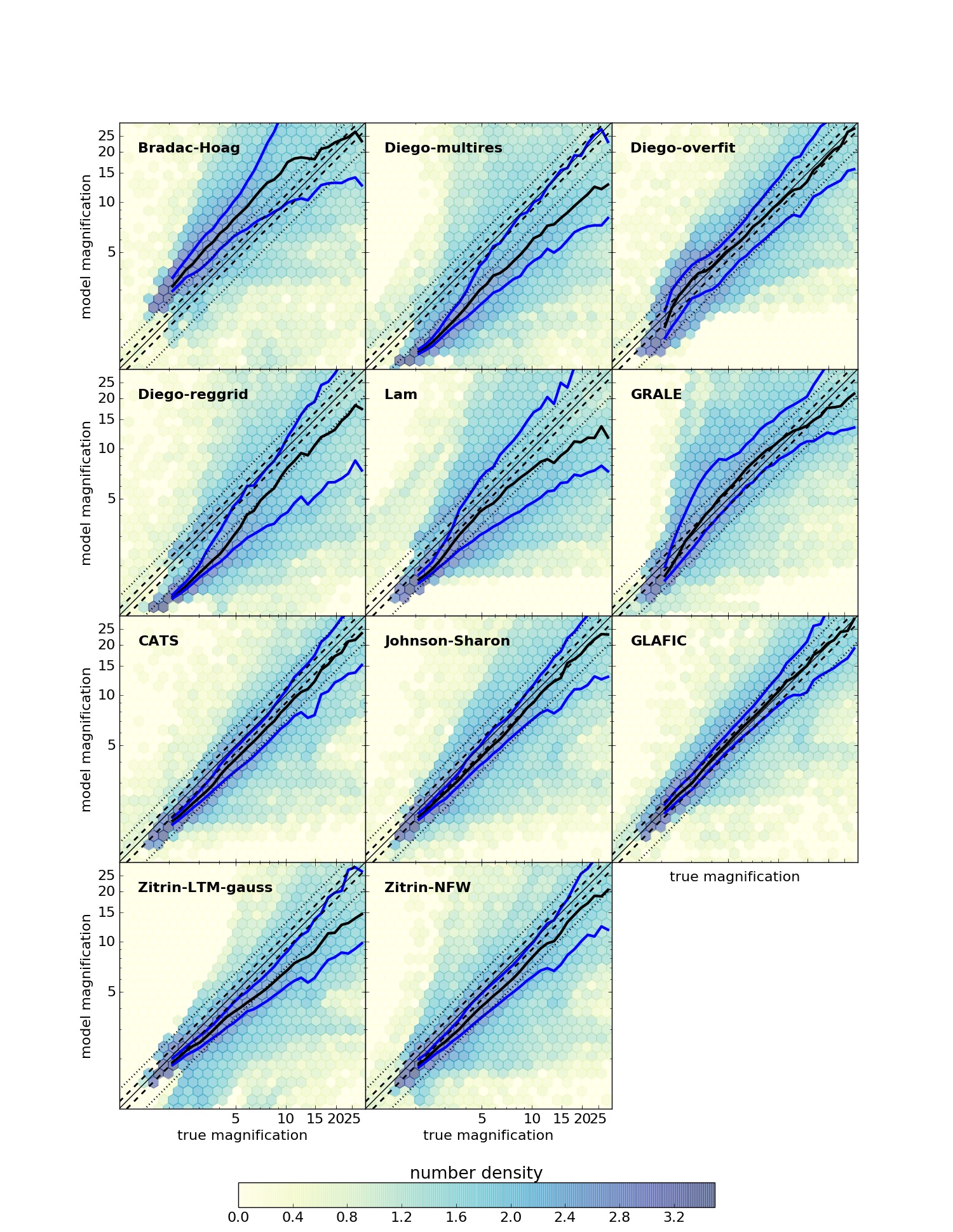}
  \caption{Model vs.~true magnifications ($z_s=9$) for {\em Hera} , as in Fig.~\ref{fig:ares_muvsmu}.}
 \label{fig:hera_muvsmu}
\end{figure*}

In Figs~\ref{fig:ares_muvsmu} and \ref{fig:hera_muvsmu}, the measured magnifications are plotted as a function of the true magnifications.  As anticipated, the scatter around the median increases as a function of the true magnification for all models.  The scatters for parametrically reconstructed models of {\em Hera} are factors of $2-3$ larger than for the corresponding models of {\em Ares}, the mock cluster that was generated parametrically. Besides, we note that {\em Hera} was inherently less well constrained as the cluster had fewer multiple images than {\em Ares}. So a slightly lower fidelity in the reconstruction was anticipated and found as expected.

\begin{figure}
 \centering
 \includegraphics[width=1.0\hsize]{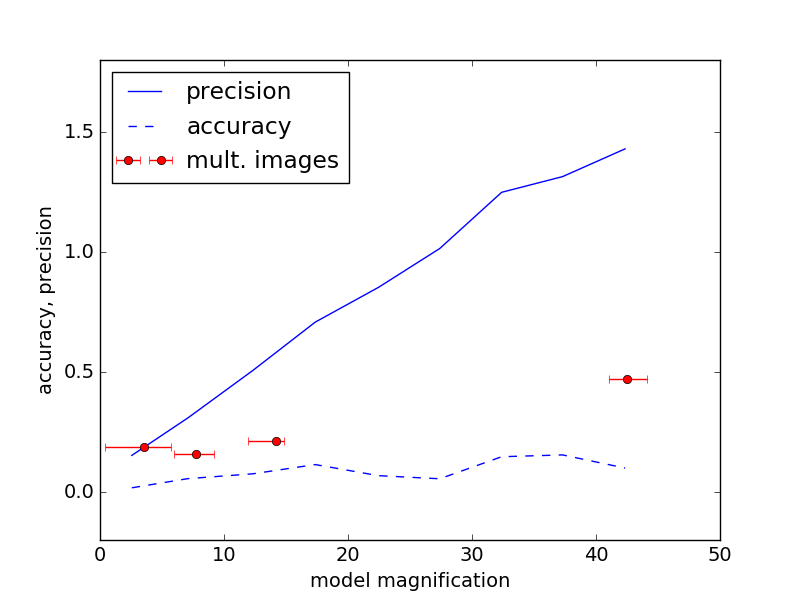}
  \caption{Magnification accuracy (dashed line) and precision (solid line) as a function of the magnification from the strong lensing constraints for the GLAFIC reconstruction of {\em Hera}. The precision is quantified by the difference between the 75-th and 25-th percentiles of the distribution of  $\mu-\mu_{true}$, sampled on a $256\times 256$  pixel grid. The accuracy is given by the median of  $\mu-\mu_{true}$. The red points show the uncertainties of the magnification measurements at the  location of the multiple images. The horizontal error-bars indicate the sizes of the  magnification bins used to estimate the precision.}
 \label{fig:mag_unc}
\end{figure}

In the best scenario obtained for {\em Hera} (i.e. the GLAFIC model, see also Fig.~\ref{fig:mag_unc}), we find very high accuracy (a few percent bias at most) and precision:  $\sim 10\%$ uncertainty for $\mu=3$, growing to $\sim 30\%$ at $\mu=10$, and increasing further at higher magnifications. 

In other cases, median magnifications are biased low or high by as much as $\sim 40-50\%$.  
Some of these biases are due to the models' inability to reproduce the correct magnification patterns interior to the tangential critical lines.  
In other cases, the gradient of the magnification around the critical lines is significantly different from that in the true magnification maps, reflecting the incorrect shape and orientation of the projected mass distribution or  the incorrect slope of the convergence profile.

Regions around substructures sometimes are characterized by large  uncertainties on magnification estimates. For example, the large substructure located south of the cluster {\em Hera} is not well constrained by any of the models, which all systematically underestimate the magnification around it. As shown in the  lower central panel of Fig.~\ref{fig:simobs}, there are no multiple images located near this substructure, which may explain why no model is able to constrain its mass properly. 

In Fig.~\ref{fig:mag_unc}, we compare  the precision of the magnification measurements over the whole map (in the case of the reconstruction provided  by the GLAFIC team for {\em Hera}; solid line) and at the observed positions of the multiple images used to build the lens model (red dots). The figure shows that the precision achieved by the model at the location of the constraints is indeed higher than in other regions with similar magnifications.  The horizontal error-bars indicate the sizes of the magnification bins used to estimate the precision of the magnification measurements.

\section{Reconstruction metrics}
\label{sect:reconmetrics}

In order to be more quantitative in estimating the ability of the different methods employed in this work to measure several relevant properties of {\em Ares} and {\em Hera}, we have defined metrics for the lens properties discussed above. More precisely, we introduce metrics for the one-dimensional radial profiles of 
\begin{itemize}
\item the 2D projected mass enclosed within radius $R$,
\item the surface mass density, or convergence $\kappa(R)$,
\item the ellipticity, as fit to iso-density contours,
\item and the orientation, as given by the position angle of the convergence contours.
\end{itemize}
We also define metrics to quantify the goodness of the reconstruction of the 2D convergence and magnification maps. 
Finally, we define a metric for the projected subhalo masses in apertures centered on the cluster galaxies.

Thus, we have seven metrics that can be used for a more quantitative comparison between the lens models of both clusters.  We can also evaluate how the performance of each algorithm changes when switching from a simulation based on a lens obtained from semi-analytic methods ( {\em Ares} ) and one obtained from a fully numerical simulation ({\em Hera}) . 

The metrics are defined as follows. Given a set of measured values $v$ and a set of true values $v_{true}$, we derive the distribution of $v/v_{true}$. Then, we compute the median, $\zeta$ and the $25$-th and $75$-th percentiles of the distribution, $p_{25}$ and $p_{75}$. The metric is finally defined as 
\begin{equation}
Q_v=\log_{10}\{[(p_{75}-p_{25})|\zeta-1|]^{-1}\} \;.
\end{equation}
By adopting this definition, the metric penalizes those reconstructions which are biased and/or affected by a large scatter.

\begin{figure*}
 \centering
 \includegraphics[width=1.0\hsize]{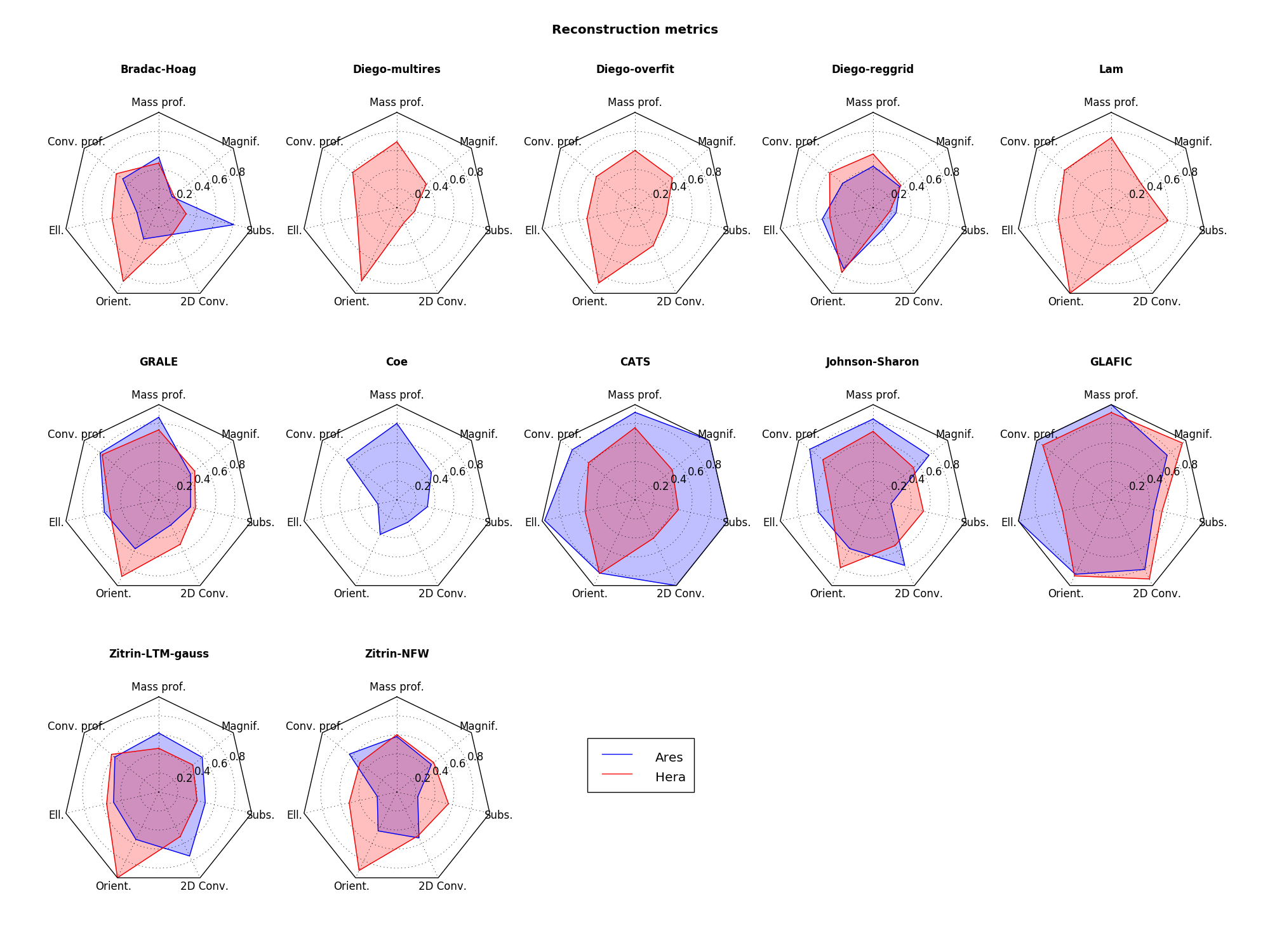}
  \caption{Radar plot showing the scores of each model for all metrics discussed in the paper.  Larger polygons correspond to better overall performance.  Each chart corresponds to a different lens model (see labels on the top) and shows results for both {\em Ares} (blue) and {\em Hera} (red), or whichever is available. The seven metrics are shown on the vertices of each chart. For each metric, the scores range from 0 (worst; plotted at the center of the chart) to 1 (best; plotted at the vertex), normalized to the maximum value recorded by all models.  A filled polygon is obtained by connecting the plotted scores of all metrics for each reconstruction.}
 \label{fig:metrics}
\end{figure*}

Of course, the metrics are not fully independent. For example, a model which is able to reproduce the convergence profile of the lens with a good accuracy will also provide a robust measurement of the mass profile. Similarly, models whose reconstructed convergence maps show little deviation from the true convergence maps will also have provide a good match with the simulation in terms of converge profile or shape (ellipticity and position angle). Nevertheless, the ranking among the models with respect to correlated lens properties is not always the same. For example, the Johnson-Sharon reconstruction of {\em Hera} ranks second in terms of convergence profiles and fourth in terms of mass profiles. In addition, the different lens properties which are discussed here are often used individually, and it may be interesting for the reader to establish which modeling technique is better suited to their scientific purposes. 

In Fig.~\ref{fig:metrics}, we show radar plots which summarize the metric values recorded by each reconstruction.  The overall performance of each model corresponds to the area of each polygon.  When one model is good at measuring some of the lens properties, but less effective at capturing others, the polygon appears elongated towards one or more of the chart vertexes.

The first eight charts correspond to free-form or hybrid methods. The remaining five charts refer to parametric techniques. As we have pointed out several times earlier, there is larger discrepancy between the performances of parametric and non-parametric methods in the case of {\em Ares} than in the case of {\em Hera}. This leads us to the conclusion that, despite our attempts to make the {\em Ares} mass distribution less ideal for parametric methods (e.g. by simulating adiabatic contraction or by introducing some twist of the iso-density contours, including some radial dependence), the simple fact that this cluster is assembled by combining mass components traced by the cluster galaxies, consistently with the basic assumptions of most parametric techniques, gives a huge advantage to these methods. The good news, in this case, is the following. First, these algorithms work as they are supposed to. Second, they provide very accurate reconstructions even if the parametrisation chosen for the lens halo density profile is not fully consistent with the true profile of the lens. For example, none of the parametric techniques, except the Zitrin-NFW method, used the NFW profile for fitting the smooth dark-matter halo components of {\em Ares}. Even so, models such as those submitted by the CATS, Johnson-Sharon, and GLAFIC teams produce an overall better fit to the input mass distribution compared to the Zitrin-NFW reconstruction. This suggests that pseudo-elliptical, cored halo models provide the right flexibility to account for most of the effects we have introduced in the simulation, such as the adiabatic contraction, which steepens the density profile in the central region of the cluster. Alternatively, these results may be interpreted as evidence for a lack of sensitivity of lensing alone to the precise share of the halo density profiles, being mostly sensitive to the mass enclosed within the Einstein radius rather than the slope of the density profile. Another possible cause may be that the Zitrin's models are calculated on a low resolution grid and perhaps  their accuracy is are limited by this resolution compared to higher resolution or completely analytic parametrizations.

When switching to a fully numerical simulation, the differences between parametric and free-form methods become weaker. At least for some of the metrics, some free-form / hybrid reconstructions of {\em Hera} (see e.g. the GRALE or Lam models) appear to be as good as the best parametric reconstructions of this cluster. This indicates that several parametric methods still cannot fully account for deviations of the mass distributions from a symmetric shape, which are, instead, more naturally captured by free-form methods. Asymmetries could be mimicked by suitable combinations of substructures in parametric models. Indeed, a degeneracy exists between these two properties of the mass distribution. However, the number of constraints in these simulations is high enough that this degeneracy is partially broken, as shown by how well the mass is constrained around the cluster galaxies in at least some of the parametric reconstructions. 

The model provided by the CATS team for {\em Hera} has significantly smaller values of all metrics (except for the cluster orientation), compared to the model submitted by the same team for {\em Ares}. The metrics agree with those of other parametric reconstructions of the same cluster (e.g. Johnson-Sharon). On the contrary, the reconstructions provided by the GLAFIC team for the two clusters have quite consistently high metric values. One feature of GLAFIC, which was enabled in the reconstruction of {\em Hera}, is the inclusion of external shear and third-order multipoles of the mass distribution. Apparently, these additional ingredients have provided the GLAFIC model extra degrees of freedom to properly account for the asymmetric mass distribution of {\em Hera}.

The comparison between the metrics of parametric and free-form methods also shows that the latter techniques are generally less accurate in reconstructing the two-dimensional maps of convergence and magnification and in measuring the mass around substructures. In fact, the spatial resolution that can be achieved with these methods is generally lower. On the contrary, radial profiles of the convergence and of the enclosed mass are measured by several of the free-form methods employed in this experiment with accuracy comparable to parametric techniques.   

\section{Limitations of this test}

We would like to remark that the tests outlined in this paper suffer of some limitations. First of all, we make the assumption that the simulations reproduce the properties of real clusters. While  some methods (e.g. the free-form ones) do not care about the correlation between dark-matter and baryons, other methods strongly rely on the assumption that light-traces mass. Both {\em Ares} and {\em Hera} implement this property, which, at least in some cases, has been questioned by observations \citep{wang15,hoag15}. In particular, the results we report on substructures are sensitive to this assumption. In a recent paper, \cite{2016MNRAS.458..660H} have explored how assuming that light traces mass in strong gravitational models can lead to systematic errors in the predicted  positions of multiple images. They find that images can be shifted by up to $\sim 1"$, assuming physically motivated offsets between dark-matter and stars. They quote a $\sim 0.5"$ rms error in the position of the multiple images due to breaking the assumption that mass traces light. 
Note, however, that, to some extent, we introduced some misalignment between matter and light in both {\em Ares} and {\em Hera}, by assigning to the observed galaxies a shape and an orientation which is not correlated with the underlying dark matter distribution.

Other limitations regard some observable properties of the galaxies in the simulated observations (e.g. luminosities and sizes) and their correlation with their halo masses.  It is known that the SAMs are not fully consistent with observations in this respect \citep[see e.g.][]{2009MNRAS.397.1254G,2015MNRAS.453.2515A,2015MNRAS.447..636X,2015arXiv151204531H}, thus the standard scaling relations adopted by some parametric techniques to translate the light into the mass or the size of the host halo might not equally applicable to observations and simulations. 

\section{Summary and conclusions}
\label{sect:summary}
In this paper we used simulated observations of two synthetic galaxy clusters to evaluate the performance of several algorithms for mass reconstruction with strong lensing. Such algorithms are currently being used to deliver to the community the lens models for the six galaxy clusters being observed in the Frontier Fields programme of the Hubble Space Telescope. 

The two clusters used in this study were obtained using very different techniques. {\em Ares} was generated using the semi-analytical  code {\tt MOKA}. {\em Hera} is instead the output of a cosmological N-body simulation at high resolution. The observable properties of the cluster galaxies are modeled using HOD and SAM techniques in {\em Ares} and {\em Hera}, respectively. In both cases, the clusters have complex mass distributions, characterized by disturbed and bi-modal morphology, similar to those of the FFI clusters. 

We used the code {\tt SkyLens} to simulate HST observations of the two mock clusters with both the ACS and the WFC3-IR camera. We produced images in all photometric bands used in the FFI, calibrating the  exposure times such to reach the depth of the FFI  observations. These HST simulated data were distributed to several groups of lens modelers for a blind analysis, i.e. without unveiling the true mass distribution of the lenses, neither the method used to simulate them.  

The simulated observations include lensing effects on a realistic distribution of background galaxies. We identified many strongly lensed galaxies and built a catalog of multiple image systems, which was delivered together with the simulated observations. The catalogs also include the redshift of all the sources. 

We complemented the HST simulations with a simulated observation in the $R_c$ band with the Subaru telescope. The main purpose of this additional simulation was to allow the inclusion of weak-lensing constraints at larger distances from the cluster center than those probed by HST. Together with the image, we also distributed a shear catalog obtained by processing the Subaru simulation through a public KSB pipeline.

We received nine reconstructions of {\em Ares} and eleven reconstructions of {\em Hera}, submitted by ten different groups. Seven groups employed their techniques to reconstruct both clusters. The remainder groups reconstructed just one of the two clusters or submitted reconstructions based on different set-ups of their methods. This is the first time that such a large number of algorithms have been tested against known mass distributions. Similar to the spirit of our experiment, in the recent collaborative effort presented in \cite{2016ApJ...817...60T}, several of the  methods used  to reconstruct the galaxy cluster MACSJ1149.5+2223 and estimate the time-delays between the multiple images of the SN ``Refsdal" were compared.  The recent re-appearance of the SN, reported by \cite{2016ApJ...819L...8K}, enabled the blind test of various model predictions, which were found to be in very accurate for several reconstructions. In addition, \citep{2015ApJ...811...70R} compared  the magnification predictions from 17 mass models of Abell 2744 using a lensed supernova of type Ia.

The methods compared here include both parametric and free-form algorithms.  We have investigated how they perform at recovering several properties of the lenses, namely: the radial profiles of the convergence and of the enclosed mass, the mass in substructures, the maps of the convergence and of the magnification. For each of these properties, we defined a metric aimed at quantifying the performance of the method.

The key results of this phase of the comparison exercise of lens mapping methodologies can be summarized as follows.
\begin{itemize}
\item Parametric methods are generally better at capturing two-dimensional properties of the lens cores (shape, local values of the convergence and of the magnification). The free-form methods are as competitive as the parametric methods to measure convergence and mass profiles. It is worth mentioning, however, that, in both  {\em Ares} and the {\em Hera}, the cluster galaxies were good tracers of the cluster mass distributions.
\item The accuracy and precision of strong lensing methods to measure the mass within the Einstein radius (or more generally within the region probed by the strong lensing constraints) is very high. The measured profiles deviate from the true profiles by only a few percent at these scales. Of course, larger deviations are found at radii larger and smaller than the Einstein radius. The determination of the mass enclosed within the Einstein radius was extremely robust for all methods.
\item The largest uncertainties in the lens models are found near substructures and around the cluster critical lines. For some of the parametric models, the total mass around substructures (identified by cluster galaxies) is constrained with an accuracy of $\sim 10\%$. However, other methods have much larger scatter. Uncertainties on the magnification grow as a function the magnification itself and are therefore more pronounced near the cluster critical lines. For the best performing methods, the accuracy in the magnification estimate is $\sim 10\%$ at $\mu_{true}=3$ and degrades to $\sim 30\%$ at  $\mu_{true}=10$. 
\item Switching from {\em Ares} to {\em Hera}, i.e. from a purely parametric to a more realistic lens mass distribution, the gap between parametric and free form methods becomes smaller. Algorithms such as that used by the GLAFIC team, which include third order multi-poles in the lens mass distribution, have extra degrees of freedom which allow them to better reproduce asymmetries. These asymmetries, and possible variations of the halo ellipticity as a function of radius, seem to be the strongest limitations of parametric methods. The adoption of an hybrid approach, where parametric and free-form methods are combined also to describe the large-scale component of the clusters, could lead to a significant improvement of the mass reconstructions.
\item Some of the participating groups used the same code but adopted different set-ups to run them. For example, two groups (CATS and Johnson-Sharon) use the public code {\tt Lenstool} with slight modifications. Similarly, Diego submitted several models of {\em Hera} using WSLAP+, which is the same code used by Lam et al. Despite using the same algorithms and making use of the same inputs (i.e. families of multiple images and redshifts), the reconstructions obtained by these groups are different, indicating that some choices made by the modelers when ingesting the data and hence set up priors influence the results.
\end{itemize}

This is the first of a series of papers in which we address the issue of the accuracy of lens modeling. In a second paper, currently in preparation, we will discuss the results of the unblinded modeling of {\em Ares} and {\em Hera}. The feedback from the unblinding was used by modelers to not only tweak their best-fits to reach the best possible match to the input mass distributions of the lenses but to also incorporate and instigate improvements in their modeling procedure. This will provide information on the  accuracy limits achievable by each method and will also give further hints on the steps that need to be taken to optimize reconstructions.

Despite their complexity and the inclusion of several observational effects, the simulations used in this paper are still idealized in many respects. For example, the lenses are isolated and no additional lensing by matter along the line of sight is included. In addition, we alleviated the work of the lens modelers by identifying the strongly lensed sources and even providing redshifts for all of them. In the case of {\em Ares}, the number of available multiple images with known redshifts exceeds by a factor of $\sim 3-4$ what is available in any of the frontier fields (e.g. MACSJ0416). We will include the uncertainties due to possible mis-identification of multiple images and photometric redshifts as well as the noise added in by the intervening matter distribution along the line of sight in the next phase of this project in future work.

\vspace{0.3cm} 
\leftline{\bf Acknowledgments} 
We thank T. Treu for the helpful discussion.
MM acknowledges support from the Italian  Ministry of Foreign Affairs and International Cooperation, Directorate General for Country Promotion, from INAF via  PRIN-INAF 2014 1.05.01.94.02, and from ASI via contract ASI/INAF/I/023/12/0.  This work was supported in part by World Premier International
Research Center Initiative (WPI Initiative), MEXT, Japan, and
JSPS KAKENHI Grant Number 26800093 and 15H05892.
AZ is supported by NASA through Hubble Fellowship grant \#HST-HF2-51334.001-A awarded by STScI, which is operated by the Association of Universities for Research in Astronomy, Inc. under NASA contract NAS~5-26555.
J.M.D acknowledges support of the consolider project CSD2010-00064 and AYA2012-39475-C02-01 funded by the Ministerio de Economia y Competitividad, Spain. 
We acknowledge the lens modeling community for enthusiastically participating in this collaborative project to compare and contrast mass models.

\bibliographystyle{mnras}
\bibliography{master}

\end{document}